\documentclass{article}

\usepackage{arxiv}

\usepackage[utf8]{inputenc} 
\usepackage[T1]{fontenc}    
\usepackage{hyperref}       
\usepackage{url}            
\usepackage{booktabs}       
\usepackage{amsfonts}       
\usepackage{nicefrac}       
\usepackage{microtype}      
\usepackage{amsmath}  
\usepackage{graphicx}
\usepackage[square,sort,comma,numbers]{natbib}
\usepackage{doi}

\title{Optimizing illumination patterns for\\ classical ghost imaging}

\author{
    Andrew M.~Kingston\thanks{Corresponding author, email andrew.kingston@anu.edu.au.}~~and Alaleh~Aminzadeh\\
	Department of Materials Physics, Research School of Physics,\\
	The Australian National University, Canberra ACT 2601, Australia\\
	\And
	Lindon~Roberts\\
	School of Mathematics and Statistics, Carslaw Building, University of Sydney,\\
	Camperdown NSW 2006, Australia\\
	\And
	Daniele~Pelliccia\\
	Instruments and Data Tools Pty Ltd, PO Box 2114, Rowville VIC 3178, Australia\\
	\And
    Imants D.~Svalbe and David M.~Paganin\\
	School of Physics and Astronomy, Monash University, Clayton VIC 3800, Australia\\
}

\renewcommand{\shorttitle}{Optimizing illumination patterns}

\hypersetup{
pdftitle={Optimizing illumination patterns for classical ghost imaging},
pdfsubject={???}, 
pdfauthor={Kingston, Roberts, Aminzadeh, Pelliccia, Svalbe, Paganin},
pdfkeywords={Structured illumination, Computational imaging, Single-pixel imaging},
}

\begin{document}
\maketitle

\begin{abstract}
	Classical ghost imaging is a new paradigm in imaging where the image of an object is not measured directly with a pixelated detector. Rather, the object is subject to a set of illumination patterns and the total interaction of the object, e.g., reflected or transmitted photons or particles, is measured for each pattern with a single-pixel or bucket detector. An image of the object is then computed through the correlation of each pattern and the corresponding bucket value. Assuming no prior knowledge of the object, the set of patterns used to compute the ghost image dictates the image quality. In the visible-light regime, programmable spatial light modulators can generate the illumination patterns. In many other regimes---such as x rays, electrons, and neutrons---no such dynamically configurable modulators exist, and patterns are commonly produced by employing a transversely-translated mask. In this paper we explore some of the properties of masks or speckle that should be considered to maximize ghost-image quality, given a certain experimental classical ghost-imaging setup employing a transversely-displaced but otherwise non-configurable mask.
\end{abstract}


\section{Introduction}


Classical ghost imaging is a computational imaging technique that can have advantages over conventional imaging in terms of signal-to-noise ratio (e.g., \cite{lane2020advantages}), achievable spatial resolution (e.g., \cite{stantchev2016noninvasive, kingston2020neutron}), and minimal imaging dose (e.g., \cite{katz2009compressive, gureyev2018}). The term arises from the quantum-optics origin of the technique \cite{Klyshko1988, Belinskii1994, Pittman1995optical, Strekalov1995} that was thought to originally rely on the {\it spooky} action at a distance of entangled photons. Subsequently it was realized that the only property of the entangled photons required is correlation \cite{bennink2002two, erkmen2008unified}, thereby enabling a classical variant of ghost imaging to be conceived \cite{Erkmen2010}. The experiment setup for classical ghost imaging and techniques for image computation from measurements are detailed in Sec.~\ref{sec:background}. For an overview of ghost imaging (GI), both quantum and classical, see Ref.~\cite{padgett2017introduction}.

Classical GI is based on structured illumination, and the set of illumination patterns employed dictates the properties of the recovered image in terms of resolution, contrast, and illumination dose required. Thus the choice of illumination patterns is an important consideration, but difficult since how these properties are influenced by the illumination patterns is not necessarily intuitive. In GI with visible light, there are many readily available methods to generate arbitrary illumination patterns. These include conventional projectors \cite{DataProjectorReference}, spatial light modulators (SLMs) \cite{SLM-reference}, and digital micro-mirror devices (DMDs) \cite{DMD-reference}. In these cases it is straightforward to explore different orthogonal bases, such as that for the Walsh-Hadamard transform \cite{HadamardBook}, or patterns with random-fractal properties \cite{SethnaBook}, etc. However, for GI probes that have been explored more recently, such as x rays \cite{pelliccia2016experimental,Yu2016fourier} and neutrons \cite{kingston2020neutron}, methods to generate patterns are much more limited to date. The key limitation, here, is the non-existence of high-resolution dynamically configurable beam-shaping elements that are the x-ray or neutron equivalent of a spatial light modulator \cite{paganin2019writing,ceddia2022a,ceddia2022b}.  While x-ray ghost-imaging experiments have been designed to use patterns based on natural variations in intensity (e.g., via hard x-ray speckles in the synchrotron radiation emitted by individual electron bunches \cite{pelliccia2016experimental}), if control is required over the properties of the patterns, then some form of rotating or translating {\it mask} is required upstream of the object \cite{Yu2016fourier,Schori2017xray,zhang2018table,pelliccia2018towards}.

Masks provide attenuation or propagation-based phase contrast \cite{Snigirev1995} to the incident illumination. They can be made from natural materials with a randomized structure or be specifically designed and fabricated. Examples in the literature for hard x-ray and neutron classical ghost imaging include copy paper \cite{Schori2017xray}, sandpaper \cite{zhang2018table}, sand grain pack \cite{pelliccia2018towards}, salt grain pack \cite{kingston2020neutron}, porous gold film \cite{Yu2016fourier}, metal foam \cite{KingstonOptica2018}, electroplated gold foil on a glass substrate to create a set of binary pixelated masks (324 pairs of $128 \times 128$ pixel patterns) \cite{he2020energy}, and gadolinium oxide (Gd$_2$O$_3$) Hadamard patterns etched onto a silicon substrate (1024 patterns of $32 \times 32$ \cite{he2021single}). 

Given no prior knowledge of the object, we seek to understand what properties of masks are important and how to measure them. We aim to understand the practical strengths and weaknesses of masks in order to identify the most appropriate choice for a given experiment.

Regarding the properties of illumination patterns used in classical GI, there are many experimental questions to consider. How many measurements should be taken? How much should the mask be transversely translated between each measurement? What mean intensity is required per measurement and how accurate should exposure time be to yield acceptable ghost images? How precise does mask translation need to be for a given spatial resolution? If a specified balance is desired for the three factors of (i) resolution, (ii) ghost-image contrast, and (iii) sample dose, what class of ghost-imaging mask will be optimal? 

In what follows we seek to provide some mask analysis tools to be able to answer some of these questions. The tools we present are based on conventional image analysis techniques such as the point-spread function (defined as the average Green's function of the system), Fourier ring correlation, and Fourier spectral power distribution, as well as some common mathematical analysis methods including singular-value decomposition, matrix rank, and perturbation theory.


We close this introduction with a brief overview of the remainder of the paper. Section~\ref{sec:background} provides the details of classical GI experiments and image reconstruction from measurements. The types of pattern explored in this paper, which may be either naturally occurring or designed, are defined and some examples are presented in Sec.~\ref{sec:patterns}. The effects of pattern properties on ghost image quality are explored in Sec.~\ref{sec:properties}, which covers ghost image resolution (Sec.~\ref{sec:psf}), the amount of independent information in the set of patterns (Sec.~\ref{sec:rank}), the size of the total mask used to create a set of patterns (Sec.~\ref{sec:footprint}), and robustness to various experimental conditions (Sec.~\ref{sec:robust}). This last subsection explores the effects of photon-shot noise (Sec.~\ref{sec:noise}), pattern misalignment (Sec.~\ref{sec:misalignment}), slowly varying illumination intensity (Sec.~\ref{sec:fluxVar}), multiscale masks (Sec.~\ref{sec:scalability}), illumination dose fractionation questions (Sec.~\ref{sec:doseFrac}), and mask fabrication imperfections (Sec.~\ref{sec:fabErrs}). The main findings are summarized in Sec.~\ref{sec:summary}. These findings are interpreted in practical terms in Sec.~\ref{sec:recommendations} to aid in mask design and/or selection given experiment conditions and required outcomes. This is followed by some concluding remarks and a presentation of future research directions in Sec.~\ref{sec:conclusion}.

\section{Classical ghost imaging}
\label{sec:background}

\begin{figure}
    \centering

    \begin{minipage}{0.3\textwidth}
        \centering
        \scriptsize{(a) pattern characterization with a pixelated detector}\\
        \includegraphics[height=0.7\textwidth]{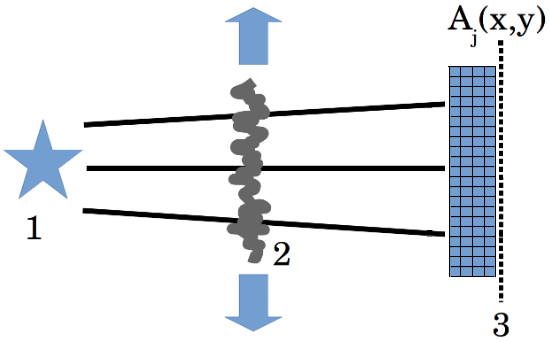}
    \end{minipage}%
    \begin{minipage}{0.7\textwidth}
        \centering
        \scriptsize{(b) bucket measurements with a single sensor}\\
        \includegraphics[height=0.3\textwidth]{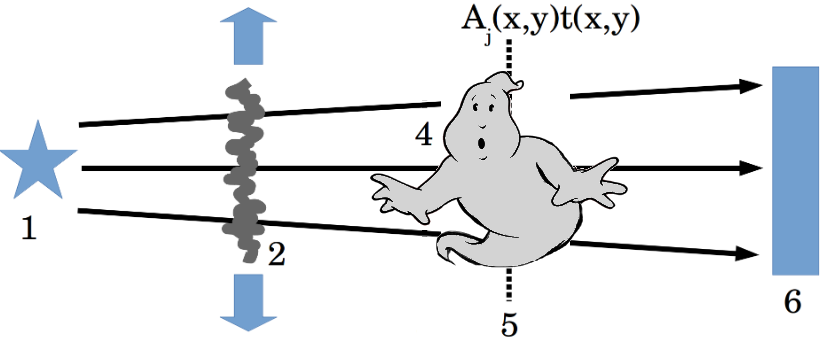}
    \end{minipage}%
\caption{Schematic for computational variant of classical ghost transmission imaging using a single non-configurable mask to create the patterned illumination.}
\label{fig:GISetUp}
\end{figure}

Figure~\ref{fig:GISetUp} shows a simple schematic for a wide class of computational variant of classical ghost-imaging experiments that employ a single non-configurable mask.  Here the experiment is broken into two stages. In the first stage, a source (1) illuminates a mask (2), to produce a pattern over the imaging plane that is pre-recorded or characterized by a pixelated detector (3). In the second stage, the sample (4) replaces the pixelated detector at the imaging plane. The previously recorded patterns illuminate the sample (5) and the transmitted signal then propagates to a ``bucket detector'' (6), namely a position-insensitive detector or sensor which merely records a single number 
that is proportional to the total exposure which falls upon it.  The mask may be transversely displaced, as indicated, with the resulting set of illumination patterns over the sample entrance surface (3) being assumed known.  
The classical ghost-imaging problem then seeks to infer the transmission function of the sample, given the set of known illumination patterns and the associated numbers measured by the bucket.    

Classical GI experiments do not directly record the image of an object. Quanta (e.g.~x-ray photons, neutrons, electrons etc.) that pass through the sample are not registered or imaged by a position-sensitive detector. GI experiments are a patterned-illumination method that divides the object measurements into two components: (i) the set of images of the patterned illuminations, and (ii) the total object interaction (e.g., transmission, scatter, fluorescence, etc.) with each patterned illumination. If the patterned illuminations are not repeatable, then measurements (i) and (ii) must be taken simultaneously, e.g., by utilizing a beamsplitter. However, for repeatable illumination patterns---e.g., from a transversely translating mask that modulates the incident illumination---measurements (i) and (ii) can be taken sequentially. The sequential variant is known as computational GI \cite{shapiro2008computational}. Here, the illumination patterns in (i) are either pre-recorded with a pixelated position sensitive detector, or are controlled and thus known {\em a priori}, e.g., employing a digital micromirror device in optical classical GI.

Assume the sample to be sufficiently thin that the projection approximation \cite{paganin2006} holds.  The intensity transmission function of the sample is therefore a well defined function of transverse coordinates $(x,y)$. Representing the transmission-function image of the object as $t(x,y)$, we can model the GI measurement process mathematically as follows: (i) is specified as a set of $J$ illumination patterns with recorded/predicted intensity $A_j(x,y)$ for $j \in \mathbb{Z}_J$; (ii) is represented as the correlation of the pattern with the object image, i.e.,
\begin{equation}
    b_j = \sum_x \sum_y A_j(x,y) t(x,y) \qquad \mbox{for}~j \in \mathbb{Z}_J. \label{eq_data_sampling}
\end{equation}
The set of total object interactions is recorded with a {\it bucket} detector and referred to as {\it bucket} values or {\it bucket} measurements. The model for this measurement process is a set of linear equations that can be represented in matrix form as
\begin{equation}
    \mathbf{At} = \mathbf{b},
\end{equation}
where $\mathbf{t}$ is the object image, $t(x,y)$ in vector format, the matrix $\mathbf{A}$ has $J$ rows with each row being the illumination pattern, $A_j(x,y)$, as a row vector, and $b$ is the set of $J$ corresponding bucket values.

The illumination patterns, $A_j(x,y)$, can be thought of as a set of basis vectors, while the corresponding bucket values, $b_j$, are the coefficients. In that sense, GI is recording the image in another space and to map the measurements back into image space is to invert the process, i.e., $\mathbf{t} = \mathbf{A}^{-1}\mathbf{b}$. As the intensity measured from our illumination patterns is always non-negative, basis vectors are not orthogonal (unless the mask is a delta function, i.e., a pin-hole).
By removing the mean value from the measured illumination patterns (as well as bucket values) they can be made closer to orthogonal. We can then rewrite the background-subtracted measurements as
\begin{equation}
    \mathbf{\widetilde{A}}\mathbf{t} = \mathbf{\widetilde{b}},
\end{equation}
where
\begin{equation}
    \widetilde{A}_j(x,y) = A_j(x,y) - \langle A(x,y) \rangle = A_j(x,y) - \frac{1}{J}\sum_k A_k(x,y) \rangle (x,y),
\end{equation}
and
\begin{equation}
    \widetilde{b}_j = b_j - \langle b \rangle = b_j - \frac{1}{J}\sum_k b_k.
\end{equation}

If this set of background-subtracted illumination patterns (or basis vectors) are orthogonal, then the inverse operator is the adjoint operator, i.e., $\mathbf{\widetilde{A}}^T = \mathbf{\widetilde{A}}^{-1}$.

A set of random illumination patterns are not orthogonal but can be considered {\it close} to orthogonal in some sense and we can approximate ghost image recovery using the adjoint operation as follows:
\begin{eqnarray}\label{eq:VanillaGI Reconstruction}
    \mathbf{t} \approx \mathbf{\widetilde{A}}^{T}\mathbf{\widetilde{b}}.
\end{eqnarray}
This expands to give the conventional ghost imaging recovery equation commonly reported in the literature (see, e.g.,~\cite{katz2009compressive, Bromberg2009ghost}):
\begin{equation}\label{eq:convGI}
    t(x,y) \approx \sum_j A_j(x,y) ( b_j - \langle b \rangle ).
\end{equation}

Rather than subtracting the mean bucket signal in the adjoint operation, the background signal can be compensated for by performing one Landweber iteration \cite{landweber1951iteration}, with the initial estimate being a constant image equal to the mean transmission of the object over all measurements. This can be efficiently calculated as
\begin{equation}\label{eq:diffGI}
    t(x,y) \approx \mu + \sum_j A_j(x,y) \left( b_j - \mu \sum_x\sum_y A_j(x,y) \right)
    \quad\quad\mbox{where}\quad \mu = \sum_j b_j \Big/ \sum_j \sum_x \sum_y A_j(x,y),
\end{equation}
which was shown in Ref.~\cite{ferri2010differential} to be more robust than the conventional adjoint method, particularly for weakly interacting objects that result in high-intensity bucket signals.

While this provides an approximate inversion, the problem itself may not have a solution or may have many solutions, depending on whether it is under constrained or over constrained. A better inversion in general may be formed using the Moore-Penrose pseudoinverse as follows:
\begin{eqnarray}
    \mathbf{t} = (\mathbf{\widetilde{A}}^{T}\mathbf{\widetilde{A}})^{-1}\mathbf{\widetilde{A}}^{T}\mathbf{\widetilde{b}}.
\end{eqnarray}
In cases where the matrices become too large to compute the inverse, algorithms such as the Kaczmarz method \cite{kaczmarz1937Angenaherte} or Landweber iteration converge to the same result in an iterative fashion, and exiting the iteration early (i.e., before convergence) serves as a regularization method. The Kaczmarz method proceeds as follows:
\begin{equation}
    \widehat{\mathbf{t}}^{k+1} = \widehat{\mathbf{t}}^{k} + \lambda \widetilde{\mathbf{A}}_j^T \frac{\widetilde{b_j} - \widetilde{\mathbf{A}}_j\widehat{\mathbf{t}}^{k}}{\| \widetilde{\mathbf{A}}_j \|^2}.
\end{equation}
Here $k$ denotes the iteration number, $\lambda$ is a relaxation parameter in $(0,1]$, $j = k (\textrm{mod}~J)$ (although often with a randomized order), and $\widetilde{\mathbf{A}_j}$ is the $j^\mathrm{th}$ row of $\mathbf{A}$ representing the $j^\mathrm{th}$ mean-corrected illumination pattern. The initial estimate of $\mathbf{t}$, i.e., $\mathbf{t}^0$, is typically the zero vector.

We can see from these equations that the properties of the illumination patterns in $\mathbf{A}$ are fundamental to the performance of classical ghost imaging. In the context of classical x-ray and neutron ghost imaging considered here---together with other forms of classical ghost imaging employing radiation and matter wavefields for which dynamic configurable beamshaping elements do not exist---these patterns may be generated by a translating or rotating a non-configurable mask. In what follows we seek to provide some mask analysis tools to be able to choose an appropriate mask for a given GI experiment.

\section{Illumination pattern categories}
\label{sec:patterns}

\begin{figure}
    \centering
    \begin{minipage}{0.2\textwidth}
        \centering
        \scriptsize{(a)}\\
        \includegraphics[width=0.8\textwidth]{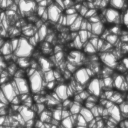}
    \end{minipage}%
    \begin{minipage}{0.2\textwidth}
        \centering
        \scriptsize{(b)}\\
        \includegraphics[width=0.8\textwidth]{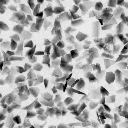}
    \end{minipage}%
    \begin{minipage}{0.2\textwidth}
        \centering
        \scriptsize{(c)}\\
        \includegraphics[width=0.8\textwidth]{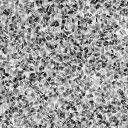}
    \end{minipage}%
    \begin{minipage}{0.2\textwidth}
        \centering
        \scriptsize{(d)}\\
        \includegraphics[width=0.8\textwidth]{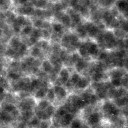}
    \end{minipage}%
    \begin{minipage}{0.2\textwidth}
        \centering
        \scriptsize{(e)}\\
        \includegraphics[width=0.8\textwidth]{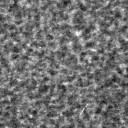}
    \end{minipage}
    \caption{Example $128 \times 128$ pixel patterns formed by natural random structures after illumination with x-rays: (a) nickel foam (pixel pitch 30$\mu$m), (b) 80 grit sandpaper (pixel pitch 28$\mu$m), (c) 120 grit sandpaper (pixel pitch 28$\mu$m); and illumination with thermal neutrons with pixel pitch of 101$\mu$m: (d) ball-bearings, nuts, bolts, washers inside concentric Al cylinders, (e) large-grained, iodized, table salt inside concentric Al cylinders.}
    \label{fig:naturalRandomMasks}
\end{figure}

\begin{figure}
    \centering
    \begin{minipage}{0.2\textwidth}
        \centering
        \scriptsize{(a)}\\
        \includegraphics[width=0.8\textwidth]{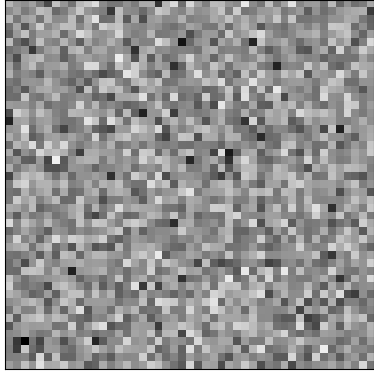}
    \end{minipage}%
    \begin{minipage}{0.2\textwidth}
        \centering
        \scriptsize{(b)}\\
        \includegraphics[width=0.8\textwidth]{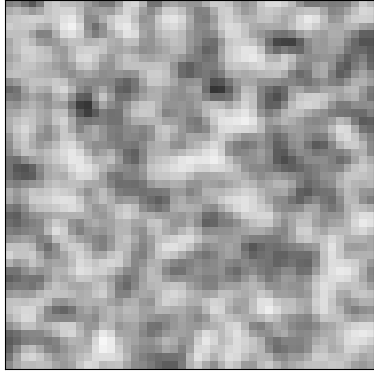}
    \end{minipage}%
    \begin{minipage}{0.2\textwidth}
        \centering
        \scriptsize{(c)}\\
        \includegraphics[width=0.8\textwidth]{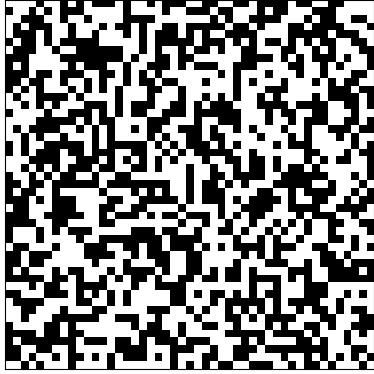}
    \end{minipage}%
    \begin{minipage}{0.2\textwidth}
        \centering
        \scriptsize{(d)}\\
        \includegraphics[width=0.8\textwidth]{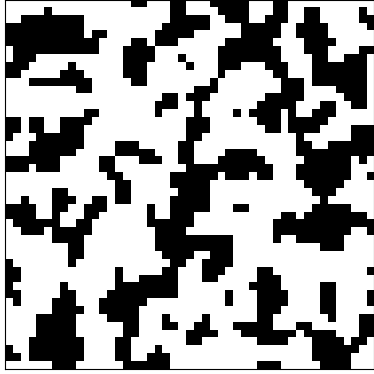}
    \end{minipage}%
    \begin{minipage}{0.2\textwidth}
        \centering
        \scriptsize{(e)}\\
        \includegraphics[width=0.8\textwidth]{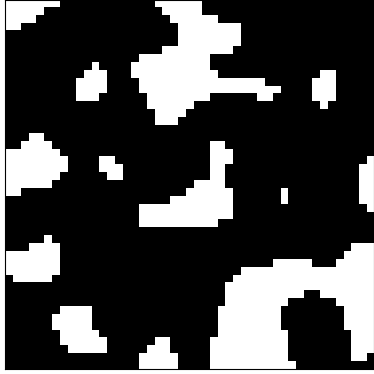}
    \end{minipage}
    \caption{Example $47 \times 47$ pixel patterns generated to simulate patterns from natural random structures. (a) Gaussian random noise, (b) Gaussian random noise convolved with a kernel with $\sigma = 1$ pixel, (c) Binary random noise, (d) Binary random noise convolved with a kernel with $\sigma = 1$ pixel and (e) 2 pixels.}
    \label{fig:simNatRandMasks}
\end{figure}

\begin{figure}
    \centering
    \begin{minipage}{0.2\textwidth}
        \centering
        \scriptsize{(a)}\\
        \includegraphics[width=0.8\textwidth]{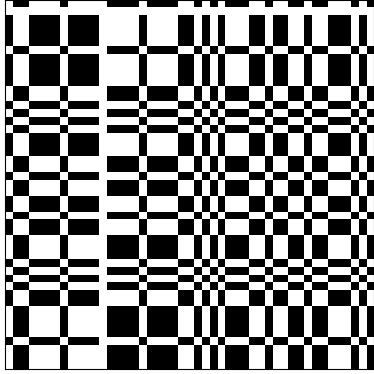}
    \end{minipage}%
    \begin{minipage}{0.2\textwidth}
        \centering
        \scriptsize{(b)}\\
        \includegraphics[width=0.8\textwidth]{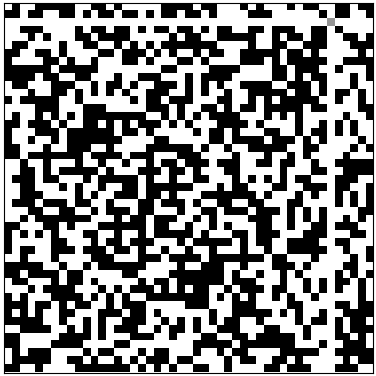}
    \end{minipage}%
    \begin{minipage}{0.2\textwidth}
        \centering
        \scriptsize{(c)}\\
        \includegraphics[width=0.8\textwidth]{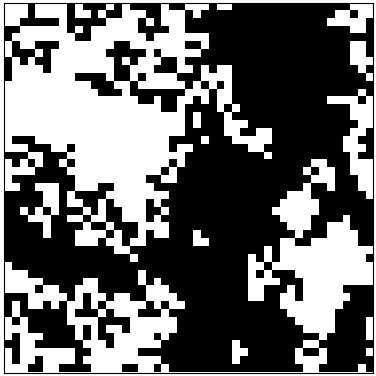}
    \end{minipage}%
    \begin{minipage}{0.2\textwidth}
        \centering
        \scriptsize{(d)}\\
        \includegraphics[width=0.8\textwidth]{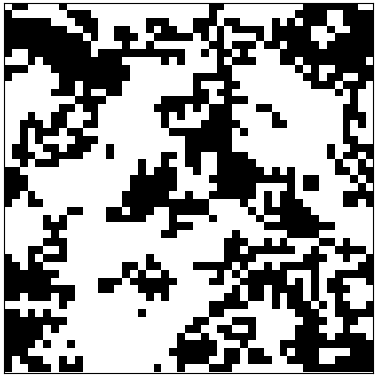}
    \end{minipage}%
    \begin{minipage}{0.2\textwidth}
        \centering
        \scriptsize{(e)}\\
        \includegraphics[width=0.8\textwidth]{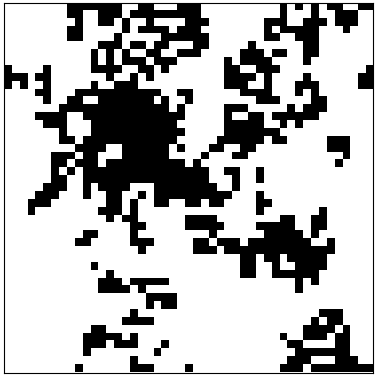}
    \end{minipage}
    \caption{Example $47 \times 47$ pixel patterns generated to form binary designed and fabricated masks. (a) a URA mask that forms part of an orthogonal set, (b) an FRT based mask that forms part of an orthogonal set, (c--e) fractal masks generated with $(\alpha,\beta) = $ (1.0,0.02), (1.1,0.02), (1.0,0.01).}
    \label{fig:simFabMasks}
\end{figure}

There are many different types of illumination pattern that can be produced, using masks that introduce either attenuation contrast or propagation-based refraction contrast \cite{Snigirev1995} to the incident illumination. Here we summarize the types of mask explored in this work. Broadly speaking, these may be partitioned into the categories of (i) masks made from natural materials with a randomized structure, and (ii) masks that are specifically designed and fabricated with certain desirable properties. Example illumination patterns from each type of mask are depicted in Figs.~\ref{fig:naturalRandomMasks}-\ref{fig:simFabMasks}.

Natural random structures are commonly found as grain distributions such as sandpaper and bead packs, or their dual structure, foams, that are typically metallic. Some example illumination patterns in this class are presented in Fig.~\ref{fig:naturalRandomMasks} from experiments with either neutron or x-ray illumination. We have included a nickel foam, coarse and finer grit sandpaper, and grain packs of steel (ball-bearings, nuts, bolts and washers) and iodized table-salt. Care must be taken when selecting and/or creating these masks to ensure some degree of randomness. For example, a foam with uniform bubble size or bead pack with uniform bead size may exhibit regular crystalline properties with too much symmetry (i.e.~too high a degree of medium-range order) and therefore yield insufficient unique patterns for GI.

Before proceeding, there are two points we wish to emphasize, regarding masks made from natural random structures: 
\begin{itemize}
    \item Random or near-random bubble networks, such as metallic foams, will typically have a projected density distribution that has a significantly smaller characteristic transverse length scale, in comparison to the characteristic diameter of the bubbles.\footnote{This statement applies to random-bubble slab-type masks whose longitudinal slab thickness is significantly larger than the bubble diameter. }  This implies that the autocorrelation of the projected density, which is closely related to the point spread function (PSF) associated with the simplest forms of computational ghost imaging \cite{paganin2019writing,pelliccia2018towards,ferri2010differential}, will be surrounded by an annular halo whose radius is on the order of the mean bubble size.  This annular halo, while often small in amplitude relative to the central auto-correlation (PSF) peak, may nevertheless be comparable in strength to the central peak in an integrated-signal (total power) sense. Such PSF rings or annular halos can cause serious problems with ghost-imaging reconstructions. To avoid such problems, the mask transverse translation between measurements, or stride, should be bigger than the uniform bubble dimension, rather than merely bigger than the characteristic size of the speckles generated by the mask, when bubble-network masks are employed for the purposes of ghost imaging.  A similar statement applies to the inverse of bubble-network masks, such as slab-type masks composed of randomly-packed spheres or ellipsoids, when the slab thickness is much larger than the characteristic diameter of the particles.  
    \item Some self-assembled random structures, such as masks associated with the Ising model in the vicinity of a thermodynamic phase transition \cite{moghadam2022}, have a hierarchy of speckle sizes that corresponds to a random fractal over a specified range of length scales \cite{SethnaBook}. Such masks are statistically self similar over a range of length scales, which has the very useful property that they are invariant with respect to a change of spatial scale, over a specified range of length scales.  These random-fractal masks, which have a range of speckle sizes, may be useful in ghost-imaging contexts where a range of spatial resolutions may be used to interrogate an object, e.g.~with a low-resolution ``scoping scan'' to identify a particular region or regions of interest, followed by a finer-resolution ghost-imaging experiment localized to the region or regions of interest. For our purposes, such random fractal masks will be taken to have an inverse-power-law dependence of their power spectrum with respect to radial spatial frequency, over a Fourier-space annulus whose inner and outer radii delineate the range of length scales over which the random-fractal mask is statistically self similar.
    \end{itemize}


In order to simulate illumination patterns from natural masks, here we have used several types of random noise:
\begin{enumerate}
    \item normalized Gaussian random noise, i.e., $A(x,y) = \min(\max(a(x,y), 0), 1)$, where $a(x,y) \overset{\mathrm{iid}}{\sim} \mathcal{N}(\mu,\sigma^2)$, with mean, $\mu$, and standard deviation, $\sigma$;
    \item uniform random noise in the range (0,1), i.e., $A(x,y) \overset{\mathrm{iid}}{\sim} \mathcal{U}(0,1)$;
    \item binary random noise (i.e.~discrete uniform noise from $\{0,1\}$), i.e., $A(x,y) \overset{\mathrm{iid}}{\sim} \mathrm{Bernoulli}(\nicefrac{1}{2}).$
\end{enumerate}
Illumination patterns with different feature sizes and/or resolution are simulated by convolution of the random masks with a Gaussian blurring kernel, i.e., $A \ast K$, where $K$ is defined as follows:
\begin{equation}
    K(x,y) = \frac{1}{2 \pi \sigma^2} \exp(-(x^2 + y^2) / 2\sigma^2).
\end{equation}
Here, the standard deviation, $\sigma$, determines the resulting feature size. Assuming that the full-width at half-maximum is a reasonable measure of the feature size, we can estimate these features to be distributed around $2.355\sigma$ pixels in diameter. The resulting pattern may also be binarized using a thresholding operation. Some examples of these simulated random illumination patterns are presented in Fig.~\ref{fig:simNatRandMasks}.

The set of illumination patterns employed for GI, i.e., $A_j(x,y)$ in Eq.~\eqref{eq_data_sampling}, is formed by taking a sample, or  subset, of the larger pattern $A$ as follows:
\begin{equation}
    A_j(x,y) = A(x_j+x, y_j+y),
\end{equation}
for some offset $(x_j,y_j)$ associated with pattern $j$. An example of offsets that form an $M \times \mbox{quot}(J,M)$ grid of mask subsets can be defined as $x_j= s_x \mbox{rem}(j,M)$ and $y_j= s_y \mbox{quot}(j,M)$ for some integer $M$ and strides $s_x,s_y\in\mathbb{Z}$). Note that the required range, or footprint, of $(x,y)$ in $A$ must be much larger than that for each mask $A_j$ and increases in proportion to the stride selected. This translating mask definition of patterns is distinct from having separate patterns $A_j$ built using the methods described above; here, the different patterns $A_j$ can have some overlap. An important benefit of this construction is that much smaller overall patterns, $A$, need to be built: using distinct $A_j$ requires a total of $J N_x N_y$ pixels (where $J$ is typically of order $N_x N_y$), but overlapping $A_j$ requires on the order of only $N_x+N_y$ pixels (depending on the strides $s_x$ and $s_y$). Where pattern construction is difficult and/or expensive, using overlapping patterns can have significant practical benefit.\footnote{We note that for classical GI with visible light, devices such as projectors and spatial light modulators make constructing non-overlapping patterns straightforward.}

In order to minimize fabrication complexity, designed masks are typically binary. The 2D binary patterns described above that simulate natural random structures can be manufactured in a reasonably straightforward manner (depending on the scale involved). Apart from ease of manufacture, a further benefit of binary masks is that an illumination pattern with a binary distribution maximizes pattern variance; this is important since the signal in ghost imaging is related to pattern variance \cite{kingston2021inherent}. Binary masks can have a random structure, be designed to form an orthogonal set, or have other desirable properties. In this work we will look at the performance of random binary masks with a range of feature sizes (as demonstrated in Fig.~\ref{fig:simNatRandMasks}c-e).

We also utilize masks that are orthogonal under translation, such as those based on uniformly redundant arrays (URA) constructed using quadratic residues \cite{Gottesman1989newFamily}, and a technique based on the finite Radon transform (FRT) \cite{cavy2015construction}. We have previously presented and explored these patterns as potential masks in Ref.~\cite{KingstonIEEE2019}. The magnitude of the Fourier transform of these URA and FRT orthogonal masks is uniform across all spatial frequencies. This means that each measurement, using these masks to generate illumination patterns, is probing all frequencies simultaneously. An example of both of these mask categories is presented in Fig.~\ref{fig:simFabMasks}a-b.

Binary masks must typically be fabricated with a specific resolution in mind and their properties (particularly the pattern variance) often degrade quickly as resolution is coarsened. Here we explore binary fractal masks that maintain statistical properties over a range of spatial frequencies and enable flexibility in resolution. A degraded ghost image reconstructed at a high-resolution can still achieve a reasonable low-resolution image in this case. One method to generate fractal masks is by convolution of a random binary mask with a blurring kernel that decays with a power law in spatial-frequency, ($k_x$,$k_y$), such as 
\begin{equation}\label{eq:fractal}
    H(k_x,k_y) = \frac{\gamma}{\sqrt{k_x^2 + k_y^2}^\alpha + \beta},
\end{equation}
with $\alpha$ around 1, $\beta$ close to zero, and $\gamma$ set to normalize the result (cf.~Refs.~\cite{PowerLawNoise1,PowerLawNoise2,PowerLawNoise3}). Convolution can then be performed in the Fourier domain, i.e., $A' = \mathcal{F}^{-1}(\mathcal{F}(A).H)$ and the result binarized by thresholding. Example illumination patterns with $\alpha \in \{1.0,1.1\}$ and $\beta \in \{0.01,0.02\}$ are presented in Fig.~\ref{fig:simFabMasks}c-e.

Another set of binary masks that should be mentioned is the Hadamard masks. These are an orthogonal set of patterns that can be considered to form a basis set for a binary form of Fourier transform. They are commonly employed in optical GI using rapidly transformable discrete structuring methods such as a projector, SLM, or DMD. They are less practical when considering mask fabrication since each mask is a unique pattern (as discussed in more detail in Sec.~\ref{sec:footprint}). By design, each pattern probes a limited range of spatial frequencies. This can be useful since it provides multiscale capabilities, but is not a good choice of basis if the object being imaged is sparse in Fourier space. We believe the properties of Hadamard masks are captured by the other masks explored in this work.

In the remainder of this paper, we explore the properties and performance of the patterns described in this section, in the context of GI.  We do so with the aim of establishing some {\it rules-of-thumb} to enable appropriate mask selection and/or design for a given physical scenario.

\section{Effect of illumination pattern properties on ghost image quality}
\label{sec:properties}

\subsection{Achievable resolution}
\label{sec:psf}

The spatial resolution of a ghost imaging system can be predicted by the properties of the point-spread-function (PSF) of that system. The PSF is the average impulse response of the entire computational imaging process. A point object, such as a pin-hole, serves as the impulse input and the PSF is defined as the average representation of that point or impulse as it is scanned over the entire imaging field-of-view (FOV). The result of imaging an object with the system is approximated as the ideal image convolved with the PSF. Thus the sharpness of the PSF determines the achievable spatial resolution of the imaging system. An example metric to specify spatial resolution, in this context, is the full-width at half-maximum (FWHM) of the PSF \cite{houston1926fine}.

Let the Green's function $G_{(x^*,y^*)}$ describe the point spread effects about the point $(x^*,y^*)$. Note that a recovered ghost image can be composed as a weighted sum over the set of Green's functions. $G_{(x^*,y^*)}$ can be simulated by applying the ghost imaging process (simulating the experimental measurements and image recovery) to the Dirac delta at $(x^*,y^*)$, i.e.~$\delta(x'-x^*, y'-y^*)$. The expected Green's function, or the average over all coaligned Green's functions, can be realized by shifting each $G_{(x^*,y^*)}$ to be about a common point. Shifting about $(x,y) = (0,0)$, the PSF is defined as
\begin{equation}
    \mbox{PSF}(x,y) = \frac{1}{N_xN_y}\sum_{x^*}\sum_{y^*} G_{(x^*,y^*)} (x+x^*,y+y^*).
\end{equation}

\begin{figure}
    \centering
    \begin{minipage}{0.2\textwidth}
        \centering
        \scriptsize{(a-i)}\\
        \includegraphics[width=0.8\textwidth]{egMaskUraScanning47x47.png}\\[1ex]
        \scriptsize{(a-ii)}\\
        \includegraphics[width=0.9\textwidth]{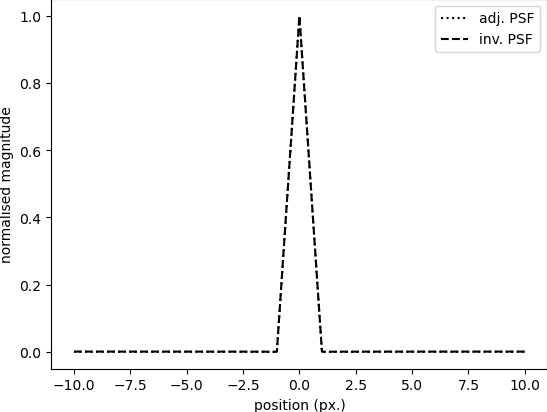}\\[1ex]
        \scriptsize{(a-iii)}\\
        \includegraphics[width=0.8\textwidth]{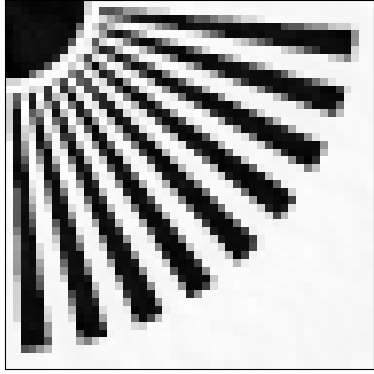}\\[1ex]
        \scriptsize{(a-iv)}\\
        \includegraphics[width=0.8\textwidth]{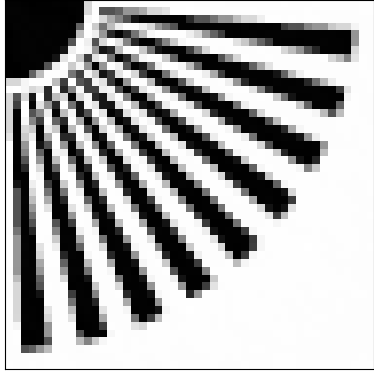}\\[1ex]
        \scriptsize{(a-v)}\\
        \includegraphics[width=\textwidth]{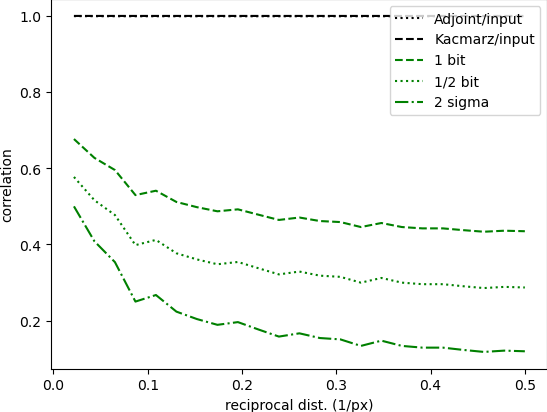}
    \end{minipage}%
    \begin{minipage}{0.2\textwidth}
        \centering
        \scriptsize{(b-i)}\\
        \includegraphics[width=0.8\textwidth]{egMaskGaussianScanning47x47.png}\\[1ex]
        \scriptsize{(b-ii)}\\
        \includegraphics[width=0.9\textwidth]{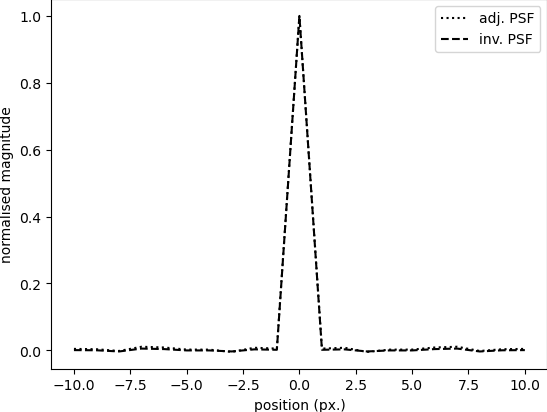}\\[1ex]
        \scriptsize{(b-iii)}\\
        \includegraphics[width=0.8\textwidth]{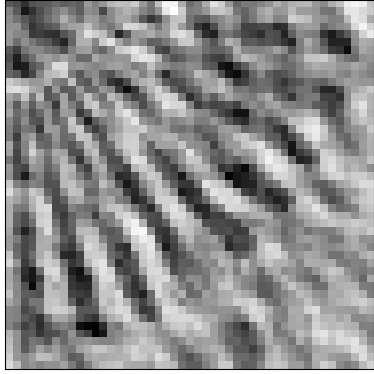}\\[1ex]
        \scriptsize{(b-iv)}\\
        \includegraphics[width=0.8\textwidth]{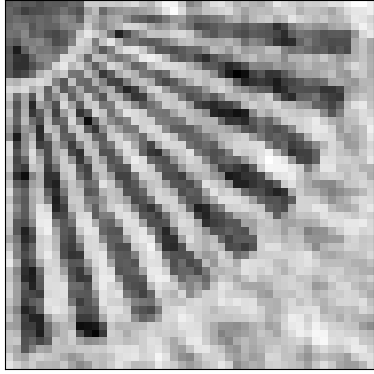}\\[1ex]
        \scriptsize{(b-v)}\\
        \includegraphics[width=\textwidth]{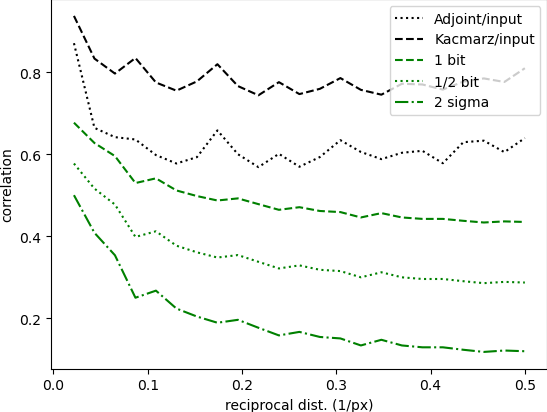}
    \end{minipage}%
    \begin{minipage}{0.2\textwidth}
        \centering
        \scriptsize{(c-i)}\\
        \includegraphics[width=0.8\textwidth]{egMaskGaussianBlur1Scanning47x47.png}\\[1ex]
        \scriptsize{(c-ii)}\\
        \includegraphics[width=0.9\textwidth]{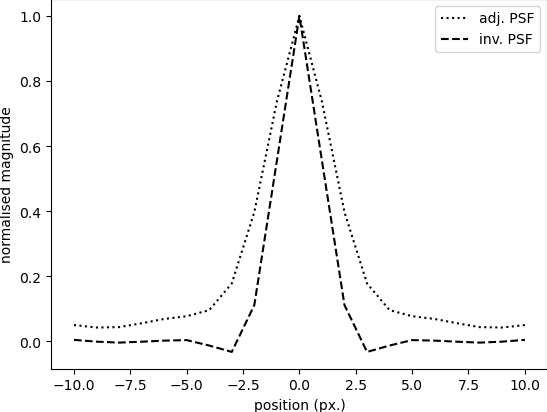}\\[1ex]
        \scriptsize{(c-iii)}\\
        \includegraphics[width=0.8\textwidth]{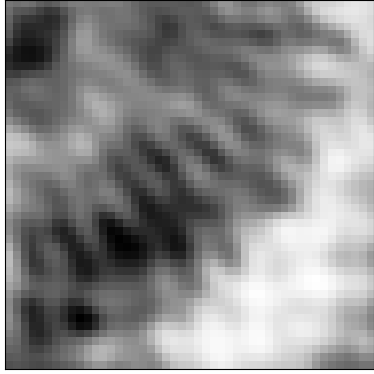}\\[1ex]
        \scriptsize{(c-iv)}\\
        \includegraphics[width=0.8\textwidth]{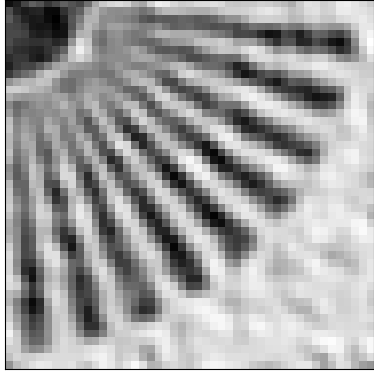}\\[1ex]
        \scriptsize{(c-v)}\\
        \includegraphics[width=\textwidth]{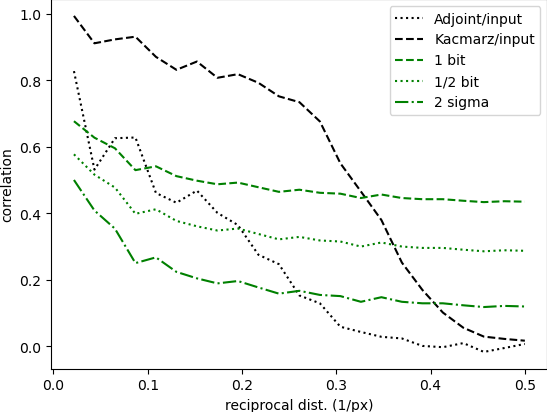}
    \end{minipage}%
    \begin{minipage}{0.2\textwidth}
        \centering
        \scriptsize{(d-i)}\\
        \includegraphics[width=0.8\textwidth]{egMaskBinaryScanning47x47.png}\\[1ex]
        \scriptsize{(d-ii)}\\
        \includegraphics[width=0.9\textwidth]{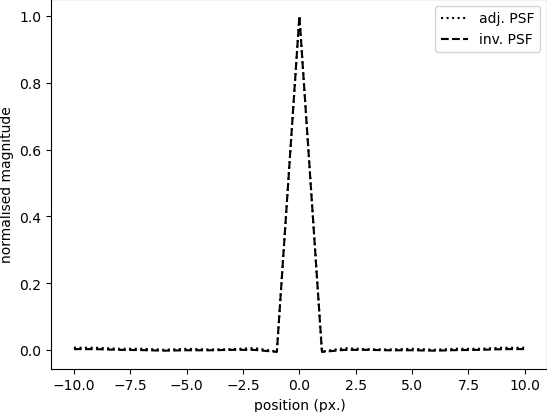}\\[1ex]
        \scriptsize{(d-iii)}\\
        \includegraphics[width=0.8\textwidth]{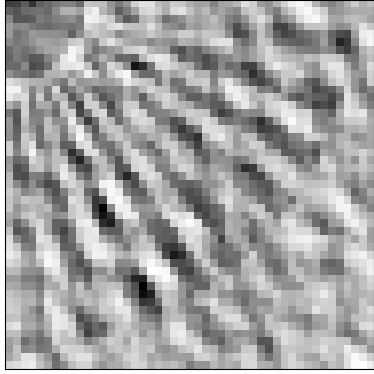}\\[1ex]
        \scriptsize{(d-iv)}\\
        \includegraphics[width=0.8\textwidth]{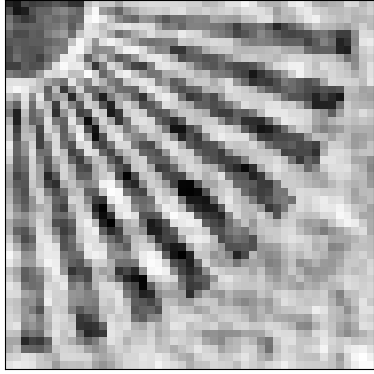}\\[1ex]
        \scriptsize{(d-v)}\\
        \includegraphics[width=\textwidth]{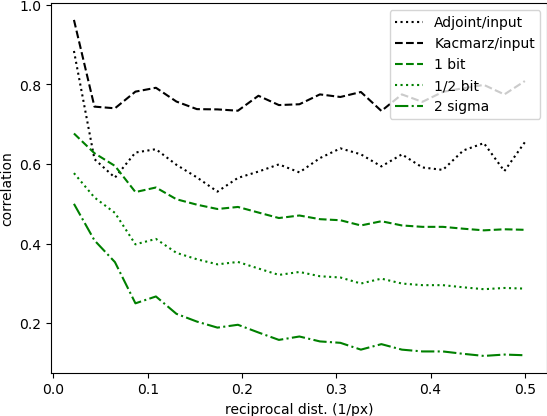}
    \end{minipage}%
    \begin{minipage}{0.2\textwidth}
        \centering
        \scriptsize{(e-i)}\\
        \includegraphics[width=0.8\textwidth]{egMaskBinaryBlur1RebinScanning47x47.png}\\[1ex]
        \scriptsize{(e-ii)}\\
        \includegraphics[width=0.9\textwidth]{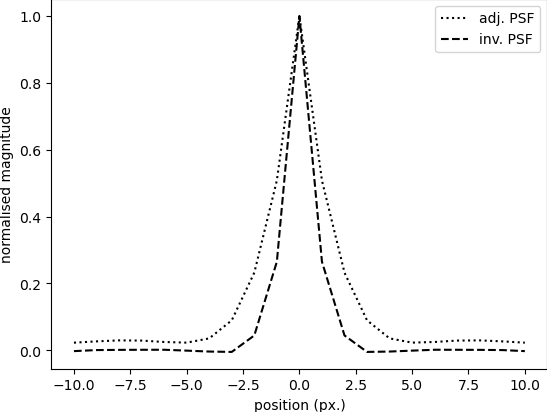}\\[1ex]
        \scriptsize{(e-iii)}\\
        \includegraphics[width=0.8\textwidth]{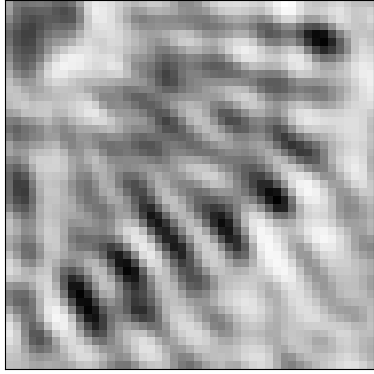}\\[1ex]
        \scriptsize{(e-iv)}\\
        \includegraphics[width=0.8\textwidth]{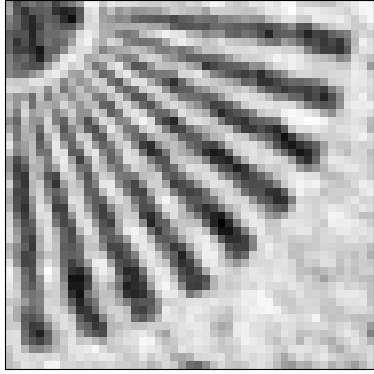}\\[1ex]
        \scriptsize{(e-v)}\\
        \includegraphics[width=\textwidth]{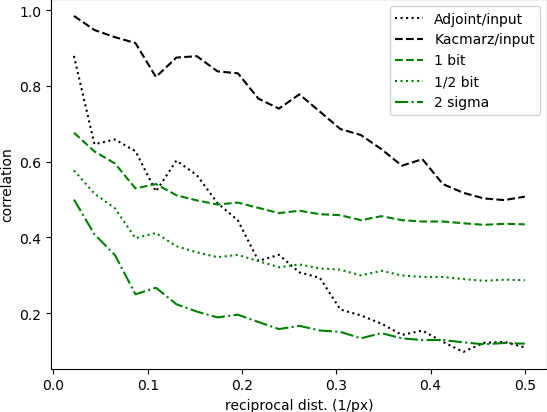}
    \end{minipage}
    \caption{Example (i) $47 \times 47$ pixel illumination patterns, (ii) their associated adjoint and inverse PSF, (iii) adjoint reconstruction, (iv) 4 iterations of Kaczmarz reconstruction, and (v) FRC analysis. All examples contain $2209$ patterns in the set and the grayscale window for the example patterns shown is [0,1]. (a) Uniformly redundant array (orthogonal under translation), (b) Gaussian random, (c) Gaussian random blurred by a Gaussian kernel with $\sigma = 1.0$ px, (d) Random binary, (e) Random binary with larger feature sizes. All patterns were generated by scanning a larger pattern in a square grid pattern with step size of 1px.}
    \label{fig:psf}
\end{figure}

Since classical ghost imaging is a computational imaging system, the PSF depends on the image reconstruction method used. In this section we discuss two image reconstruction schemes: (1) 
well-conditioned ghost image recovery using the differential adjoint of the imaging system (Eq.~(\ref{eq:diffGI})),
and (2) an approximation to the pseudoinverse using Kaczmarz iteration \cite{kaczmarz1937Angenaherte}. We denote the PSF associated with each of these methods as the {\it adjoint PSF} and {\it inverse PSF} respectively. The adjoint and inverse PSF for several types of illumination patterns are presented in Fig.~\ref{fig:psf}.

Note that for orthogonal sets of illumination patterns, the adjoint PSF is equal to the inverse PSF (as seen in Fig.~\ref{fig:psf}a) and has a FWHM of 1.0 pixel (px). However, in general, the adjoint PSF will be more diffuse than the inverse PSF. Observe from the examples in Fig.~\ref{fig:psf} that for the Gaussian and Binary random masks (b and d respectively) that the adjoint PSF is as sharp as the inverse PSF but with non-zero tails; here the FWHM is 1px for both adjoint and inverse PSF. Looking at the example adjoint and Kaczmarz iteration reconstructed images in rows (iii-iv) we see that both have a similar sharpness but the adjoint has more artifacts from the aforementioned tails of the PSF. When the patterns are blurred or their feature size increases (c and e respectively) then the adjoint PSF becomes significantly degraded (FWHM of (c) 3.4px and (e) 2.5px) while the inverse PSF is only slightly affected (FWHM of (c) 2.5px and (e) 1.0px). 

Another method to determine image resolution is Fourier ring correlation (FRC) \cite{saxton1982correlation, vanHeel1982structure}. FRC estimates resolution by correlating two independent images (or measurements) of the same object over a range of spatial frequencies, (i.e., rings in Fourier space). Low correlation occurs when the signal is dominated by noise or artifacts and indicates the limit to measurement resolution. We can see from the plots in Figs.~\ref{fig:psf}c and \ref{fig:psf}e, where the adjoint curves cross the 2$\sigma$ threshold at 0.28/px and 0.40/px respectively. This corresponds to a FWHM of 3.6px and 2.5px, respectively, which matches the adjoint PSF analysis. Also we observe that, for the Kaczmarz curve in Figs.~\ref{fig:psf}(c-v), resolution can be estimated as 0.40/px = 2.5px; this again matches the inverse PSF analysis.

In light of the remarks above, this section explores the factors that affect the PSF and thus ghost imaging resolution.


\subsubsection{Limitation of pattern characterization resolution}

A factor that affects the sharpness of both the adjoint and inverse PSF, and sets the upper limit to ghost imaging sharpness, is the resolution to which the illumination patterns can be characterized. No matter how much addition and subtraction of illumination patterns is performed, high spatial frequencies that are not measured, cannot be recovered.

A simple example of this limitation can be seen by comparing Figs.~\ref{fig:psf}b and \ref{fig:psf}c. Here the illumination pattern in both cases is created from a scanned random Gaussian mask. In Fig.~\ref{fig:psf}b the mask was characterized with 1px resolution, while in Fig.~\ref{fig:psf}c the imaging system had a blurring kernel with $\sigma = 1.0$px (corresponding to a resolution of 2.4px). The degradation of both the adjoint and inverse PSF can be observed. The FWHM of the inverse PSF and adjoint PSF should be 2.4px and $2.4 \sqrt{2} = 3.3$px respectively and corresponds to 0.42/px and 0.3/px in FRC analysis. These estimates are consistent with the values measured by simulation.

It can also be seen from the resolution star images reconstructed in Fig.~\ref{fig:psf}d (when compared with those from Fig.~\ref{fig:psf}b, where pattern characterization was ideal) that the Kaczmarz iterations, while able to considerably improve the image quality, do not alter the limitation that high frequency information remains inaccessible.

Note that, to ensure robustness, the mask translation (or stride) used between each pattern subset illuminated should be selected to be greater than the resolution to which the mask is characterized. For the blurred Gaussian pattern example in Fig.~\ref{fig:psf}c, the resolution could be estimated as the FWHM of the Gaussian used to blur the image, i.e., 2.35px. The results when using patterns with a 3px step size in mask translation are presented in Fig.~\ref{fig:featureSize}b. Here we observe that the tails on the adjoint PSF are significantly improved and the corresponding adjoint image recovered is much improved in comparison to the 1px translation case. FRC analysis shows similar resolution, however, the effects from artifacts cause less degradation at lower spatial frequencies.

\subsubsection{Effect of minimum feature size in patterns}

When selecting an appropriate natural material for use as a mask, such as a foam or grain pack, it is useful to understand the importance of feature size and characteristic lengths. What size grains should be selected? What mask transverse-translation step size should be used between each pattern used?

\begin{figure}
    \centering
    \begin{minipage}{0.166\textwidth}
        \centering
        \scriptsize{(a-i)}\\
        \includegraphics[width=0.8\textwidth]{egMaskGaussianBlur1Scanning47x47.png}\\[1ex]
        \scriptsize{(a-ii)}\\
        \includegraphics[width=0.9\textwidth]{adjInvNormPsfGaussianBlur1Scanning47x47.png}\\[1ex]
        \scriptsize{(a-iii)}\\
        \includegraphics[width=0.8\textwidth]{adjReconGaussianBlur1Scanning47x47.png}\\[1ex]
        \scriptsize{(a-iv)}\\
        \includegraphics[width=0.8\textwidth]{invReconGaussianBlur1Scanning47x47.png}\\[1ex]
        \scriptsize{(a-v)}\\
        \includegraphics[width=\textwidth]{adjInvFrcGaussianBlur1Scanning47x47.png}
    \end{minipage}%
    \begin{minipage}{0.166\textwidth}
        \centering
        \scriptsize{(b-i)}\\
        \includegraphics[width=0.8\textwidth]{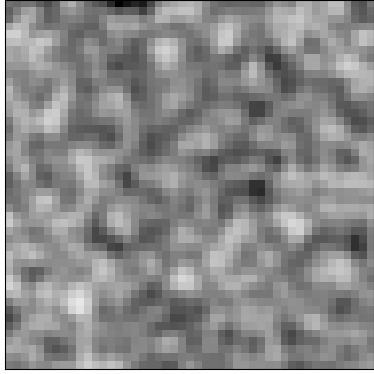}\\[1ex]
        \scriptsize{(b-ii)}\\
        \includegraphics[width=0.9\textwidth]{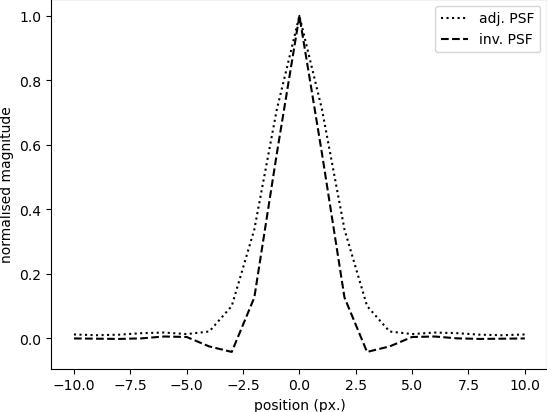}\\[1ex]
        \scriptsize{(b-iii)}\\
        \includegraphics[width=0.8\textwidth]{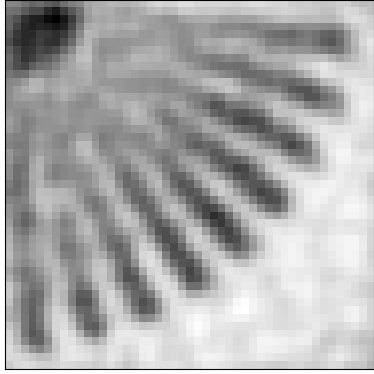}\\[1ex]
        \scriptsize{(b-iv)}\\
        \includegraphics[width=0.8\textwidth]{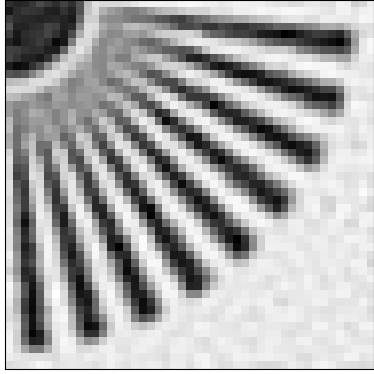}\\[1ex]
        \scriptsize{(b-v)}\\
        \includegraphics[width=\textwidth]{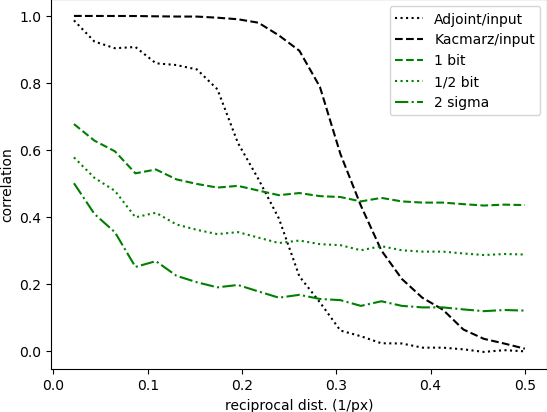}
    \end{minipage}%
    \begin{minipage}{0.166\textwidth}
        \centering
        \scriptsize{(c-i)}\\
        \includegraphics[width=0.8\textwidth]{egMaskBinaryBlur1RebinScanning47x47.png}\\[1ex]
        \scriptsize{(c-ii)}\\
        \includegraphics[width=0.9\textwidth]{adjInvNormPsfBinaryBlur1RebinScanning47x47.png}\\[1ex]
        \scriptsize{(c-iii)}\\
        \includegraphics[width=0.8\textwidth]{adjReconBinaryBlur1RebinScanning47x47.png}\\[1ex]
        \scriptsize{(c-iv)}\\
        \includegraphics[width=0.8\textwidth]{invReconBinaryBlur1RebinScanning47x47.png}\\[1ex]
        \scriptsize{(c-v)}\\
        \includegraphics[width=\textwidth]{adjInvFrcBinaryBlur1RebinScanning47x47.png}
    \end{minipage}%
    \begin{minipage}{0.166\textwidth}
        \centering
        \scriptsize{(d-i)}\\
        \includegraphics[width=0.8\textwidth]{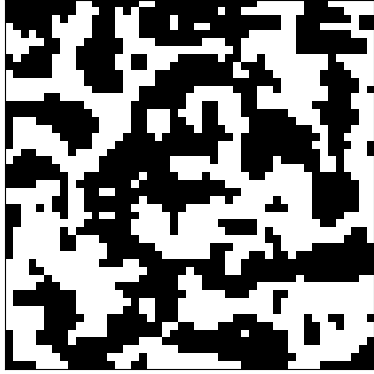}\\[1ex]
        \scriptsize{(d-ii)}\\
        \includegraphics[width=0.9\textwidth]{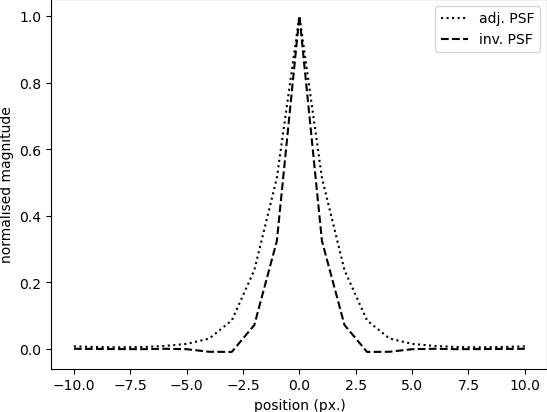}\\[1ex]
        \scriptsize{(d-iii)}\\
        \includegraphics[width=0.8\textwidth]{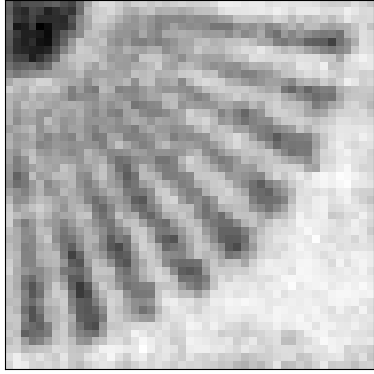}\\[1ex]
        \scriptsize{(d-iv)}\\
        \includegraphics[width=0.8\textwidth]{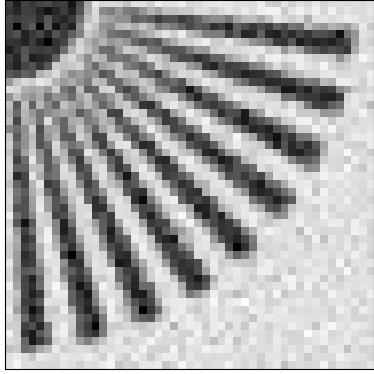}\\[1ex]
        \scriptsize{(d-v)}\\
        \includegraphics[width=\textwidth]{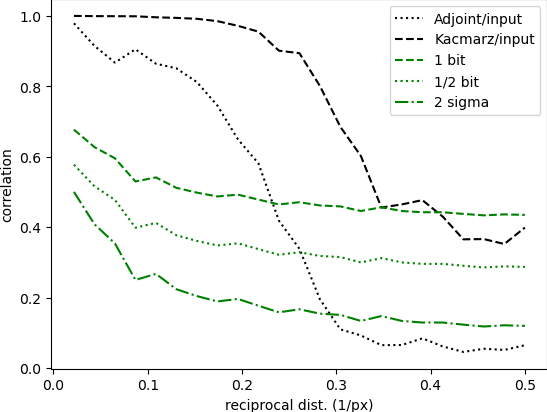}
    \end{minipage}%
    \begin{minipage}{0.166\textwidth}
        \centering
        \scriptsize{(e-i)}\\
        \includegraphics[width=0.8\textwidth]{egMaskBinaryBlur2RebinScanning47x47.png}\\[1ex]
        \scriptsize{(e-ii)}\\
        \includegraphics[width=0.9\textwidth]{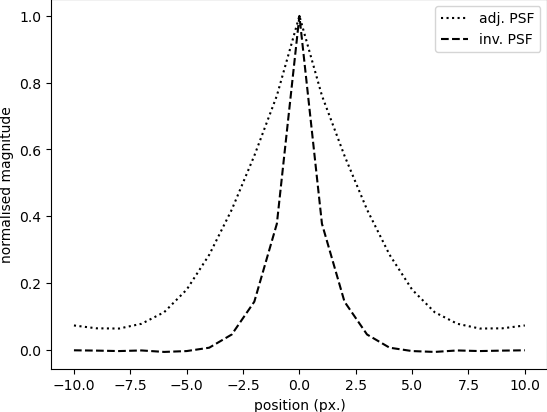}\\[1ex]
        \scriptsize{(e-iii)}\\
        \includegraphics[width=0.8\textwidth]{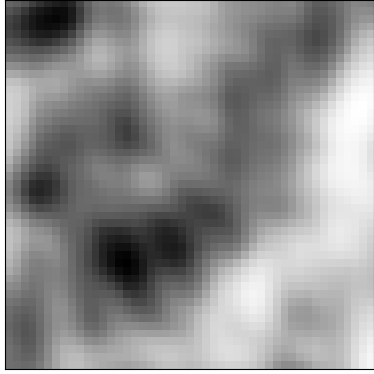}\\[1ex]
        \scriptsize{(e-iv)}\\
        \includegraphics[width=0.8\textwidth]{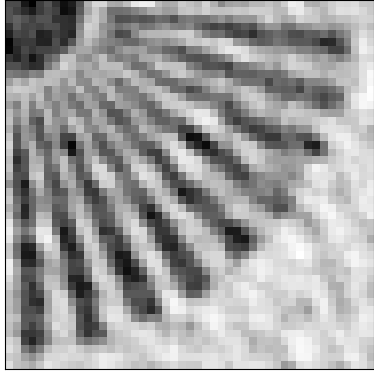}\\[1ex]
        \scriptsize{(e-v)}\\
        \includegraphics[width=\textwidth]{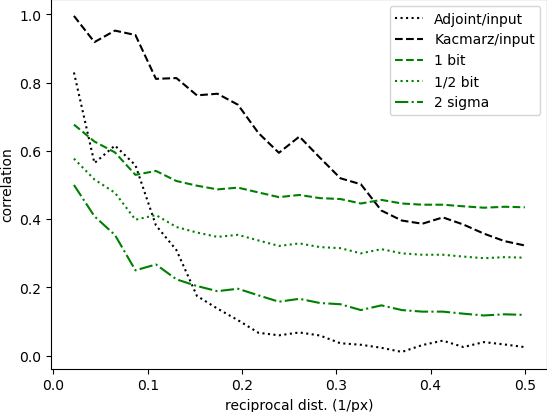}
    \end{minipage}%
    \begin{minipage}{0.166\textwidth}
        \centering
        \scriptsize{(f-i)}\\
        \includegraphics[width=0.8\textwidth]{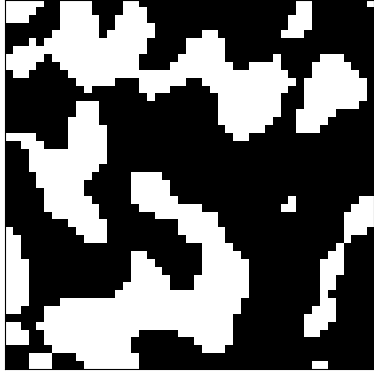}\\[1ex]
        \scriptsize{(f-ii)}\\
        \includegraphics[width=0.9\textwidth]{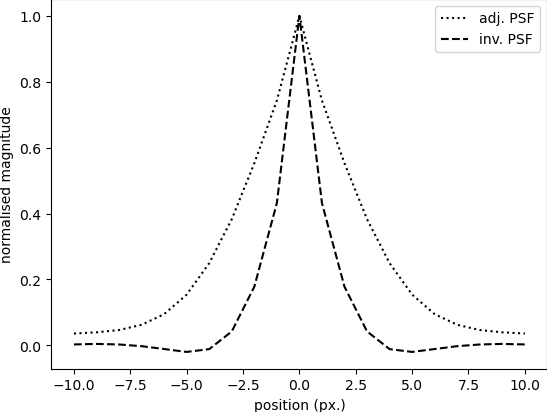}\\[1ex]
        \scriptsize{(f-iii)}\\
        \includegraphics[width=0.8\textwidth]{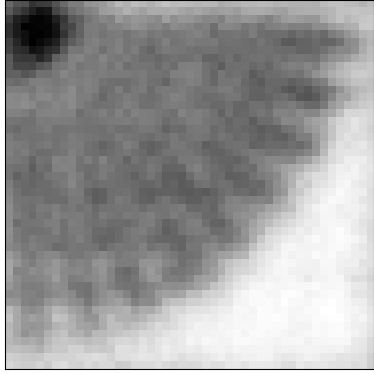}\\[1ex]
        \scriptsize{(f-iv)}\\
        \includegraphics[width=0.8\textwidth]{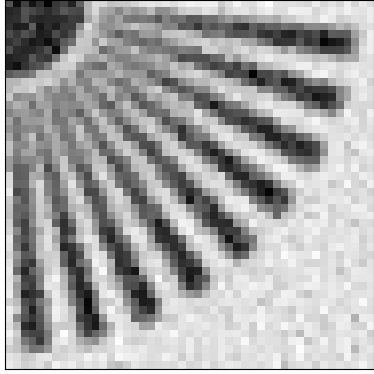}\\[1ex]
        \scriptsize{(f-v)}\\
        \includegraphics[width=\textwidth]{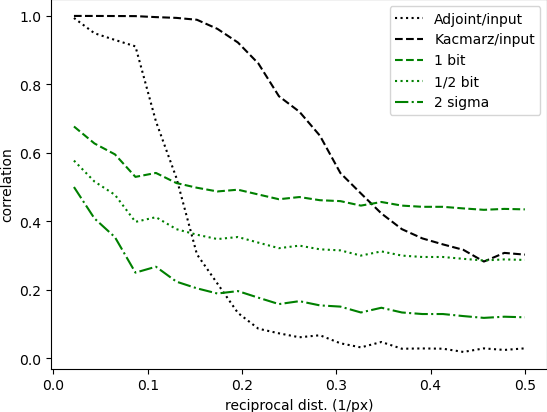}
    \end{minipage}
    \caption{Example (i) $47 \times 47$ pixel illumination patterns, (ii) their associated adjoint and inverse PSF, (iii) adjoint reconstruction, (iv) 4 iterations of Kaczmarz reconstruction, and (v) FRC analysis. All examples contain $2209$ patterns in the set and the greyscale window for the example patterns shown is [0,1]. (a) Gaussian distribution blurred by Gaussian with $\sigma = 1.0$px (same as Fig.~\ref{fig:psf}c) --- 1px stride, (b) Blurred Gaussian distribution --- 3px stride, (c) Random binary with larger feature sizes (same as Fig.~\ref{fig:psf}e) --- 1px stride, (d) Random binary with larger feature sizes --- 3px stride, (e) Random binary with double feature size --- 1px stride, (f) Random binary with double feature size --- 6px stride. All patterns were generated by scanning a larger pattern in a square grid pattern with step size as indicated by {\it stride}.}
    \label{fig:featureSize}
\end{figure}

Some examples of binary patterns with different feature sizes and different pattern strides (translation step sizes) are presented in Fig.~\ref{fig:featureSize} for comparison. In all cases the patterns are characterized to the maximum, 2 pixel, resolution. We observe that the adjoint PSF has a triangular shape resulting from the convolution of step functions of binary features with the specified size. It can also be observed from the inverse PSF profiles, the Kaczmarz reconstructed images, and the FRC plots, that Kaczmarz iteration can attain maximum resolution. In an ideal scenario, feature size in the masks is not important, rather the critical factor is the sharpness of the features in the patterns.


In practice, however, these systems are ill-conditioned and iterative inversion schemes are highly unstable (subject to noise, pattern misalignment, etc.). It is prudent not to rely on these schemes to extract the ghost-imaging information. Conversely, the adjoint is a robust operation. Orthogonal masks are ideal in this situation, as the adjoint is then equal to the inverse. However for other types of mask, we can improve the adjoint reconstruction by making the set of patterns closer to orthogonal. This can be achieved by setting the pattern transverse-translation step to be greater than or equal to the minimum feature size. The results in Figs.~\ref{fig:featureSize}b and \ref{fig:featureSize}d show that the adjoint image is significantly improved by this strategy. The tails of the adjoint PSF approach zero in these cases; it is these non-zero tails in (a) and (c) that can cause the low-frequency artifacts. We also note that the artifacts in the Kacmarz-reconstructed images are higher frequency (more noise like). In the FRC plots of Fig.~\ref{fig:featureSize} we observe that the lower-frequency artifacts in (a) and (c), with 1px step size pattern translation, cause significant degradation of the correlations at the coarse scale, making it difficult to even resolve the image, while for (b) and (d), with larger step size pattern translation, we observe a more typical FRC trend with high correlation at low frequencies, up to a relatively sharp transition into the levels of noise.


\subsubsection{Optimizing pattern translation stride}

We observe in Fig.~\ref{fig:featureSize} that pattern translation stride, or step size, can affect the robustness and quality of ghost image recovery depending on the feature sizes present, as well as the resolution of pattern characterization. In this section we present three methods to estimate an appropriate transverse-translation step size, to optimize GI performance.

\begin{figure}
    \centering
    \begin{minipage}{0.33\linewidth}
        \centering
        \scriptsize{(a)}\\
        \includegraphics[width=\linewidth]{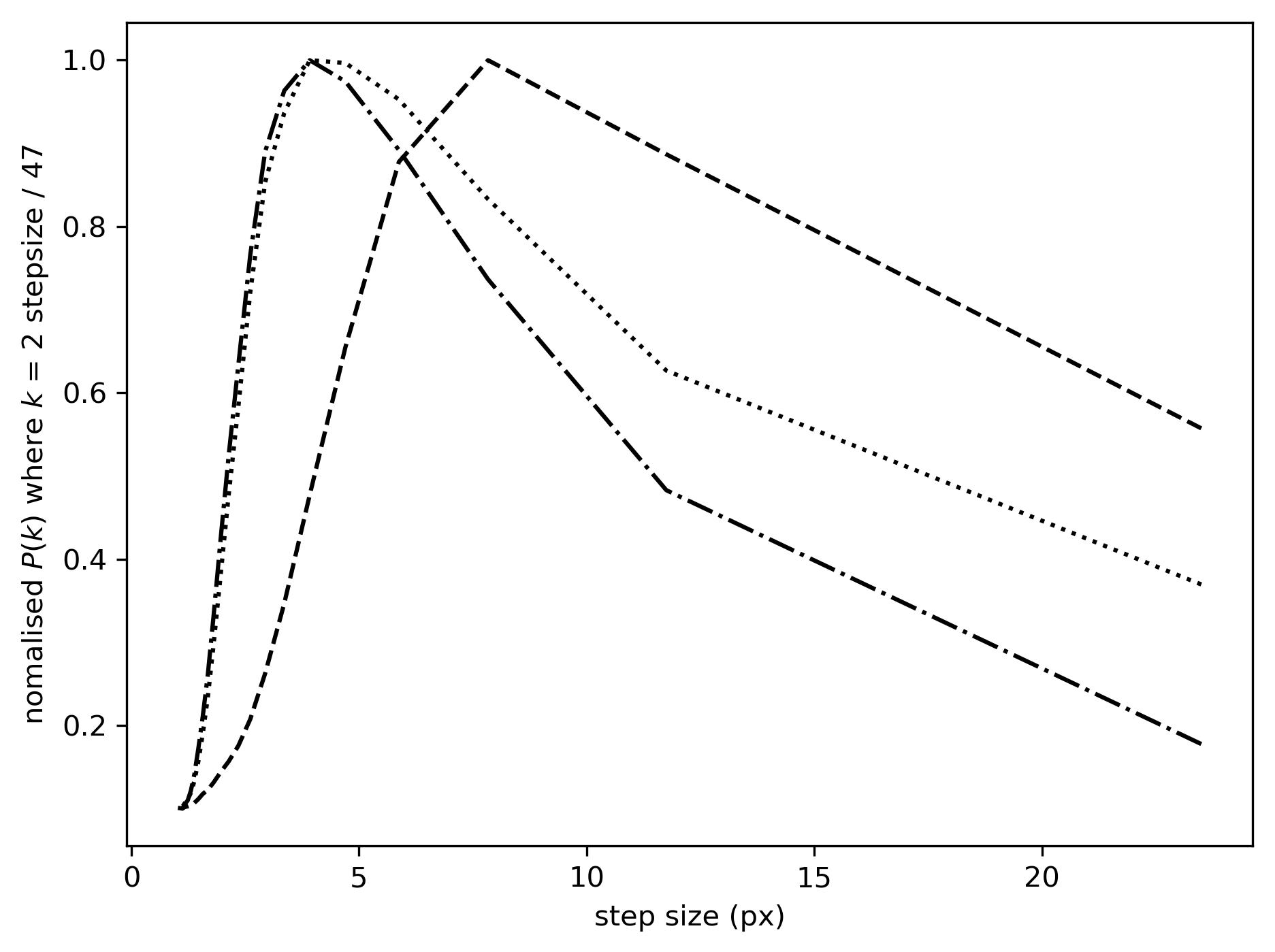}
    \end{minipage}%
    \begin{minipage}{0.33\linewidth}
        \centering
        \scriptsize{(b)}\\
        \includegraphics[width=\linewidth]{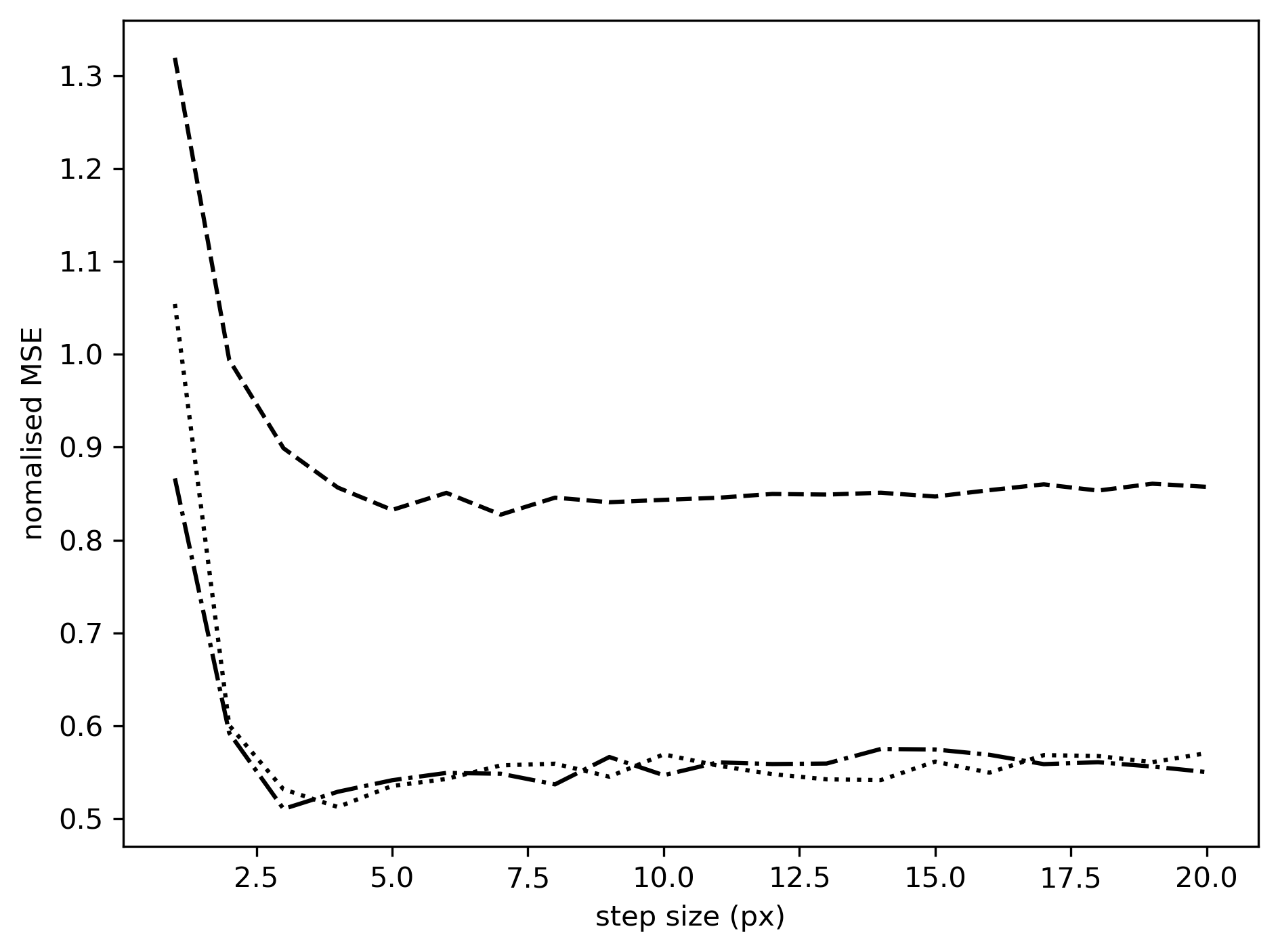}
    \end{minipage}%
    \begin{minipage}{0.33\linewidth}
        \centering
        \scriptsize{(c)}\\
        \includegraphics[width=\linewidth]{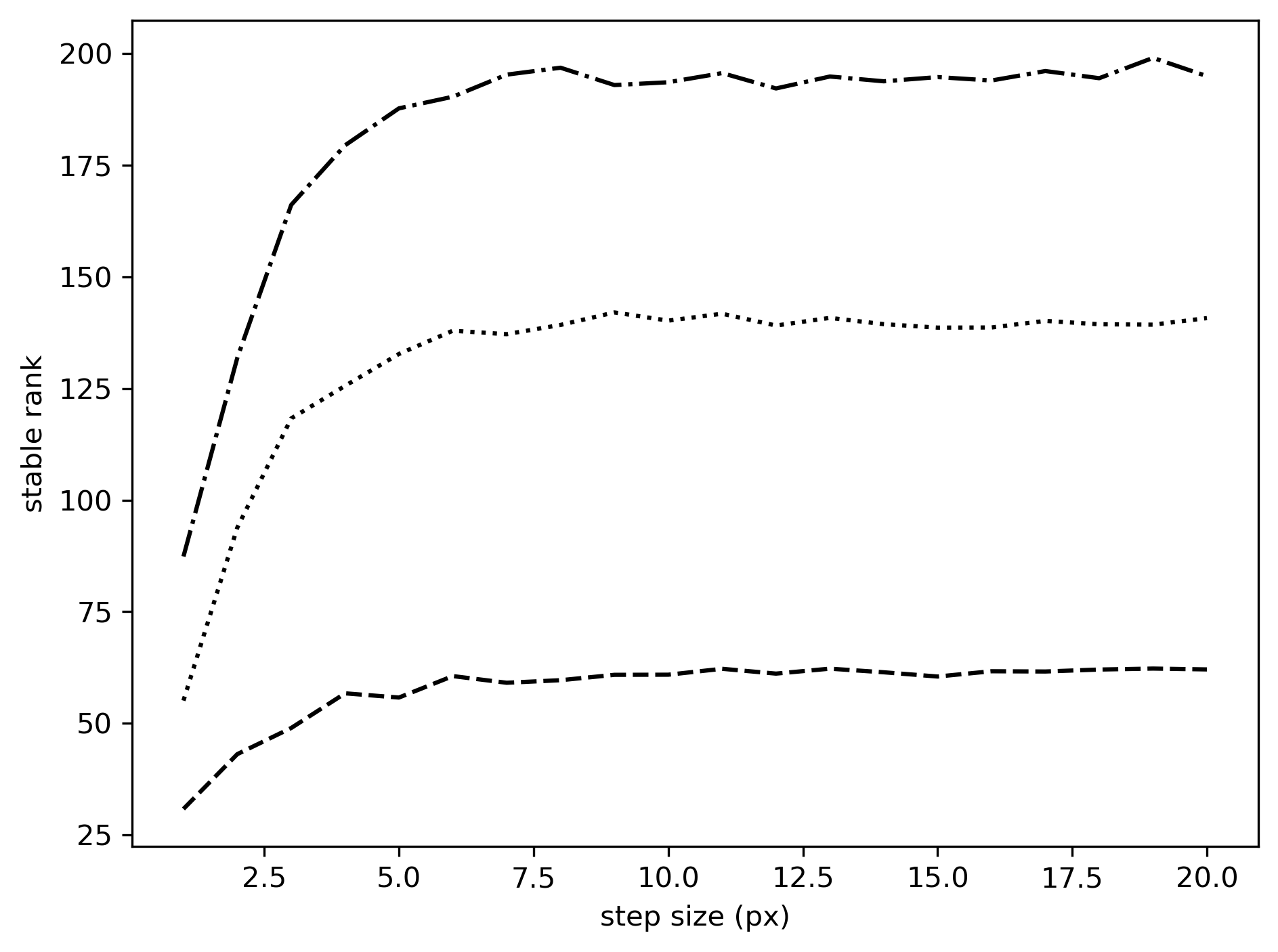}
    \end{minipage}
    \caption{Methods to determine translation step size. Each demonstration is applied to $47 \times 47$ pixel masks with [...] a Gaussian distribution blurred by a Gaussian with $\sigma = 1.0$, [---] a binary distribution with a medium feature size (approx. 2.6px), and [-.-] a binary distribution with a large feature size (approx. 4.8px). (a) normalized mask power spectrum with step size defined as half a wavelength, i.e., $47/2k$ for spatial frequency $k$; (b) normalized mean square error (MSE) as a function of step size; (c) stable rank as a function of step size.}
    \label{fig:stepSize}
\end{figure}

\paragraph{Method 1} Analyzing the Fourier spectral properties of the patterns, we can identify a characteristic length scale (if present\footnote{We here allude to the fact that, for fractal masks, a characteristic length scale does not exist.}). A stride equal to the characteristic length (or feature size) is the minimum required to avoid pattern redundancy. The power spectra of Gaussian and Binary $47 \times 47$ pixel masks with non-trivial properties used earlier are presented in Fig.~\ref{fig:stepSize}a. The angular-averaged power spectrum gives the signal strength or variance of the features at each resolution (or spatial frequency). It can be defined as follows:
\begin{equation}
    P(k) = C \sum_{k_x} \sum_{k_y} |\mathcal{F}\{A\}(k_x,k_y)|^2 \delta\left(k-\sqrt{k_x^2 + k_y^2}\right).
\end{equation}
Here $\mathcal{F}\{f\}(k_x,k_y)$ denotes the 2D discrete Fourier transform (DFT) of an image, $f(x,y)$, $k$ is the radial spatial frequency (presented in units of cycles/image), and $C$ is a normalization constant that is chosen to make the maximum value equal unity. Note that in Fig.~\ref{fig:stepSize}a the $x$-axis has been modified to be a function of step size, defined as half a wavelength or $47/2k$ for spatial frequency $k$. We have investigated: (i) Gaussian distributions blurred by a Gaussian with $\sigma = 1$ (as used in Fig. \ref{fig:psf}c); (ii) Binary distributions with medium-sized features (approximately 2.6px, similar to that for (i)); (iii) Binary distributions with large-sized features (approximately 4.8px). We observe in this figure that the modified power spectrum of both (i) and (ii) peak at a step size or stride of 4px. The power spectrum of (iii) peaks at a step size of approximately 8px.

\paragraph{Method 2} Given a {\it master mask}, $A$, this method simulates GI using patterns from subsets of this mask that are selected with different strides, investigating the normalized mean square error (MSE) of the adjoint image. Here the mean is removed from the input and adjoint images and they are scaled to have a standard deviation of 1.0 before calculating the mean squared difference. A plot of MSE with pattern stride has been presented in Fig.~\ref{fig:stepSize}b for the set of masks under investigation here. Again we observe that (i) and (ii) show similar trends, with MSE reducing rapidly until the stride is 3px. The plot for (iii) seems to decrease up to approximately a 6px stride, beyond which it is flat.

\paragraph{Method 3} Again given a {\it master mask}, $A$, and a field-of-view, the stable rank (see Eq.~(\ref{eq:stableRank}) in Sect. \ref{sec:rank}) as a function of pattern stride stabilizes once stride is sufficiently large. Stable rank has been plotted for the example patterns in Fig.~\ref{fig:stepSize}c. The plots exhibit a more gradual transition than MSE (seen in Fig.~\ref{fig:stepSize}b), and a more conservative estimate of required stride emerges. Here, (i) and (ii) both seem to peak at approximately 5px, while (iii) peaks at approximately 8px.

Each of the presented stride-determination methods has advantages and disadvantages and could be preferred in different scenarios. Moreover, other methods exist. There are four length scales to consider in the determination of mask stride: (i) desired resolution of the object, (ii) characteristic length scale of the object, (iii) resolution of the mask/pattern, and (iv) characteristic length of mask features. Method 1 requires the minimum amount of information (even just a single pattern image could be enough) and is the simplest computationally, requiring just analysis of the Fourier transform of the pattern. It is therefore ideal for quick analysis of natural mask patterns. However, not all mask patterns contain a characteristic length scale, e.g., fractal masks and masks that are orthogonal under translation (since the latter have a flat Fourier power spectrum); in this case we would require a stride that is greater than all remaining length scales (i)-(iii) above, since a stride that is too small collects redundant information. Method 2 is more generally applicable and potentially more robust than method 1, but is computationally more expensive, not user-friendly for experimentalists, and requires knowledge of the full master mask. Since GI is simulated in this method, it requires an object/image; while this makes the method less general, it does have the advantage that it can also include object properties (i) and (ii) in the determination. It could also be used to validate or verify the results from method 1. Method 3 is similar to method 2 in complexity and information required. It is more general, being purely based on properties of the masks, however, does not probe length scales (i) and (ii) and it does not seem to be as precise. Finally, the field-of-view and/or number of masks used in the calculation must be limited in scope, to be able to perform singular value decomposition of matrix $\mathbf{A}$.

\subsection{Recorded information}
\label{sec:rank}

While the number of measurements taken in ghost imaging certainly dictates experiment time, it may or may not also dictate the image quality. More bucket values recorded with new illumination patterns will generally yield a superior ghost image. However, if a new illumination pattern is a copy of a previous one, or can be constructed as a linear combination of two or more previous patterns, then the new measurement is completely redundant. At best it merely contributes to the signal quality of previous measurements, and so can be designated a wasted measurement.

A set of maximally independent illumination patterns is required to minimize these wasted measurements. The number of effectively unique measurements (or nondegenerateness) can be defined as the {\it rank} of the set of illumination patterns in matrix form, $\mathbf{A}$. Here, $\mathbf{A}$ has the number of rows equal to the number of patterns and the number of columns equal to the number of pixels representing each pattern. Theoretically the rank can be defined as the number of non-zero singular values of $\mathbf{A}$, however, practically speaking it is defined as the number of singular values above some noise floor.

Let $\mathbf{A = U \Sigma V ^T}$ be the singular value decomposition of $\mathbf{A}$ where the columns of $\mathbf{U}$ and $\mathbf{V}$ are the left- and right-singular vectors of $\mathbf{A}$ respectively. $\mathbf{\Sigma}$ is a diagonal matrix with its entries being non-negative real numbers that are the singular values of $\mathbf{A}$. These singular values when ordered from largest to smallest are denoted $\sigma_1 \geq \cdots \geq \sigma_n$. Here $n$ is the minimum of the number of patterns and the number of pixels. 

\begin{figure}
    \centering
    \begin{minipage}{0.5\linewidth}
        \centering
        \scriptsize{(a)}\\
        \includegraphics[width=\linewidth]{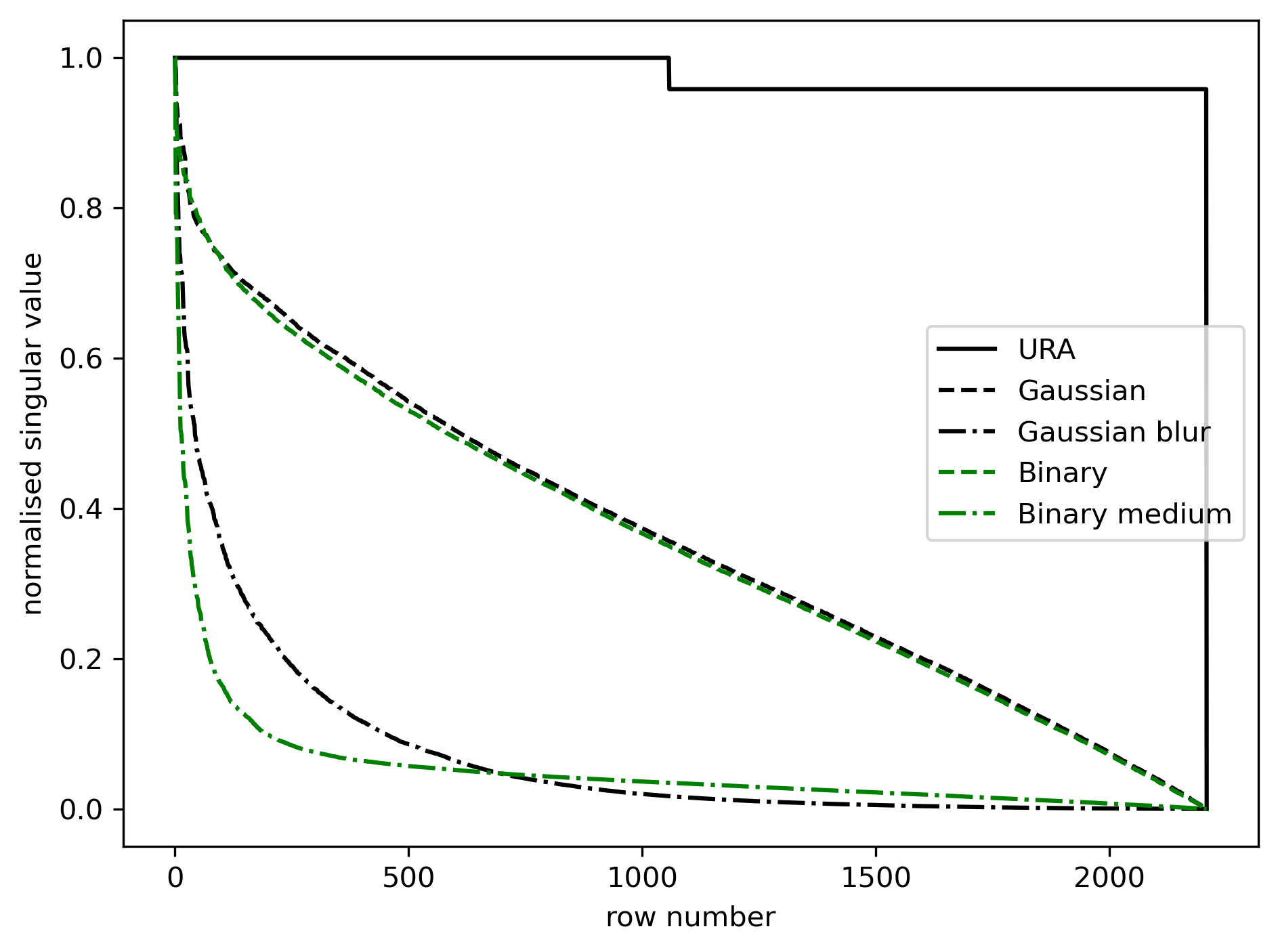}
    \end{minipage}%
    \begin{minipage}{0.5\linewidth}
        \centering
        \scriptsize{(b)}\\
        \includegraphics[width=\linewidth]{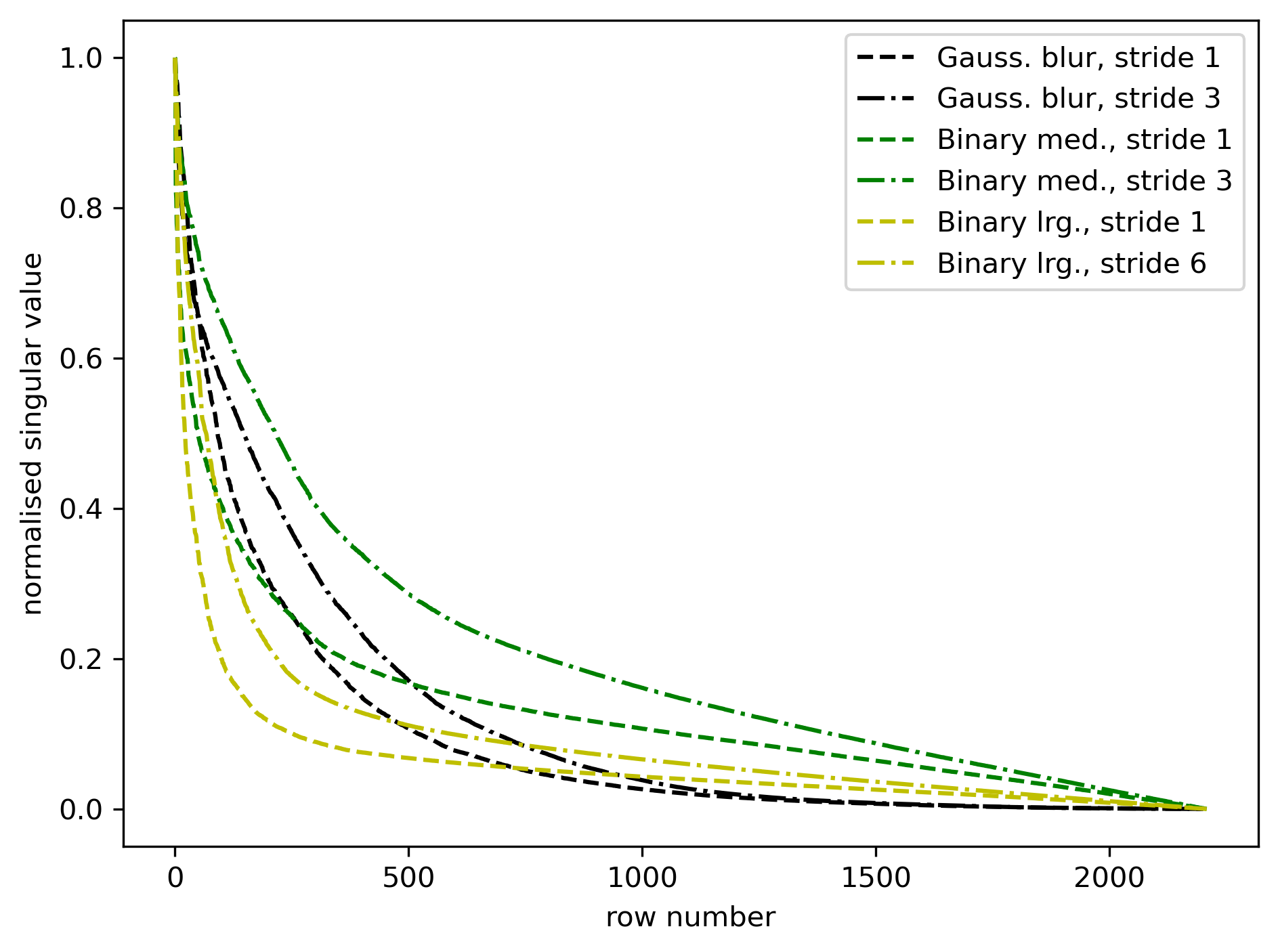}
    \end{minipage}
    \caption{(a) The normalized singular values plotted in descending order of magnitude for (a) the set of examples included in Fig.~\ref{fig:psf}, and (b) the set of examples included in Fig.~\ref{fig:featureSize}}\label{fig:SvdStride}
    \label{fig:SVD}
\end{figure}

To avoid the instability of rank determination, we recommend using the {\it numerical rank} proposed by Rudelson and Vershynin \cite{rudelson2007sampling}. It is defined as the Frobenius norm of $\mathbf{A}$ divided by the spectral norm of $\mathbf{A}$, i.e.,
\begin{equation}\label{eq:stableRank}
    \mathrm{rank}(\mathbf{A}) = \frac{\|\mathbf{A}\|^2_F}{\|\mathbf{A}\|^2_2} = \frac{\sum_j \sigma_j^2}{\sigma_1^2}.
\end{equation}
This is also know as the {\it stable rank} since it avoids specifying an arbitrary cut-off noise level as required to compute the exact rank of a matrix. It always assumes a value less than or equal to the exact rank (but exactly equal if $\mathbf{A}$ is orthogonal or has exact rank 1) and is stable under small perturbations of $\mathbf{A}$ such as from the presence of noise. The normalized singular vales for the example illumination patterns in Fig.~\ref{fig:psf} have been plotted in Fig.~\ref{fig:SVD}a. From these singular values we can calculate the stable rank according to Eq.~(\ref{eq:stableRank}).%


For the data presented in Fig.~\ref{fig:SVD}a, the stable ranks are as follows: URA patterns -- $2110$; Gaussian patterns -- 361; blurred Gaussian patterns -- 59.1; binary patterns -- 391; binary patterns with medium sized features --- 92.1. The full rank of all these systems is $2209$, so we observe that (apart from the orthogonal case of the URA patterns) the stable rank significantly underestimates the rank. However, looking at the results in Fig.~\ref{fig:psf} (particularly the adjoint recovered image), it does give a good indication of the relative performance of the sets of patterns. We therefore propose that the stable rank can be used to compare patterns.

To investigate the effect of increasing pattern translation step size, or stride, between measurements, the normalized singular values for the examples in Fig.~\ref{fig:featureSize} have been presented in Fig.~\ref{fig:SvdStride}b. For the Gaussian-blur case, the rank is 77.4 for a stride of 1px; this becomes a rank of 104 with a stride of 4px. For the binary patterns with medium sized features, the rank is 70.0 when the stride is 1px, while rank increases to 170 when stride is 3px. For the binary patterns with large sized features, the rank is 26.0 when the stride is 1px, while rank increases to 60 when stride is 6px.

From the shape of the plots in Fig.~\ref{fig:SvdStride}b, the reported values of rank, and the results in Fig.~\ref{fig:featureSize}, we observe that increasing the step size to be similar to feature size improves all metrics. We observe from the plots in Fig.~\ref{fig:SvdStride}b that the normalized singular values are improved with a larger stride for $\sigma$ values greater than 0.003. This is not too important since, beyond this, information can only be extracted in cases with extremely high signal-to-noise ratio measurements.







Note that here we use normalized singular values (scaled so $\sigma_1=1$) rather than the absolute singular values. This is because we are not considering the absolute level of information in each mask, which is affected by mask material properties such as mean and variance in mask transmission (opacity). Our concern is the relative amount of extra information provided by each mask pattern or, conversely, the relative amount of redundancy in each extra mask pattern. We conclude by noting that the actual (non-normalized) singular values are used in Sec.~\ref{sec:noise} when considering the robustness of masks to noise.

\subsection{Information density, or mask footprint}
\label{sec:footprint}

Assuming that a mask is being used to generate the set of patterned illuminations in GI, then ideally the smallest mask possible is preferable, without sacrificing reconstructed ghost image quality. A smaller mask requires less translation, (which generally allows cheaper or more precise translation stages), less material and less fabrication per mask. We want to minimize the footprint of the total set of patterns, or maximize information content achievable for given mask size. There are two categories of considerations: those for designed masks and those for natural masks.

For natural masks, typically (but not necessarily) composed of grains or foams, the step size, or stride, between illumination patterns generated is one of the primary properties dictating the required total mask dimensions. Section \ref{sec:psf} demonstrates that the optimal step size is determined by the feature (grain/bubble) size; a mask composed of smaller features is therefore preferable by this metric. However, smaller features typically result in less pattern contrast and perhaps a trade off must be made, e.g., when choosing between sandpaper grades, or perhaps a different grain material could be used, e.g., metal filings instead of sand grains.

For designed masks there are more choices to make; one would ideally select an orthogonal set of masks. One of the most commonly used masks for this case is the Hadamard basis set. However, this requires an entirely unique pattern for each measurement and requires an extremely large mask footprint. Given a set of $N \times N$ pixel patterns, the total mask size scales with $N^4$ (since we require $N^2$ unique patterns each of $N \times N$ pixels). Even for modest image dimensions like $32 \times 32$ pixels, a fabricated mask that is 1000 times larger than the imaging FOV is required. Masks that are orthogonal under translation are preferable in this sense, since these can be fabricated as a $(2N-1)^2$ pixel array and so scale with $N^2$ rather than $N^4$. Some example masks in this category include uniformly redundant arrays (such as those generated by quadratic residue \cite{Gottesman1989newFamily}), or masks based on the finite Radon transform (e.g., \cite{cavy2015construction}).
Many of these masks should also be good candidates for compressed sensing since they have restricted isometry in many compressible transform spaces (\cite{candes2008introduction} provides an excellent introduction to compressed sensing).

Another question that arises in designed masks is that of {\it differential masks} \cite{ferri2010differential,Sun--DifferentialComputationalGhostImaging2013,welsh2013fast}. These require a pair of masks for each illumination pattern: one to encode the positive information of the pattern and one to encode the negative information. This doubles the mask footprint required for a given set of patterns. Positive/negative mask pairs are certainly a preferable option when implementing ternary, $\{-1,0,1\}$, patterns that contain many zeros, e.g., Haar wavelet patterns, and the overall experiment objective is to minimize dose. They are also useful for removing the effects of a slowly varying background signal (see Sec.~\ref{sec:fluxVar}). However, if mask footprint is a major consideration then this could also be achieved by monitoring the background signal with an additional independent sensor. Note that, from the perspective of photon shot-noise, binary differential masks with 50\% coverage provide no advantage over conventional binary masks, however, they can be advantageous experimentally when an unstable but slowly varying illumination flux and/or profile is unavoidable; this point is discussed in more detail in Sec.~\ref{sec:fluxVar}.

\subsection{Robustness to experimental limitations}
\label{sec:robust}

\subsubsection{Robustness to photon shot-noise}
\label{sec:noise}

Consider experimental limitations related to measurement noise that can compromise ghost-image quality. These include: 1) limited experiment time which, due to photon shot-noise, limits the signal-to-noise ratio (SNR) of the measurements; 2) the exposure time, e.g., controlled by a shutter, may have limited accuracy; or 3) the flux of the illumination may not be constant and have time dependent component. Illumination patterns that are robust to these limitations are preferable. What property of the patterns can be used to predict or compare performance in this respect? Here we will first present a technique to demonstrate and compare robustness by simulation, then we will show that the decay rate of the singular values of the set of patterns is a good predictor of robustness.

Photon shot-noise becomes significant for limited exposure times or photon flux, i.e., low dose, and is modeled by the Poisson distribution. The standard deviation in the number of photons measured equals the expected number of photons. We can explore the quality of ghost image reconstruction as a function of measured signal-to-noise ratio (i.e., up-stream photons per pixel per measurement) by simulation.

\begin{figure}
    \centering
    \begin{minipage}{0.5\linewidth}
        \centering
        \scriptsize{(a)}\\
        \includegraphics[width=\linewidth]{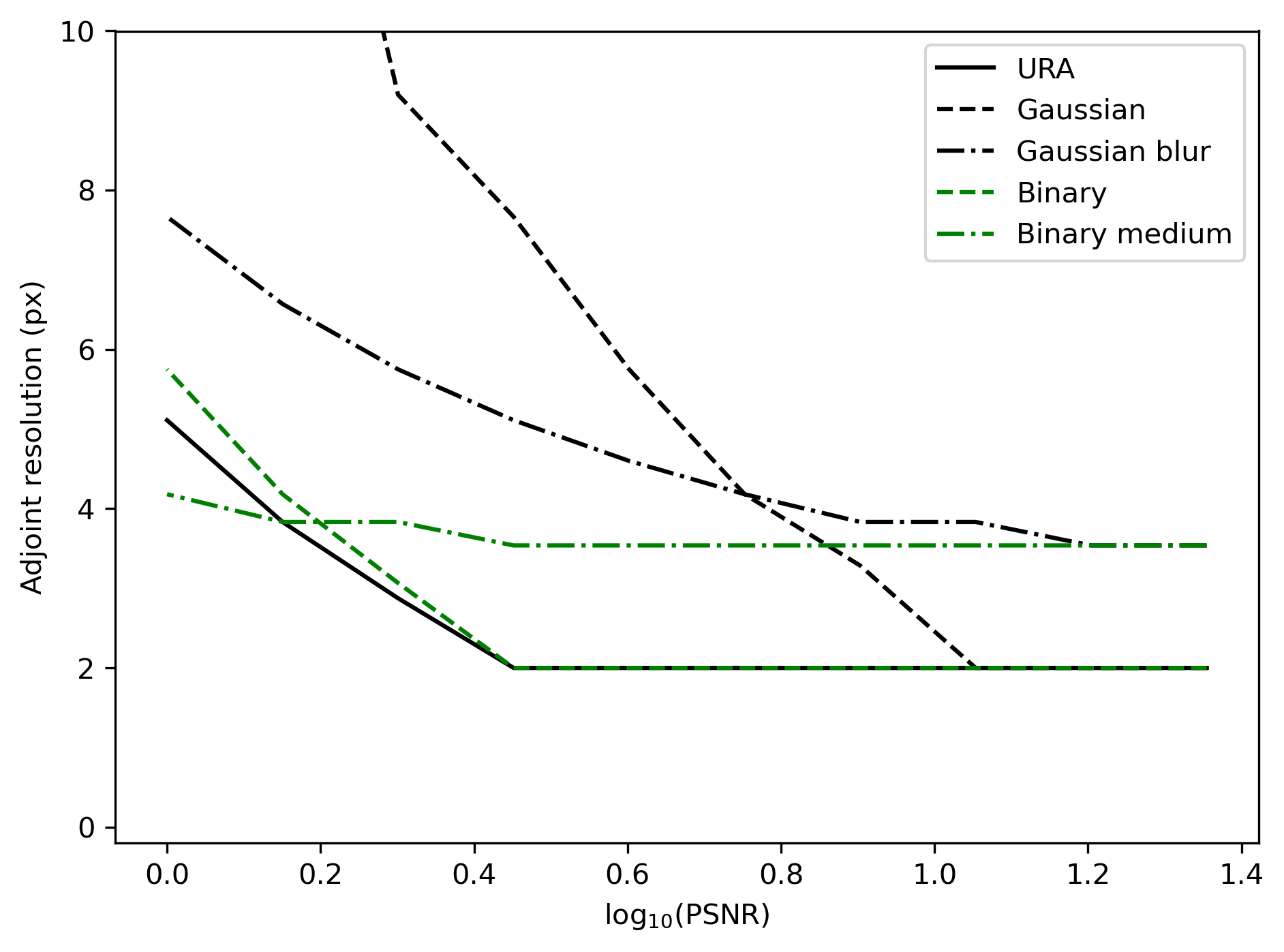}
    \end{minipage}%
    \begin{minipage}{0.5\linewidth}
        \centering
        \scriptsize{(b)}\\
        \includegraphics[width=\linewidth]{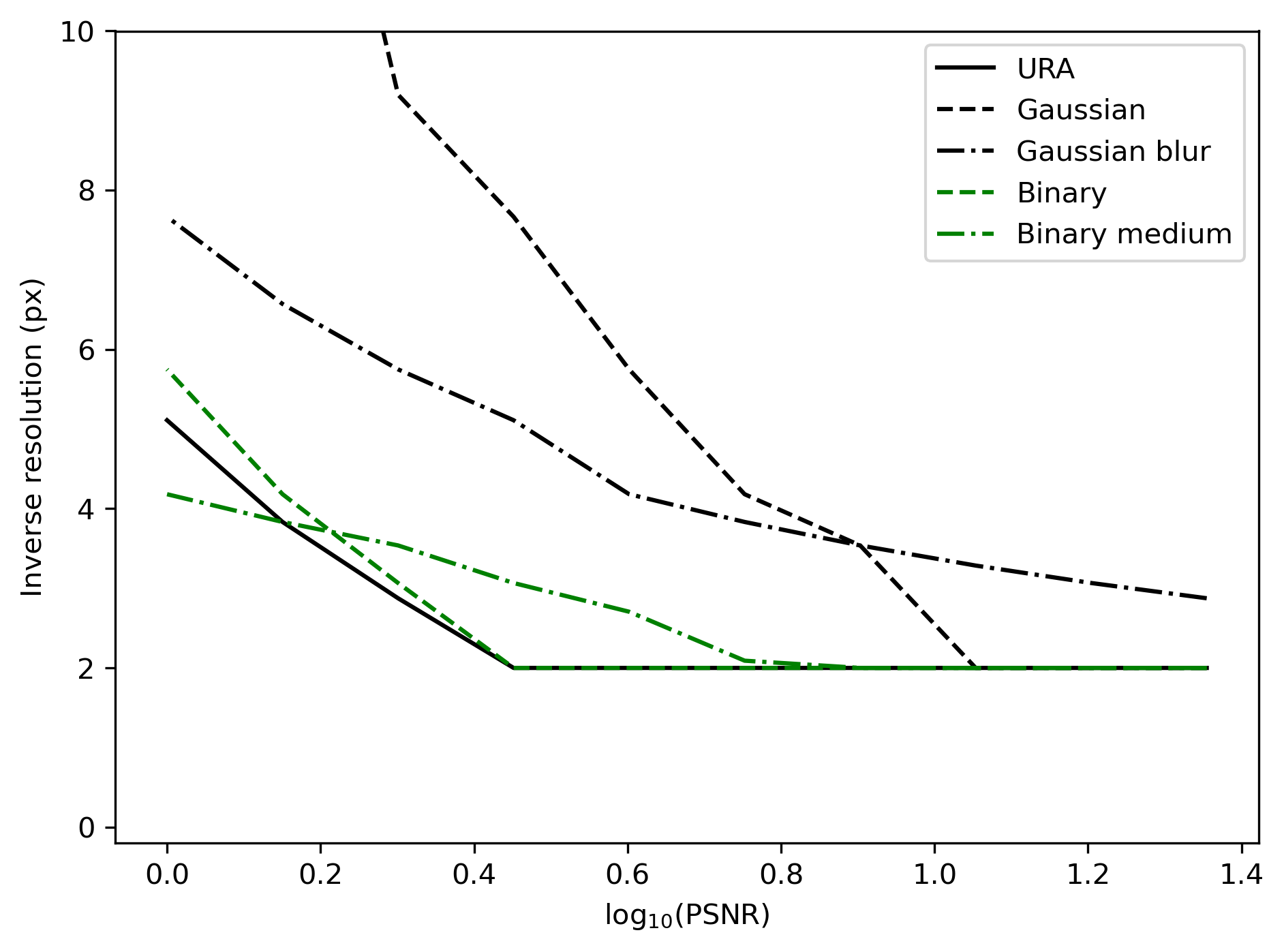}
    \end{minipage}
    \caption{Achievable GI resolution as a function of signal-to-noise ratio (modified through input x-ray flux) for the set of mask examples included in Fig.~\ref{fig:psf}}
    \label{fig:resNoise}
\end{figure}

In Sec.~\ref{sec:psf} we found that the FWHM of the PSF (either adjoint or inverse) is not a good metric of resolution since it is the tails of the PSF that can degrade image quality. Instead we propose that resolution determined by FRC be used as a metric. In what follows, for each photon flux under investigation, we have averaged the FRC plots for 100 binary images of circles with uniform random radii and center positions. We have compared both the adjoint and inverse (4 Kaczmarz iterations, $\lambda=0.5$) reconstructed images with the input images and define resolution where the FRC plots cross the 1-bit reference line. The results for the example illumination patterns in Fig.~\ref{fig:psf} are presented in Fig.~\ref{fig:resNoise}.

We observe in Fig.~\ref{fig:resNoise} that under these noisy conditions, Kaczmarz iteration does little to improve the resolution over adjoint reconstruction for the masks with small feature sizes, i.e., URA, Gaussian, and binary patterns. However, it can improve things for those with larger feature sizes, i.e., blurred Gaussian and binary mask with medium sized features. Looking at when each plot breaks from full resolution, we see that the URA and binary masks provide full resolution under the most noisy conditions and have very similar performance; the binary mask with medium feature size is next best, followed by the Gaussian mask; the blurred Gaussian mask has the worst performance. We also note that the rate of resolution degradation is much slower for the masks with larger feature size; they cross the plots of small=feature masks and become preferable in certain scenarios.

From the normalized singular value decomposition (SVD) curves in Fig.~\ref{fig:SVD}a, we cannot predict these results. We expect to see that the URA mask provides the best resolution; the small-feature Gaussian and binary masks should be next and have very similar performance; the large-feature masks should give the worst resolution, with the Gaussian mask degrading at a higher noise level but improving, relatively, as noise increases.

\begin{figure}
    \centering
    \begin{minipage}{0.5\linewidth}
        \centering
        \scriptsize{(a)}\\
        \includegraphics[width=\linewidth]{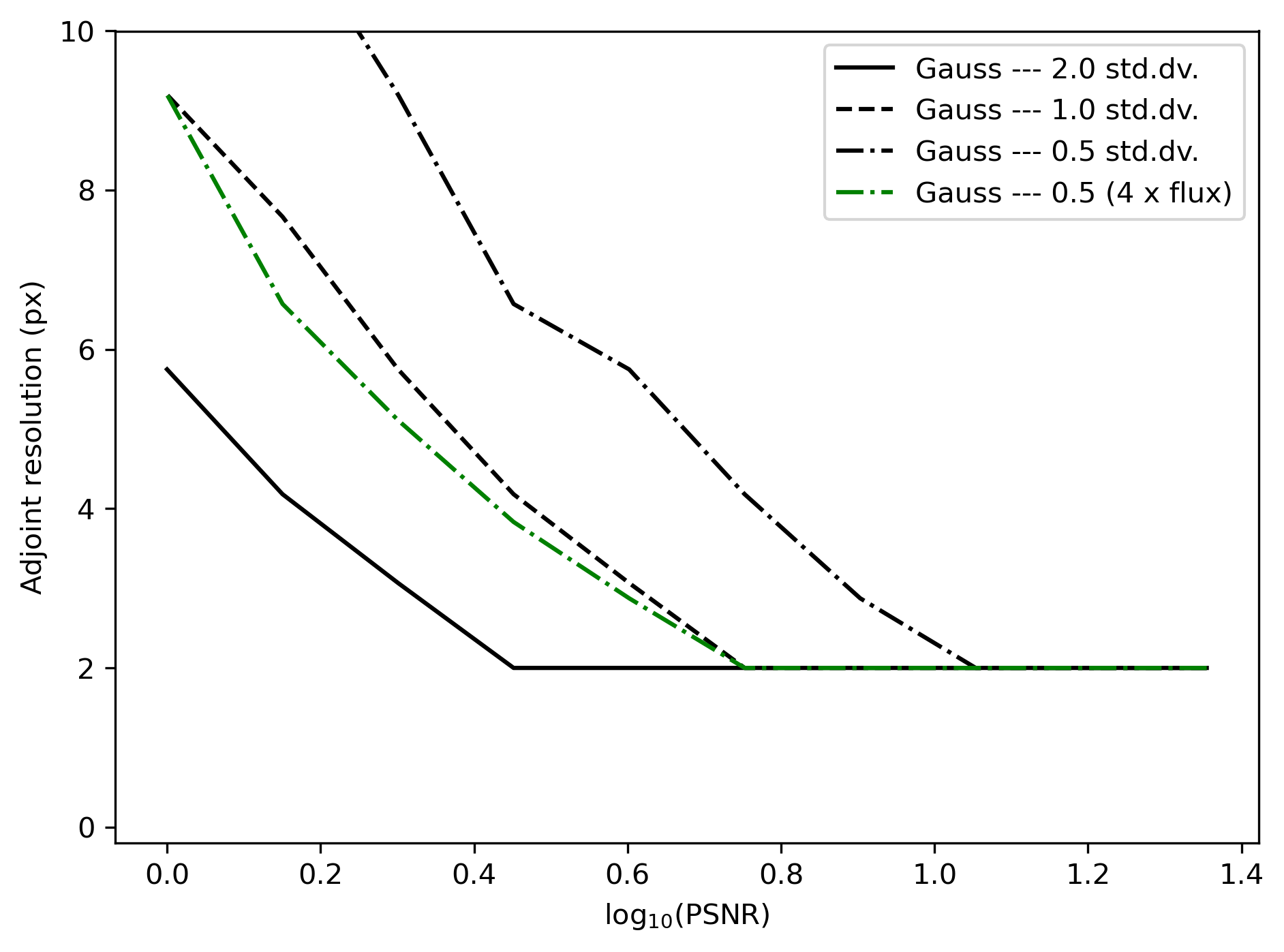}
    \end{minipage}%
    \begin{minipage}{0.5\linewidth}
        \centering
        \scriptsize{(b)}\\
        \includegraphics[width=\linewidth]{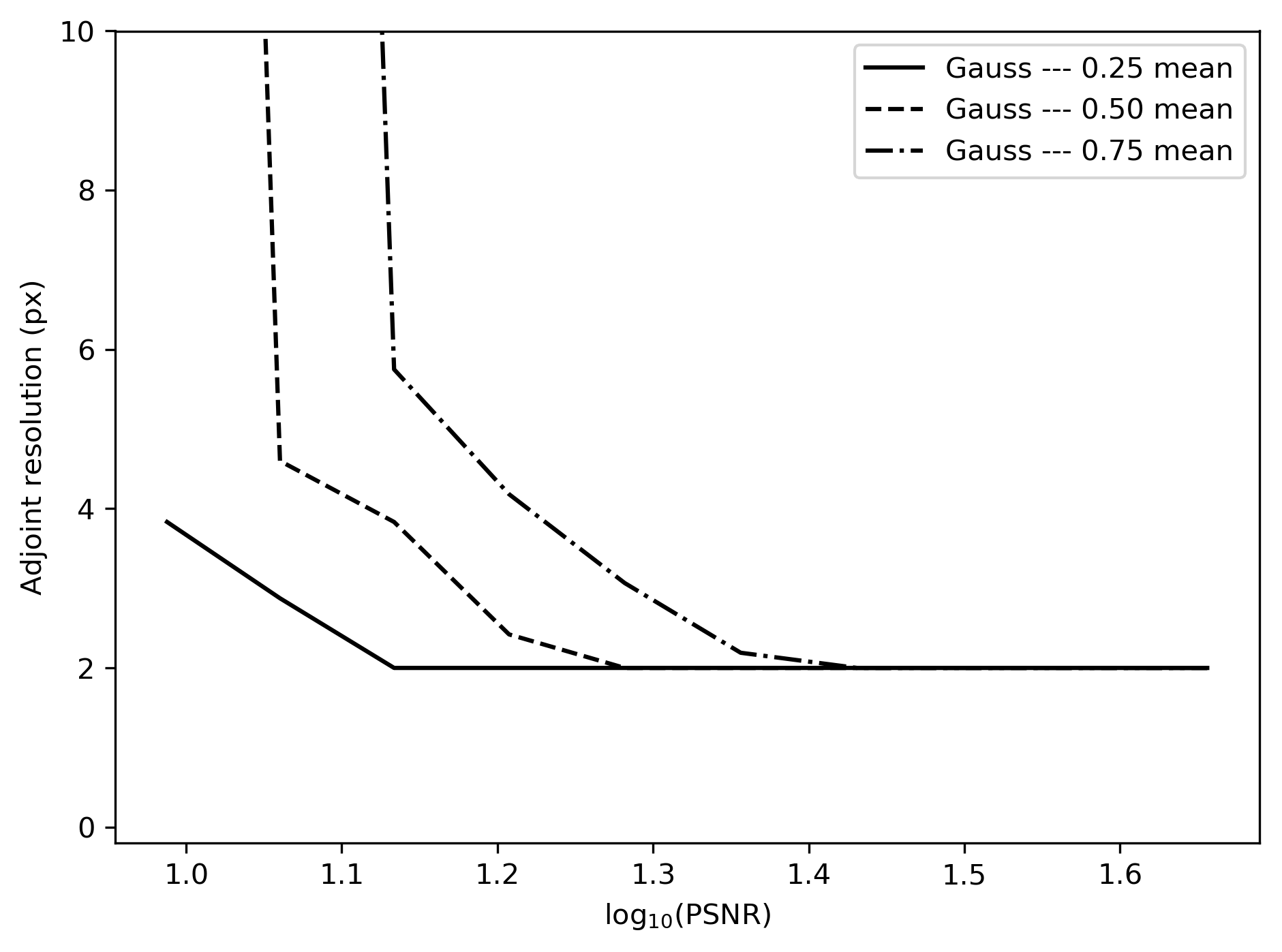}
    \end{minipage}
    \caption{Achievable GI resolution as a function of signal-to-noise ratio (modified through input x-ray flux) for Gaussian masks. (a) constant mean mask transmission of 0.5 with different levels of mask variance and exposure time; (b) constant mask transmission variance of 0.25 with different mask mean transmission. Note that only results for the adjoint GI reconstruction are presented; Kaczmarz iteration gave no significant improvement in performance.}
    \label{fig:resNoiseVarMean}
\end{figure}

This can be understood more clearly when considering the factors affecting GI performance under low-dose conditions. From Ref.~\cite{kingston2021inherent} we note that the MSE of GI reconstruction is proportional to the pattern mean and inversely proportional to the pattern variance. Assuming that an increase in MSE corresponds to a decrease in resolution, we expect that increasing pattern variance or decreasing pattern mean will improve resolution for a given illumination dose. These relationships have been demonstrated in Fig.~\ref{fig:resNoiseVarMean}.

\begin{figure}
    \centering
    \begin{minipage}{0.5\linewidth}
        \centering
        \includegraphics[width=\linewidth]{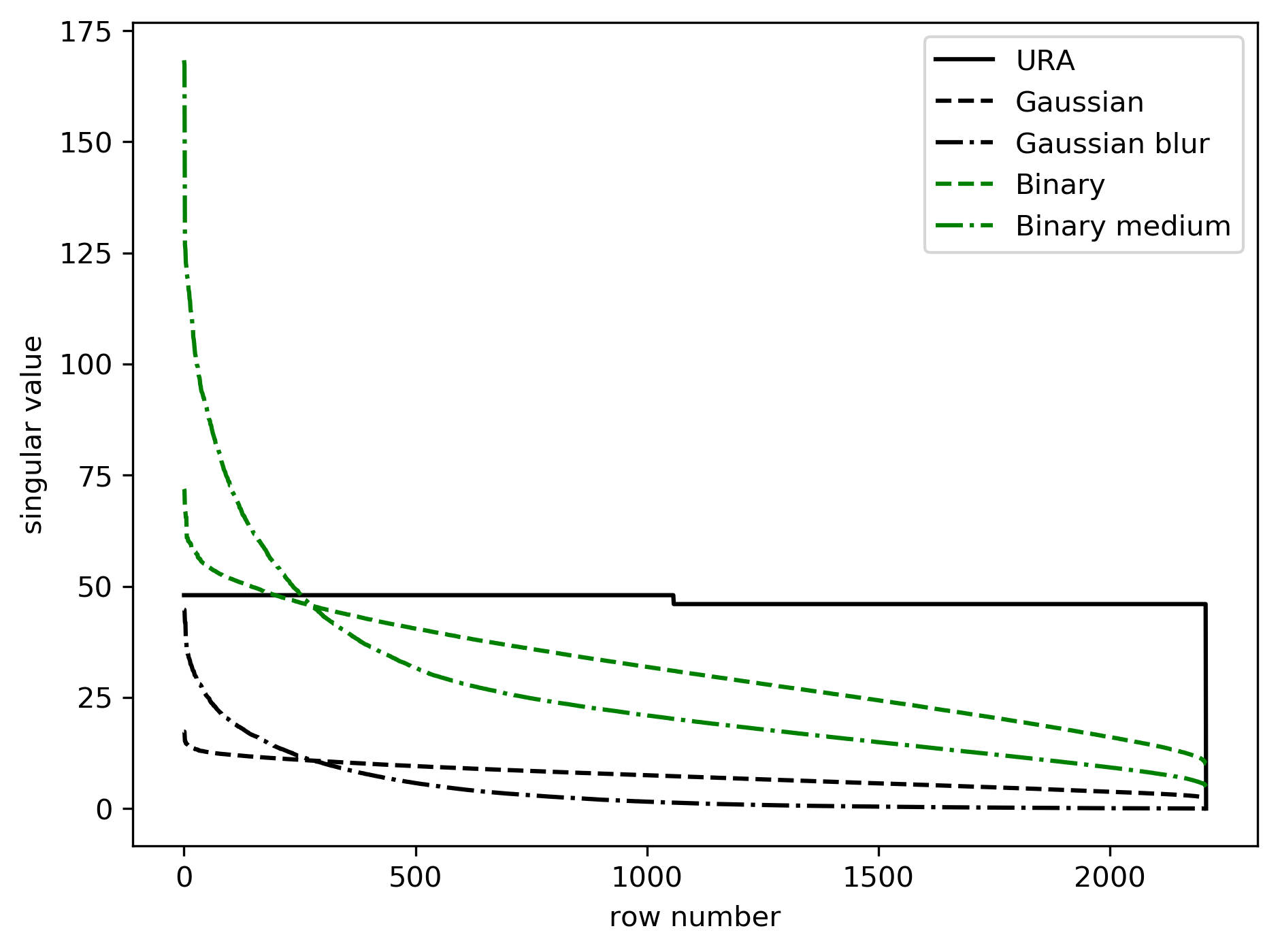}
    \end{minipage}
    \caption{The unnormalized singular values plotted in descending order of magnitude, for the set of examples included in Fig.~\ref{fig:psf}.}
    \label{fig:SVDunnorm}
\end{figure}

The normalized SVD plots have been scaled by a quantity related to the pattern variance\footnote{Specifically, the scaling for the plots is $\sigma_1$, and the variance of the (mean-adjusted) patterns is $\|A\|_F^2 = \sum_j \sigma_j^2$.}. This pattern-dependent scaling means that any absolute noise floor would assume a different value for each pattern type. In Fig.~\ref{fig:SVDunnorm} we have plotted the unnormalized SVD values. Here we can assume a consistent scale of noise across the pattern types. We can predict GI reconstruction resolution as a function of illumination dose by determining how many singular values drop below a noise floor as it is increased. Looking at this figure, we can qualitatively predict that the size of the larger singular values (on the left hand side of the plot) predicts performance in a high signal-to-noise setting. Larger singular values yields better performance, so we can understand why the larger-feature masks degrade more slowly with increased noise than small-feature masks. By contrast, the size of the smaller singular values (on the right hand side of the plot) predicts performance in a low signal-to-noise setting. Again, larger values yields better performance and predicts the order in which the plot of each pattern type deviates from full resolution as noise is increased.

Note that in order to calculate the SVD, the mean pattern transmission has been removed from the mask images. If two candidate masks have a similar performance according to the SVD plots, the mask with the lower mean transmission should be selected.

\subsubsection{Robustness to misalignment}
\label{sec:misalignment}

When optical dynamic beam-shaping techniques such as a data projector, SLM, or DMD are not available, structured illumination is typically produced using a {\it mask} that is translated to produce different patterns. Computational GI involves characterizing these masks to the required resolution and estimating the illumination pattern based on the current mask position, or pre-recording the illumination patterns for a set of mask positions with a pixelated detector and then repeating these mask positions for the bucket measurements using a single pixel camera and the object in place. In this section we consider the case where the set of expected patterns associated with each bucket value is misaligned, due to factors such as positioning inaccuracies, mask distortion, thermal expansion, etc. In particular, we seek to understand which properties of the mask patterns are more robust to such experimental limitations.

Of relevance here is the perturbation theory of linear least-squares problems, described by Golub and Van Loan in Chapter 5 of Ref.~\cite{vanLoan1996matrix}, for example. This theory can be applied to the misalignment of patterns during the ghost imaging measurement process, to estimate bounds on the accuracy of the reconstructed ghost image. Suppose we intend to solve the {\it perfect} least-squares ghost imaging problem stated earlier, i.e., $\mathbf{At} = \mathbf{b}$, where $\mathbf{A} \in \mathbb{R}^{m \times n}$ has $m \ge n$ and full column rank (i.e.,~overdetermined with linearly independent columns). However, because of misalignment and measurement errors, in practice we solve the {\it real} least-squares problem
\begin{equation}
    \mathbf{A'}\mathbf{t'} = \mathbf{b'}.
\end{equation}
Provided the misalignment errors of pattern placement are not too severe (e.g., they are on the order of the pattern characterization resolution), we can say that $\mathbf{A'} \approx \mathbf{A}$ and $\mathbf{b'} \approx \mathbf{b}$ (yielding a good reconstructed ghost image $\mathbf{t'} \approx \mathbf{t}$). The final residuals are $\mathbf{r} = \mathbf{b} - \mathbf{A}\mathbf{t}$ and $\mathbf{r'} = \mathbf{b'} - \mathbf{A'}\mathbf{t'}$.

From Ref.~\cite[Theorem 5.3.1]{vanLoan1996matrix} we know that if $m \ge n$ and $\mathbf{b} \ne 0$, and the original problem satisfies
\begin{equation}
    \sin(\theta) = \frac{\| \mathbf{r} \|_2}{\| \mathbf{b} \|_2} \ne 1,
\end{equation}
and our relative measurement errors satisfy
\begin{equation}
    \epsilon := \max \left( \frac{\| \mathbf{A'} - \mathbf{A} \|_2}{\| \mathbf{A} \|_2}, \frac{\| \mathbf{b'}-\mathbf{b} \|_2}{\| \mathbf{b} \|_2} \right) < \frac{1}{\kappa(\mathbf{A})},
\end{equation}
then we have
\begin{eqnarray}
    \frac{\| \mathbf{t'}-\mathbf{t} \|_2}{\| \mathbf{t} \|_2} & \le & \left( \frac{2\kappa(\mathbf{A})}{\cos(\theta)} + \tan(\theta)\kappa(\mathbf{A})^2 \right)\epsilon + \mathcal{O}(\epsilon^2) \label{eq_perturb} \\
    \frac{\| \mathbf{r'}-\mathbf{r} \|_2}{\| \mathbf{b} \|_2} & \le & (2\kappa(\mathbf{A}) + 1)\min(1,m-n)\epsilon + \mathcal{O}(\epsilon^2).
\end{eqnarray}

Note that the relative error in the measured residual, $\mathbf{r'}$, scales as $\mathcal{O}(\kappa(\mathbf{A})\epsilon)$, while the relative error in the recovered signal, $\mathbf{t'}$, scales as $\mathcal{O}(\kappa(\mathbf{A})^2\epsilon)$. However, there is an exception if the original problem is consistent (i.e., $\mathbf{r} = 0$, as can be assumed for the noise-free ghost imaging problem), since in this case the error is only $\mathcal{O}(\kappa(\mathbf{A})\epsilon)$. Note that this last fact comes from the identity
\begin{equation}
    \tan(\theta) = \frac{\| \mathbf{r} \|_2}{\sqrt{ \| \mathbf{b} \|^2_2 - \|\mathbf{r}\|^2_2 }}.
\end{equation}
Lastly, we note that the condition number, $\kappa(\mathbf{A})$, is defined as $ \sigma_1(\mathbf{A})/\sigma_n(\mathbf{A})$, i.e., the ratio of largest to smallest singular values. Alternatively, if $\mathbf{A}$ has full column rank, then
\begin{equation}
    \kappa(\mathbf{A})
    = \| \mathbf{A} \|_2 \| \mathbf{A}^\dagger \|_2
    = \| \mathbf{A} \|_2 \| (\mathbf{A}^T\mathbf{A})^{-1}\mathbf{A}^T \|_2.
\end{equation}

We also find in Ref.~\cite[Theorem 5.7.1]{vanLoan1996matrix} that for underdetermined systems with full rank, the relative error in $\mathbf{t'}$ also has a magnitude of $\mathcal{O}(\kappa(\mathbf{A})\epsilon)$, provided we take the minimal-norm solution for both the original and perturbed systems, and $\epsilon<\sigma_m(\mathbf{A})$.

The accuracy of ghost imaging under pattern misalignment errors is therefore directly related to both the condition number of the patterns and the relative magnitude of errors introduced by pattern misalignment, i.e., $\| \mathbf{A'} - \mathbf{A} \|_2 / \| \mathbf{A} \|_2$. This can be estimated in a straightforward manner as the relative gradient magnitude of the patterns.

\begin{figure}
    \centering
    \begin{minipage}{0.2\textwidth}
        \centering
        \scriptsize{(a-i)}\\
        \includegraphics[width=0.8\textwidth]{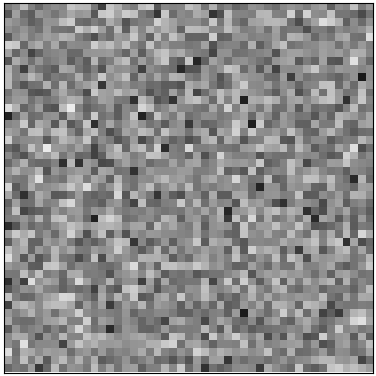}\\[1ex]
        \scriptsize{(a-ii)}\\
        \includegraphics[width=0.8\textwidth]{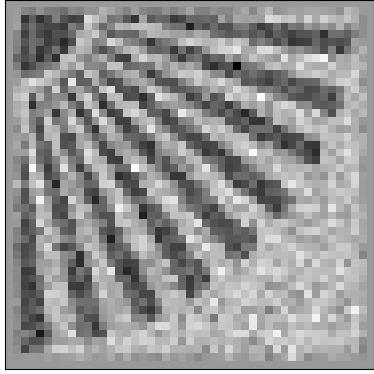}\\[1ex]
        \scriptsize{(a-iii)}\\
        \includegraphics[width=0.8\textwidth]{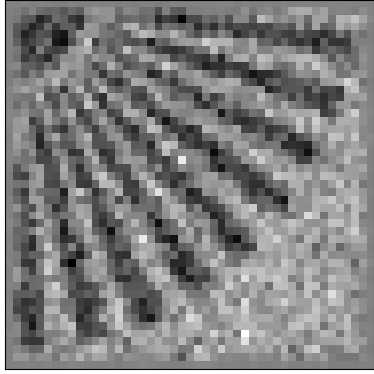}\\[1ex]
        \scriptsize{(a-iv)}\\
        \includegraphics[width=0.8\textwidth]{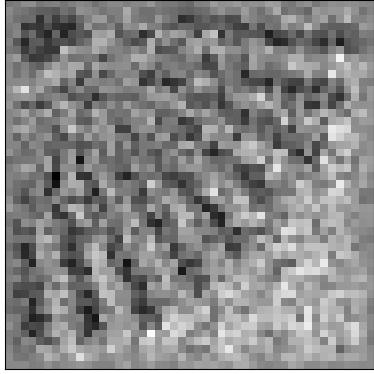}
    \end{minipage}%
    \begin{minipage}{0.2\textwidth}
        \centering
        \scriptsize{(b-i)}\\
        \includegraphics[width=0.8\textwidth]{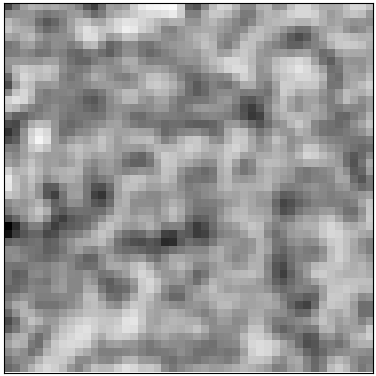}\\[1ex]
        \scriptsize{(b-ii)}\\
        \includegraphics[width=0.8\textwidth]{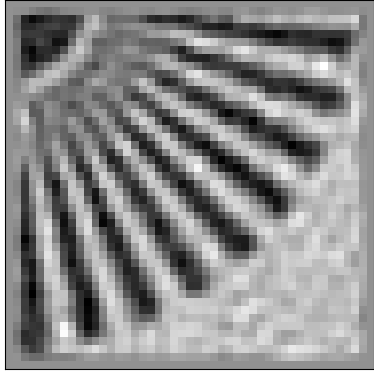}\\[1ex]
        \scriptsize{(b-iii)}\\
        \includegraphics[width=0.8\textwidth]{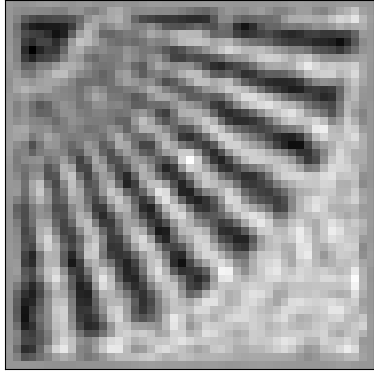}\\[1ex]
        \scriptsize{(b-iv)}\\
        \includegraphics[width=0.8\textwidth]{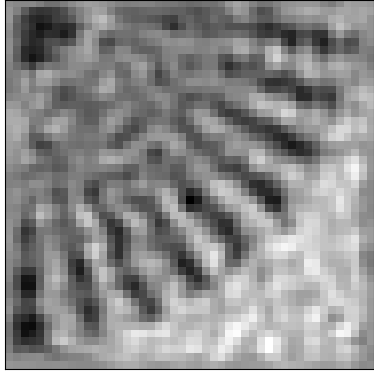}
    \end{minipage}%
    \begin{minipage}{0.2\textwidth}
        \centering
        \scriptsize{(c-i)}\\
        \includegraphics[width=0.8\textwidth]{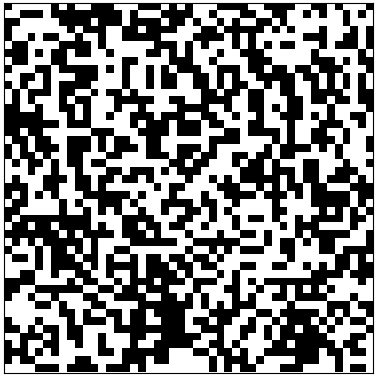}\\[1ex]
        \scriptsize{(c-ii)}\\
        \includegraphics[width=0.8\textwidth]{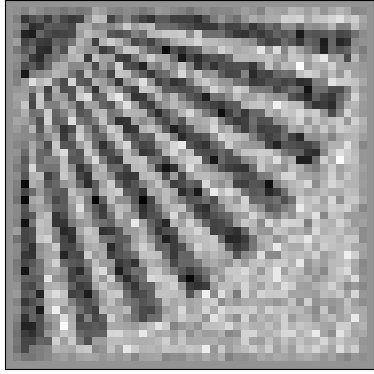}\\[1ex]
        \scriptsize{(c-iii)}\\
        \includegraphics[width=0.8\textwidth]{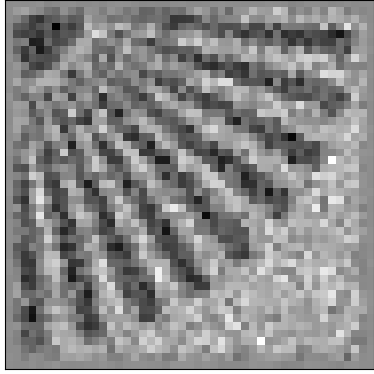}\\[1ex]
        \scriptsize{(c-iv)}\\
        \includegraphics[width=0.8\textwidth]{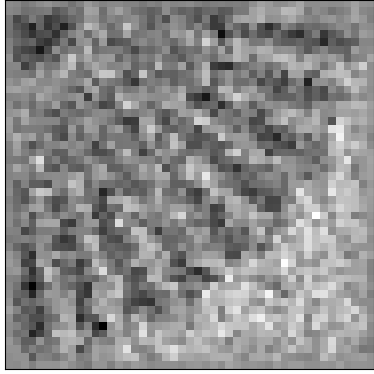}
    \end{minipage}%
    \begin{minipage}{0.2\textwidth}
        \centering
        \scriptsize{(c-i)}\\
        \includegraphics[width=0.8\textwidth]{egMaskBinaryBlur1RebinScanning47x47.png}\\[1ex]
        \scriptsize{(c-ii)}\\
        \includegraphics[width=0.8\textwidth]{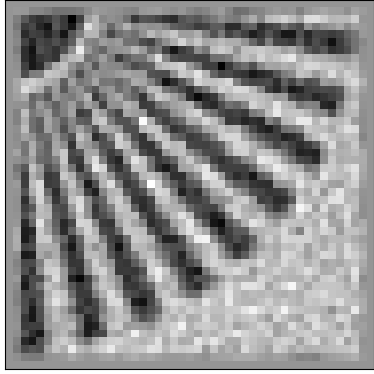}\\[1ex]
        \scriptsize{(c-iii)}\\
        \includegraphics[width=0.8\textwidth]{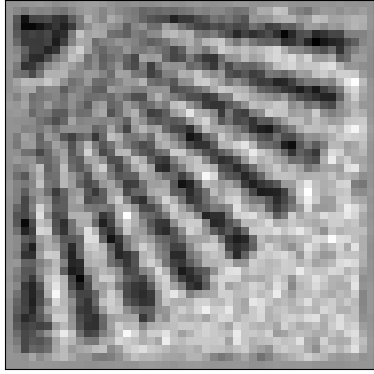}\\[1ex]
        \scriptsize{(c-iv)}\\
        \includegraphics[width=0.8\textwidth]{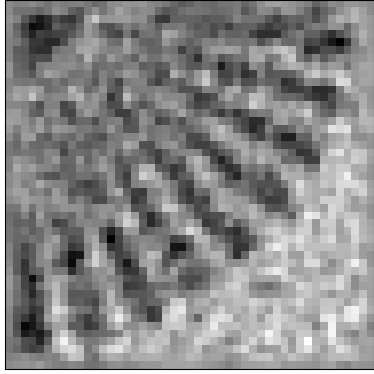}
    \end{minipage}%
    \begin{minipage}{0.2\textwidth}
        \centering
        \scriptsize{(c-i)}\\
        \includegraphics[width=0.8\textwidth]{egMaskBinaryBlur2RebinScanning47x47.png}\\[1ex]
        \scriptsize{(c-ii)}\\
        \includegraphics[width=0.8\textwidth]{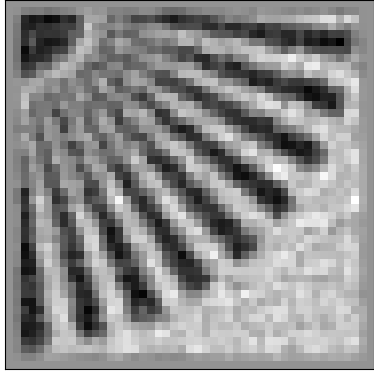}\\[1ex]
        \scriptsize{(c-iii)}\\
        \includegraphics[width=0.8\textwidth]{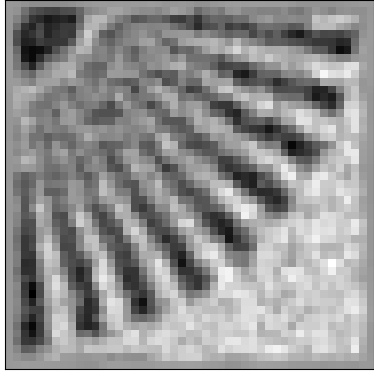}\\[1ex]
        \scriptsize{(c-iv)}\\
        \includegraphics[width=0.8\textwidth]{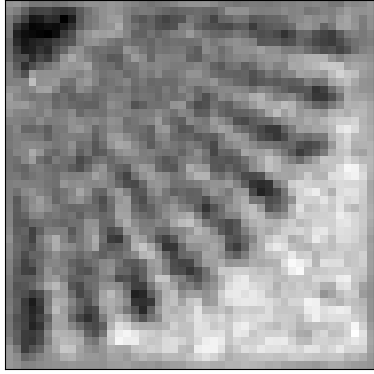}
    \end{minipage}
    \caption{Example (i) $47 \times 47$ pixel illumination patterns, (ii-iv) 4 iterations of Kaczmarz reconstruction with masks misaligned randomly with shifts $(x,y)$ having a normal distribution with $\sigma = 0.25$px, $0.5$px, and $1.0$px. All examples contain 2209 patterns in the set, and the grayscale window for the example patterns shown is [0,1]. (a) Gaussian random, (b) Gaussian distribution blurred by Gaussian with $\sigma = 1.0$px, (c) Binary random, (d) Binary random with medium feature sizes, (e) Binary random with large feature size. All patterns were generated as unique sets of random patterns, so mask {\it stride} had no effect, and a two pixel boundary was set to 0.5 transmission to minimize edge effects.}
    \label{fig:misalign}
\end{figure}


Figure \ref{fig:misalign} shows the effect of pattern misalignment on the ghost image achievable through Kaczmarz iteration. In these examples, each pattern was perturbed in both the $x$- and $y$-directions, with a random Gaussian distribution having $\sigma = 0.25$px, 0.5px, and 1.0px. The relative gradient magnitude of these sets of patterns are as follows: (a) 2.0, (b) 0.98, (c) 2.0, (d) 1.3, and (e) 0.99. Observe from the images reconstructed in (i-iii) that patterns with a higher relative gradient magnitude, namely (a) and (c), produce sharper but more noisy images that degrade significantly as the magnitude of misalignment increases. For (a) the average normalized MSE for the images exemplified in (iii) over 100 random perturbations degrades from (i) 0.29, (ii) 0.44, to (iii) 0.80, while for (c) NMSE degrades from (i) 0.24, (ii) 0.45, to (iii) 0.79. The patterns with a lower relative gradient magnitude, namely (b), (d), and (e), have less noise and better contrast that is maintained as the magnitude of misalignment increases. The most robust to misalignment according to the NMSE metric over 100 random perturbations is (b), with NMSE degrading from (i) 0.22, (ii) 0.31, to (iii) 0.56. The NMSE for the binary patterns with larger features was similar with (d) as (i) 0.23, (ii) 0.34, to (iii) 0.61, and NMSE for (e) as (i) 0.24, (ii) 0.36, (iii) 0.63. This correlates very well with the relative gradient magnitude and indicates that this is a useful metric to predict pattern performance under perturbation or misalignment.

\subsubsection{Robustness to slowly temporally varying illumination}
\label{sec:fluxVar}

\begin{figure}
    \centering
    \begin{minipage}{0.2\textwidth}
        \centering
        \scriptsize{(a-i)}\\
        \includegraphics[width=0.8\textwidth]{egMaskBinaryScanning47x47.png}\\[1ex]
        \scriptsize{(a-ii)}\\
        \includegraphics[width=0.8\textwidth]{egMaskBinaryScanning47x47.png}\\[1ex]
        \scriptsize{(a-iii)}\\
        \includegraphics[width=0.8\textwidth]{egMaskUraScanning47x47.png}
    \end{minipage}%
    \begin{minipage}{0.2\textwidth}
        \centering
        \scriptsize{(b-i)}\\
        \includegraphics[width=0.8\textwidth]{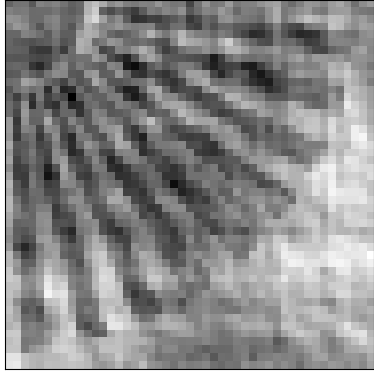}\\[1ex]
        \scriptsize{(b-ii)}\\
        \includegraphics[width=0.8\textwidth]{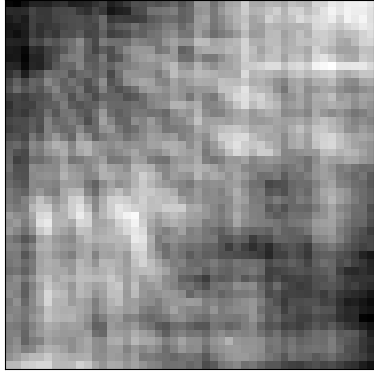}\\[1ex]
        \scriptsize{(b-iii)}\\
        \includegraphics[width=0.8\textwidth]{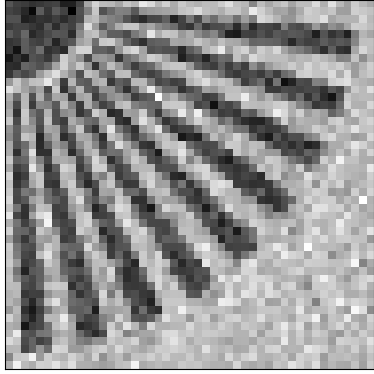}
    \end{minipage}%
    \begin{minipage}{0.2\textwidth}
        \centering
        \scriptsize{(c-i)}\\
        \includegraphics[width=0.8\textwidth]{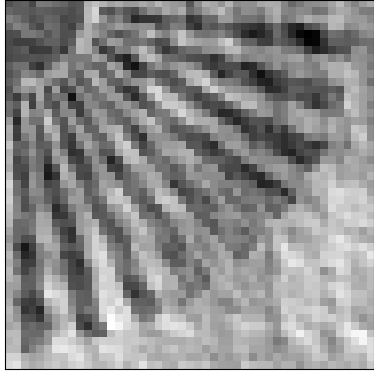}\\[1ex]
        \scriptsize{(c-ii)}\\
        \includegraphics[width=0.8\textwidth]{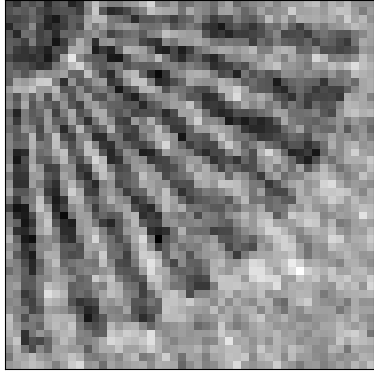}\\[1ex]
        \scriptsize{(c-iii)}\\
        \includegraphics[width=0.8\textwidth]{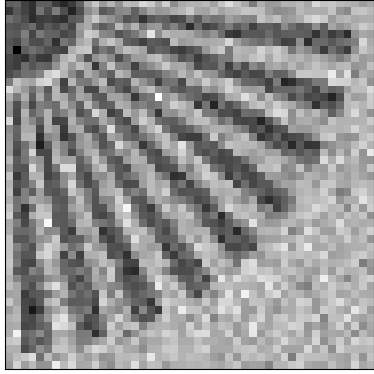}
    \end{minipage}%
    \begin{minipage}{0.2\textwidth}
        \centering
        \scriptsize{(d-i)}\\
        \includegraphics[width=0.8\textwidth]{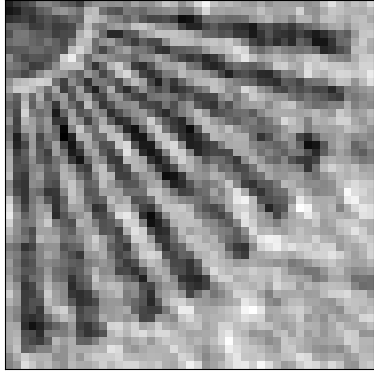}\\[1ex]
        \scriptsize{(d-ii)}\\
        \includegraphics[width=0.8\textwidth]{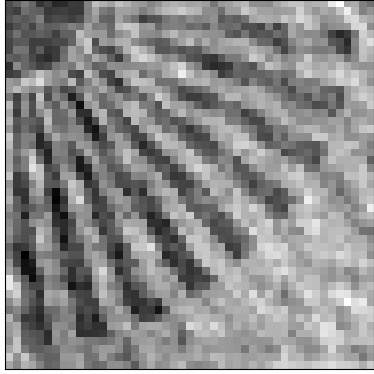}\\[1ex]
        \scriptsize{(d-iii)}\\
        \includegraphics[width=0.8\textwidth]{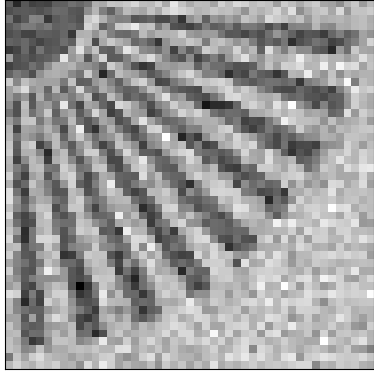}
    \end{minipage}%
    \begin{minipage}{0.2\textwidth}
        \centering
        \scriptsize{(e-i)}\\
        \includegraphics[width=0.8\textwidth]{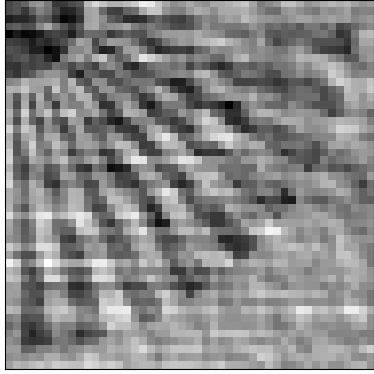}\\[1ex]
        \scriptsize{(e-ii)}\\
        \includegraphics[width=0.8\textwidth]{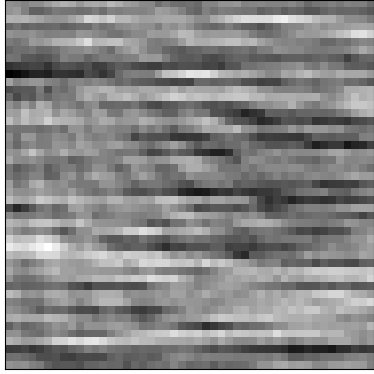}\\[1ex]
        \scriptsize{(e-iii)}\\
        \includegraphics[width=0.8\textwidth]{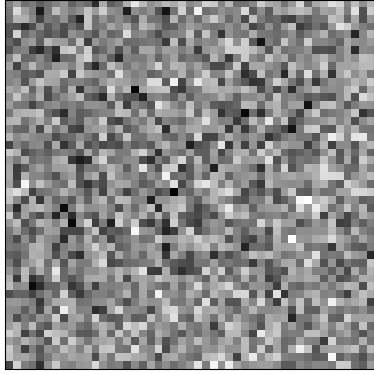}
    \end{minipage}%
    \caption{A $47 \times 47$ pixel demonstration of differential GI \cite{ferri2010differential} for (i) low-transmission object (0.0-0.25) with a scanning random binary mask (4418 patterns), (ii) high-transmission object (0.75-1.0) with a scanning random binary mask (4418 patterns), (iii) high-transmission object (0.75-1.0) with a scanning URA mask that is orthogonal under translation ($2209$ patterns). (a) Example mask pattern, (b) the result of adjoint GI with constant illumination, (c) the result of differential GI (or a single Landweber iteration given a constant initial estimate equal to the mean transmission of the object) with constant illumination, (d) the result of differential GI with slowly varying illumination intensity (patterns recorded simultaneously), (e) the result of differential GI with slowly varying  illumination intensity (patterns pre-recorded). All measurements had a Poisson distribution with an expected flux of 1000 photons/pixel/measurement. Intensity variation was had a standard deviation of approximately 1\%.}
    \label{fig:diffGI}
\end{figure}

A common experimental problem in optical GI is that of background signal from ambient light fluctuations. A technique that is commonly used to tackle this problem is that of differential ghost imaging (DGI). Originally it was observed that the background signal can be compensated for in a classical GI setup that uses a beamsplitter \cite{ferri2010differential}. In this case, given the measured illumination pattern, $A_j(x,y)$, at the pixelated detector, and assuming illumination $\alpha A_j(x,y)$ is incident on the object, the mean transmission of the object can be estimated as $b_j / ( \alpha \sum_x\sum_y A_j(x,y))$ for bucket measurement $b_j$. The background signal can then be compensated for by performing one Landweber iteration, with the initial estimate being a constant image equal to the mean transmission of the object over all measurements. DGI was shown to be superior to the conventional adjoint image recovery, particularly for high-signal images (or highly transmitting objects, see row (ii) of Fig.~\ref{fig:diffGI}). For constant experiment flux, this is only true when the mean transmission of the patterns varies from pattern to pattern; for sets of patterns that have a constant transmission, such as the URA and FRT orthogonal masks, DGI yields no improvement (see row (iii) of Fig.~\ref{fig:diffGI}). However, measuring the illumination pattern simultaneously to the bucket measurement means that any variation in background signal, illumination intensity, or a translating incident beam profile is captured to some degree, and DGI can compensate for this. A demonstration for variation in illumination intensity is given in column (d) of Fig.~\ref{fig:diffGI}, for both random binary and URA illumination patterns.

For computational GI variants, the illumination patterns, $A_j(x,y)$, are pre-recorded and are subject to different background signal, illumination intensity, or beam profile conditions than during the collection of bucket values. This can cause significant artifacts in the recovered ghost images. This has been demonstrated by simulation for illumination intensity variations in column (e) of Fig.~\ref{fig:diffGI}. In this scenario, it has been reported in the literature that employing differential patterns (i.e., positive and negative pairs of patterns) is beneficial, e.g., \cite{welsh2013fast, Sun--DifferentialComputationalGhostImaging2013,yu2016compressive, he2020energy}. Using differential patterns as a pair effectively enables {\it zero mean} patterns that minimize the effect of background signal and illumination flux variations. This also enables ternary (and even grayscale) patterns such as Haar wavelets to be implemented with maximum SNR or minimum dose.

\begin{figure}
    \centering
    \begin{minipage}{0.2\textwidth}
        \centering
        \scriptsize{(a-i)}\\
        \includegraphics[width=0.8\textwidth]{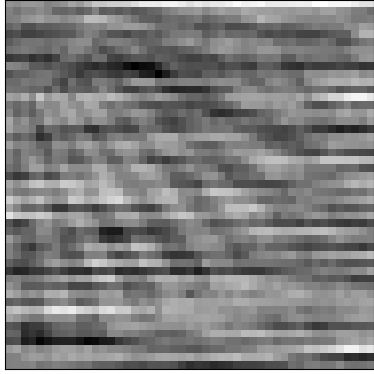}\\[1ex]
        \scriptsize{(a-ii)}\\
        \includegraphics[width=0.8\textwidth]{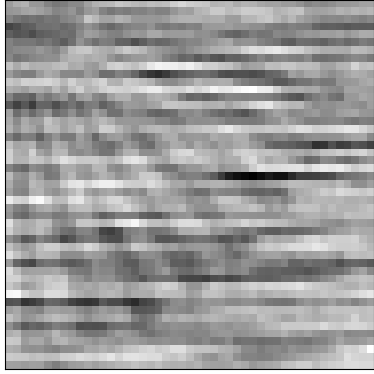}\\[1ex]
        \scriptsize{(a-iii)}\\
        \includegraphics[width=0.8\textwidth]{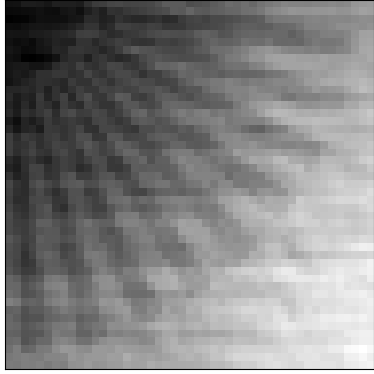}
    \end{minipage}%
    \begin{minipage}{0.2\textwidth}
        \centering
        \scriptsize{(b-i)}\\
        \includegraphics[width=0.8\textwidth]{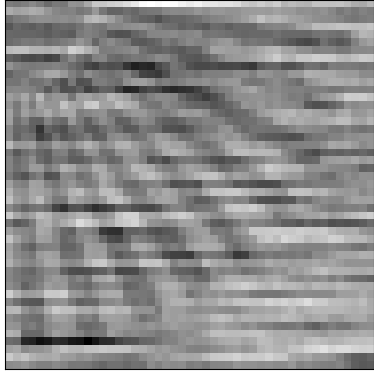}\\[1ex]
        \scriptsize{(b-ii)}\\
        \includegraphics[width=0.8\textwidth]{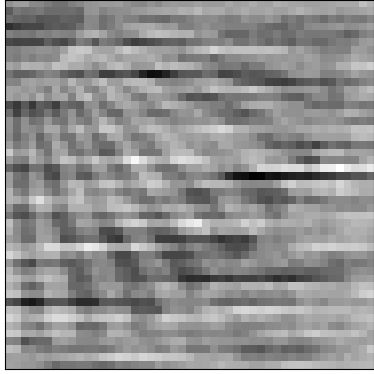}\\[1ex]
        \scriptsize{(b-iii)}\\
        \includegraphics[width=0.8\textwidth]{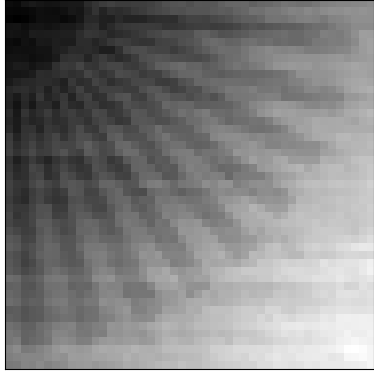}
    \end{minipage}%
    \begin{minipage}{0.2\textwidth}
        \centering
        \scriptsize{(c-i)}\\
        \includegraphics[width=0.8\textwidth]{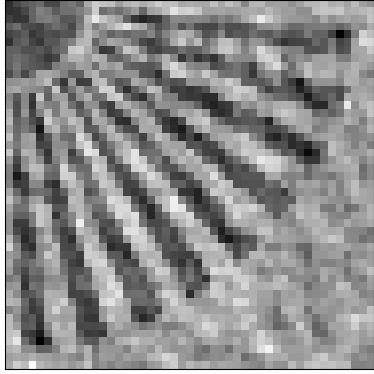}\\[1ex]
        \scriptsize{(c-ii)}\\
        \includegraphics[width=0.8\textwidth]{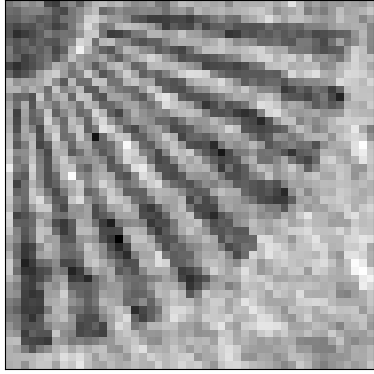}\\[1ex]
        \scriptsize{(c-iii)}\\
        \includegraphics[width=0.8\textwidth]{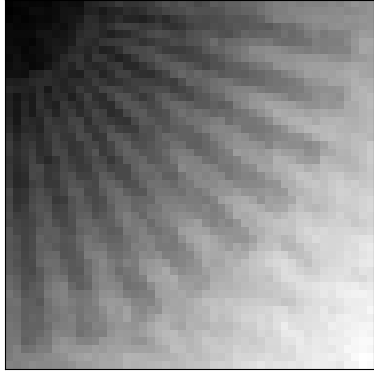}
    \end{minipage}%
        \begin{minipage}{0.2\textwidth}
        \centering
        \scriptsize{(d-i)}\\
        \includegraphics[width=0.8\textwidth]{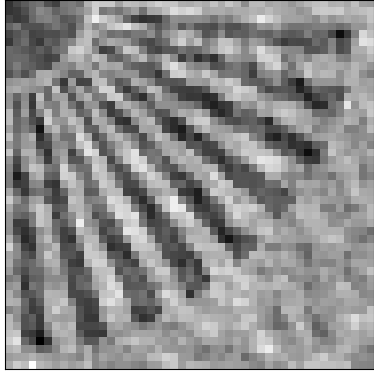}\\[1ex]
        \scriptsize{(d-ii)}\\
        \includegraphics[width=0.8\textwidth]{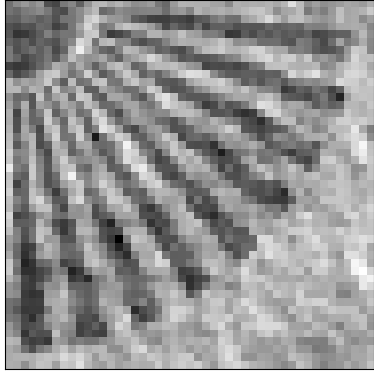}\\[1ex]
        \scriptsize{(d-iii)}\\
        \includegraphics[width=0.8\textwidth]{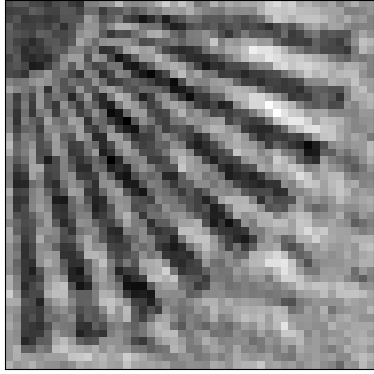}
    \end{minipage}%
    \caption{A demonstration of the effect of positive/negative mask pattern pairs \cite{Sun--DifferentialComputationalGhostImaging2013,welsh2013fast} using random binary masks of $47 \times 47$ pixels and given (i) slowly varying flux (standard deviation of 1\%); (ii) slowly varying background signal (same magnitude as in (i)); (iii) Gaussian beam profile with $\sigma = 47$px translating between in measurement with $\sigma=3$px in each direction. (a) GI using $6903$ unique patterns, (b) GI using $13\,806$ unique patterns, (c) GI using the $6903$ positive patterns from (i) and the corresponding $6903$ negative patterns, (d) GI using $6903$ positive/negative patterns with bucket measurements determined as difference of signal with positive and negative patterns. Each positive/negative mask pair was imaged in sequence. All measurements had a Poisson distribution with an expected flux of $1000$ photons/pixel/measurement.}
    \label{fig:fluxVariation}
\end{figure}

The advantages and requirements of using differential masks has been presented in Fig.~\ref{fig:fluxVariation}. Here the effect of several scenarios of incident and background illumination variations have been simulated. The result of computational GI with 6903 binary patterns contains severe artifacts (Fig.~\ref{fig:fluxVariation}a). Doubling the number of unique masks makes very little improvement (Fig.~\ref{fig:fluxVariation}b), however, adding $6903$ inverse patterns instead provides a significant advantage (Fig.~\ref{fig:fluxVariation}c-d). Here the positive/negative patterns are measured in sequence, i.e., as a pair. The differential masks create {\it zero mean} patterns removing the effect of flux variation. Under conditions of constant flux, no advantage is gained using differential masks; in fact, using more unique masks gives better performance. Also note that, if the measurements of a pattern and its inverse are separated over time, then there is no advantage.

While differential patterns are ideal for binary masks, and remove sensitivity to variations in illumination intensity, background signal, and even beam profile, there are several considerations when using a translating mask for this technique. Problems can arise in the differencing result if positive/negative masks are not aligned accurately. A mask with a lower gradient (as explored in Sec.~\ref{sec:misalignment}) may be a preferable choice here. A scanning mask (and its negative) would minimize the differential mask footprint, however, the positive/negative mask pairs must be imaged in sequence, i.e., close in time; this would require rapid and repeatable switching from one mask to the inverse that places strict requirements on translation stage speed, repeatability, and robustness.

To create positive/negative mask pairs for differential ghost imaging \cite{ferri2010differential,Sun--DifferentialComputationalGhostImaging2013,welsh2013fast}, one could employ microlithography techniques to manufacture masks whose transmission functions add to unity at each transverse location.  While feasible, such an approach has the disadvantage of requiring two distinct masks that need to be placed with sufficiently accurate relative positioning in the illuminating beam. A possible single-mask means for creating positive/negative mask pairs, in the x-ray domain, employs the phenomenon of magnetic x-ray circular dichroism (MXCD) \cite{LoveseyCollinsBook}.  MXCD refers to the sample-magnetization dependence of the absorption of circularly polarized x-rays.  Since left and right circularly polarized x-rays yield complementary absorption profiles for magnetic materials, a thin self-organized magnetic film with randomly-oriented domains will yield the required positive/negative transmission profiles, when  sequentially illuminated with left and right circularly polarized x rays \cite{eisebitt2004}.  Alternatively, a magnetic thin film composed of certain transition metals can be illuminated with fixed-helicity circularly-polarized x rays, and the illumination energy tuned in the vicinity of $L_{2,3}$ edges \cite{stohr1993,denbeaux2001}, to yield the required contrast reversal.  This gives a potential means for rapid switching from positive to negative illumination patterns, in the context of differential ghost imaging.  In this way, randomly-oriented magnetic domains in magnetic thin films can give non-fractal positive/negative random-mask pairs \cite{eisebitt2004, stohr1993,denbeaux2001}.  Also, if a magnetic thin film is rapidly quenched at an initial temperature that corresponds to the critical point for a thermodynamic phase transition, e.g.~if the magnetic thin film is well described by the two-dimensional Ising model at its critical temperature \cite{SethnaBook}, then the resulting random-fractal mask \cite{moghadam2022} can be employed to give positive/negative mask pairs for differential ghost imaging.

Regarding the utility of positive--negative mask pairs, it has been pointed out (in the context of differential ghost imaging \cite{ferri2010differential}) that these may be used to compute a differential bucket signal that can increase the signal-to-noise ratio of the resulting GI reconstruction \cite{Sun--DifferentialComputationalGhostImaging2013,welsh2013fast}.  Interestingly, our simulations suggest that, when a set of $2N$ random masks is replaced with $N$ pairs of positive--negative masks, with all other key parameters (such as the number of photons used to illuminate each mask) being kept fixed, but with the illumination flux varying slowly with time, the resulting GI SNR increases even when the resulting GI analysis does not specifically employ a differential-GI strategy.  From a geometric perspective, positive--negative mask pairs yield function-space vector pairs that are anti-parallel to one another for all but the constant-offset degrees of freedom, which improves the noise robustness of Eq.~(\ref{eq:VanillaGI Reconstruction}) since the vector sum, implied by this expression, contains pairs of parallel function-space vectors whose parallelism reduces the reconstruction's sensitivity with respect to noise and other imperfections that will necessarily be present in the GI data.

\subsubsection{Scalability}
\label{sec:scalability}

We define {\it scalability} as the ability to trade image resolution for robustness. For example, given noisy measured bucket values that generate a low-quality ghost image, can one improve the reconstructed image quality by increasing the pixel size and reducing the total number of pixels required to represent the masks and image?

The signal produced from GI with adjoint recovery is an increasing function of both the number of masks, $J$, and the mask variance, $\sigma_A^2$ (as confirmed in Sec.~\ref{sec:noise}) \cite{paganin2019writing, kingston2021inherent, ceddia2022a}. Intuitively, this is because GI works with the mean-corrected masks (as shown in Sec.~\ref{sec:background}) and the useful component of the bucket measurements is unrelated to the mask mean, $\mu_A$. In a multiscale setting we therefore have two competing effects when reducing pattern and imaging resolution: (1) there is an increase in measured information relative to data requirements since reducing the mask resolution reduces the unknowns in image recovery; (2) mask variance reduces when mask resolution is reduced and increases the data requirements for image recovery. To achieve scalability, we must minimize this second effect.

\begin{figure}
    \centering
    \begin{minipage}{0.2\textwidth}
        \centering
        \scriptsize{(a-i)}\\
        \includegraphics[width=0.8\textwidth]{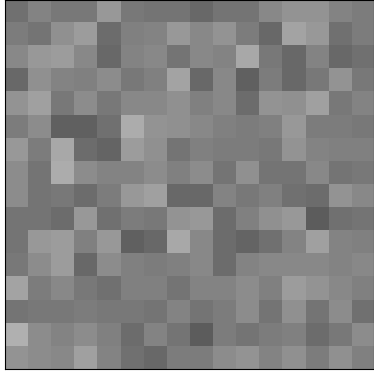}\\[1ex]
        \scriptsize{(a-ii)}\\
        \includegraphics[width=0.8\textwidth]{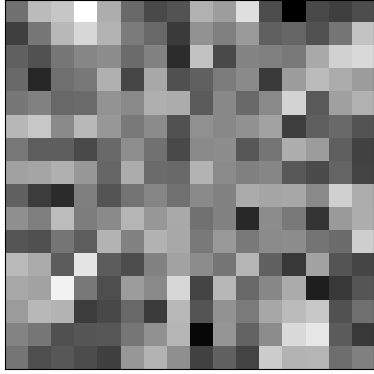}\\[1ex]
        \scriptsize{(a-iii)}\\
        \includegraphics[width=0.8\textwidth]{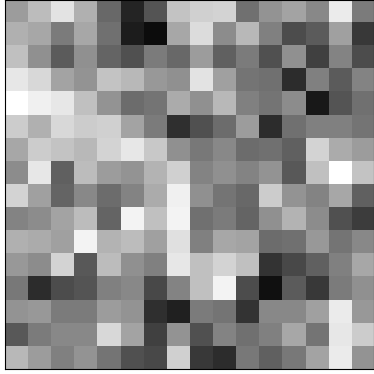}\\[1ex]
        \scriptsize{(a-iv)}\\
        \includegraphics[width=0.8\textwidth]{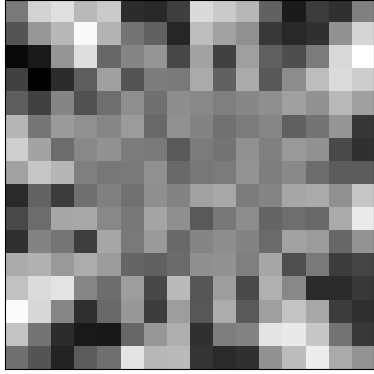}
    \end{minipage}%
    \begin{minipage}{0.2\textwidth}
        \centering
        \scriptsize{(b-i)}\\
        \includegraphics[width=0.8\textwidth]{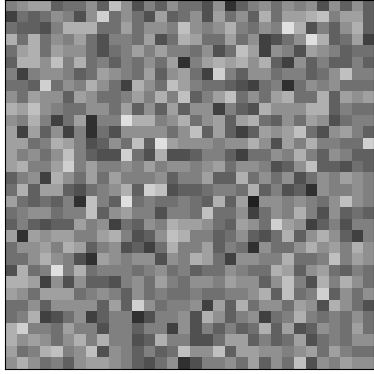}\\[1ex]
        \scriptsize{(b-ii)}\\
        \includegraphics[width=0.8\textwidth]{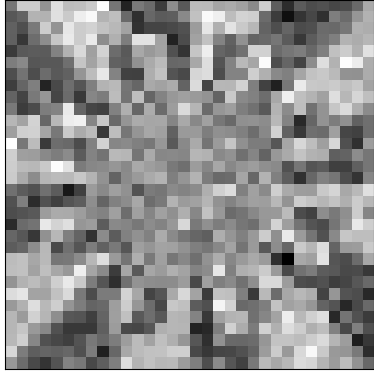}\\[1ex]
        \scriptsize{(b-iii)}\\
        \includegraphics[width=0.8\textwidth]{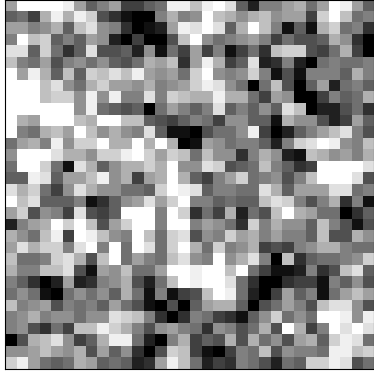}\\[1ex]
        \scriptsize{(b-iv)}\\
        \includegraphics[width=0.8\textwidth]{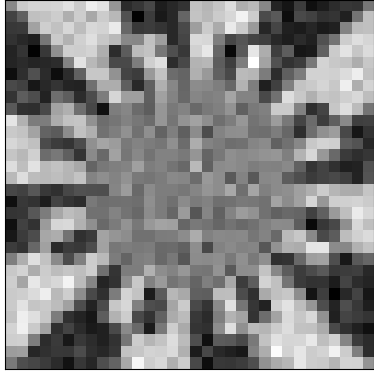}
    \end{minipage}%
    \begin{minipage}{0.2\textwidth}
        \centering
        \scriptsize{(c-i)}\\
        \includegraphics[width=0.8\textwidth]{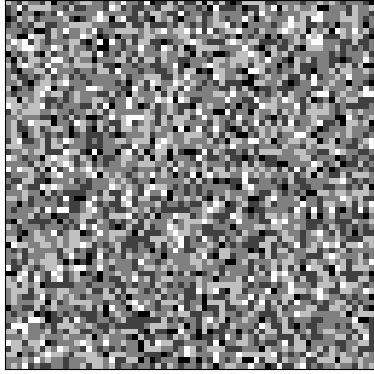}\\[1ex]
        \scriptsize{(c-ii)}\\
        \includegraphics[width=0.8\textwidth]{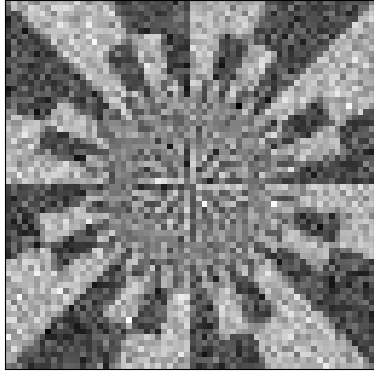}\\[1ex]
        \scriptsize{(c-iii)}\\
        \includegraphics[width=0.8\textwidth]{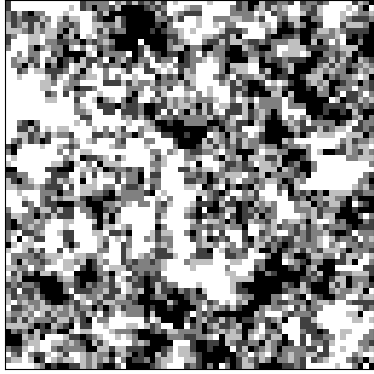}\\[1ex]
        \scriptsize{(c-iv)}\\
        \includegraphics[width=0.8\textwidth]{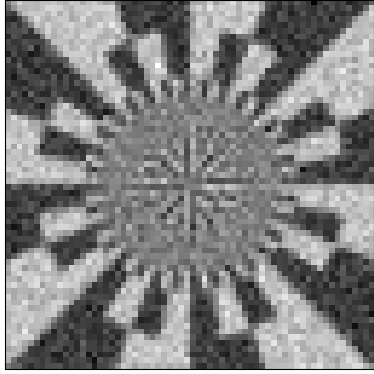}
    \end{minipage}%
    \begin{minipage}{0.2\textwidth}
        \centering
        \scriptsize{(d-i)}\\
        \includegraphics[width=0.8\textwidth]{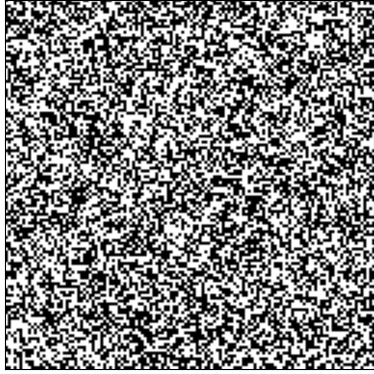}\\[1ex]
        \scriptsize{(d-ii)}\\
        \includegraphics[width=0.8\textwidth]{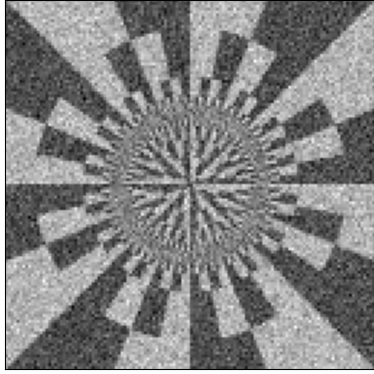}\\[1ex]
        \scriptsize{(d-iii)}\\
        \includegraphics[width=0.8\textwidth]{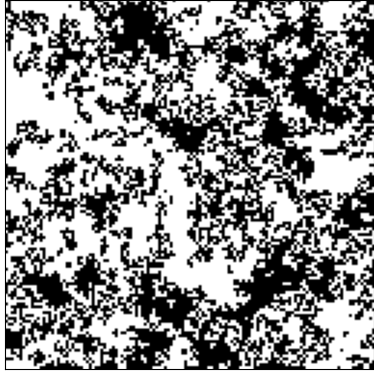}\\[1ex]
        \scriptsize{(d-iv)}\\
        \includegraphics[width=0.8\textwidth]{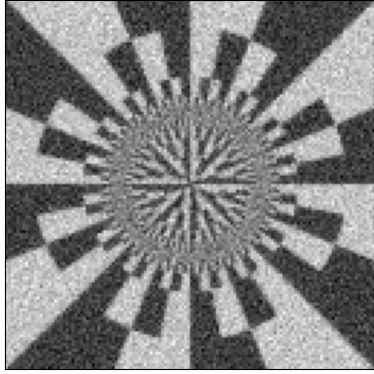}
    \end{minipage}
    \caption{Progressive multiscale GI recovery during an experiment with downsampling: (a) $8 \times 8$ pixel binning after 2048 measurements, (b) $4 \times 4$ pixel binning after $8192$ measurements, (c) half scale, i.e., $2 \times 2$ pixel binning after $32\,768$ measurements, and (d) full scale with $131\,072$ measurements. Two types of scanning masks with a stride of 32 pixels have been demonstrated: a binary mask (rows i-ii); a fractal mask (rows iii-iv). Example masks are given in rows (i) and (iii). The corresponding recovered ghost images using 4 Kaczmarz iterations ($\alpha = 0.25$) are presented in rows (ii) and (iv). Here a an incident flux of 8 photons/pixel was used per measurement.}
    \label{fig:multiscaleDemo}
\end{figure}

According to Parseval's theorem, the variance over an illumination pattern is equal to the sum of the variance of the discrete Fourier transform (DFT) of the pattern. Note that reducing resolution by pixel binning truncates the high-frequency DFT coefficients. We can approximate the effect by taking the volume under the DFT up to each binned frequency, which is a measure of how fast the DFT decays with radius. As an example, consider the case of random binary masks. A low resolution instance of the patterns generated by these masks can be obtained by binning each $2 \times 2$ pixel block into 1 larger pixel. In Fourier space, this equates to selecting the quadrant around the origin. Since a binary random signal has statistically uniform DFT coefficient magnitude, the binned mask variance is reduced by a factor 4. This directly compensates for the 4 times reduction in information required, resulting in no net gain. This has been demonstrated in row (i) of Fig.~\ref{fig:multiscaleDemo}.

To construct {\it scalable} patterns, we must weight the DFT coefficients relative to proximity $\sqrt{k_x^2 + k_y^2}$ to the origin. One straightforward technique is to use the random fractal masks presented in Sec.~\ref{sec:patterns}. Fractals are ideal for this since the statistical coefficient weighting decays with $1/\sqrt{k_x^2 + k_y^2}$ raised to a positive power (see Eq.~(\ref{eq:fractal})) (cf.~Refs.~\cite{PowerLawNoise1,PowerLawNoise2,PowerLawNoise3}). In this case, as high-frequency DFT coefficients are truncated with binning, the majority of signal (and thus mask variance) persists in the low-frequency region that remains. This has been demonstrated in row (iii) of Fig.~\ref{fig:multiscaleDemo}.

The benefit of this property is exemplified in Fig.~\ref{fig:multiscaleDemo}, which compares inverse GI using binary and fractal patterns. As the resolution is halved each time, 4 times less measurements are used. Observe that, since mask variance degrades much more slowly for fractal masks, the recovered GI quality also degrades more slowly relative to that for the binary mask. This can also be viewed as the ability to provide  progressive multiscale GI capability. As an experiment is being conducted, we can reconstruct higher and higher resolution images of the object. In the figure we have given GI results after measuring $2048$, $8192$, $32\,768$, and $131\,072$ measurements.

In this sense, scalable masks can be can be thought of as a lens or magnifier. When imaging an unknown object, fractals provide built in magnification options. One can scan the object with a small set of masks and recover a low resolution ghost image. If greater resolution is required, more mask positions and bucket measurements can be added until the desired resolution is achieved.

\begin{figure}
    \centering
    \begin{minipage}{0.2\textwidth}
        \centering
        \scriptsize{(a-i)}\\
        \includegraphics[width=0.8\textwidth]{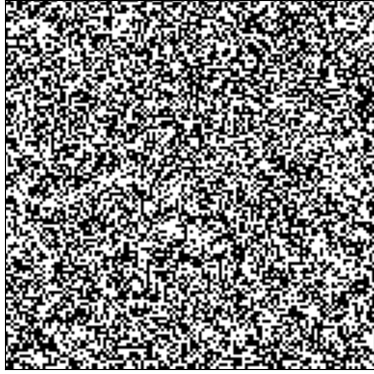}\\[1ex]
        \scriptsize{(a-ii)}\\
        \includegraphics[width=0.8\textwidth]{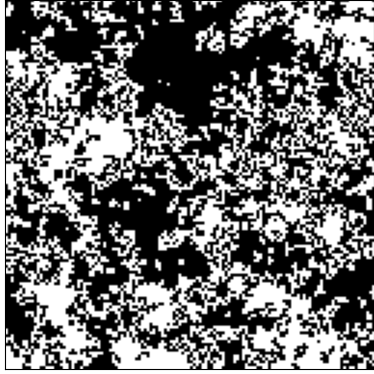}
    \end{minipage}%
    \begin{minipage}{0.2\textwidth}
        \centering
        \scriptsize{(b-i)}\\
        \includegraphics[width=0.8\textwidth]{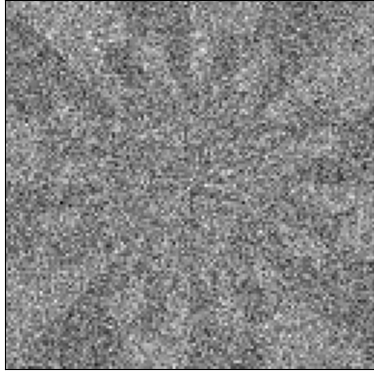}\\[1ex]
        \scriptsize{(b-ii)}\\
        \includegraphics[width=0.8\textwidth]{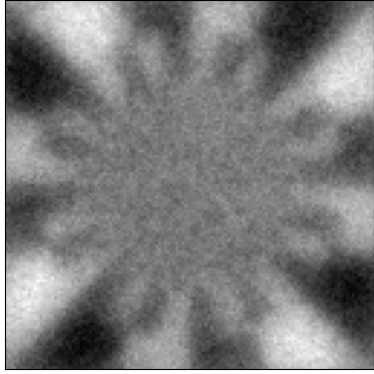}
    \end{minipage}%
    \begin{minipage}{0.2\textwidth}
        \centering
        \scriptsize{(c-i)}\\
        \includegraphics[width=0.8\textwidth]{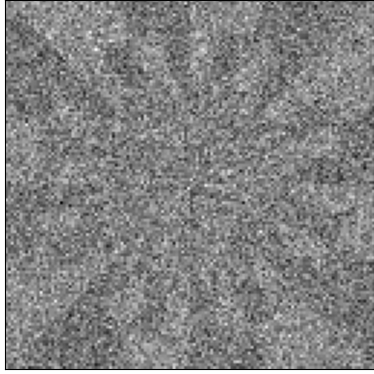}\\[1ex]
        \scriptsize{(c-ii)}\\
        \includegraphics[width=0.8\textwidth]{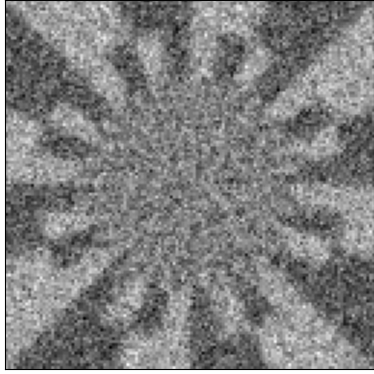}
    \end{minipage}
    \caption{$256 \times 256$ pixel GI recovery under extremely noisy circumstances: (a) example illumination patterns, (b) adjoint GI, (c) inverse GI through 100 Kaczmarz iterations ($\alpha = 0.001$). Two types of scanning masks with a stride of 32 pixels have been demonstrated: a binary mask (row i); a fractal mask (rows ii). Here an incident flux of only 0.08 photons/pixel was used per measurement.}
    \label{fig:multiscaleDemo2}
\end{figure}

Another application of scalability arises in scenarios with extremely low SNR. It is very difficult for the {\it unstable} iterative GI inversion schemes to improve on the {\it stable} GI adjoint image recovery result. See row (i) of Fig.~\ref{fig:multiscaleDemo2} for example. However, scalable masks have traded resolution for robustness, i.e., high-resolution object features are lost in noise more quickly than for random masks, but the low-resolution features remain even in extremely noisy scenarios. The relative over-representation of low-frequency information means adjoint GI images are extremely robust and it is possible to extract higher-frequency information beyond this. This has been demonstrated for fractal masks in row (ii) of Fig.~\ref{fig:multiscaleDemo2}.

Fractal masks have many useful properties that are ideal for the field to develop practical experimental techniques and protocols. Random fractals give a single fabricated mask that can be used for a range of spatial scales, enabling GI at progressively higher resolutions as an experiment continues.  Moreover, such masks are extremely robust to noise and other experiment deficiencies (such as mask misalignment). As the field matures it is likely that more application-specific masks will be adopted, however, scalable masks are very appealing until that time. Perhaps they could be used in tandem with dedicated designed masks to enable a broad scope of the object to determine which designed mask is most suitable.  They might also be used to zoom in to regions of interest, to identify where to focus high-resolution GI.

As we have already pointed out, natural materials with fractal-like structures do exist. Some potential mechanisms that produce these properties were discussion in Sec.~\ref{sec:patterns}. A topic for future research would be to explore the possibility of fractal masks that are orthogonal under translation. We should also point out that, by design, the Hadamard masks are also ideal for this scalable property, having patterns ranging from low to high spatial frequencies.

\subsubsection{Optimizing dose fractionation}
\label{sec:doseFrac}

The question of {\it dose fractionation} is as follows. Given a total experiment time, or total illumination dose, how many measurements, $J$, should the time (or dose) be divided into, i.e., what fraction of the total dose should be assigned to each measurement? The term originates from computed tomography (CT) \cite{hegerl_zn_1976} where it was determined that many measurements with low dose is preferable to few high-dose measurements. The situation is similar for GI, with a trade-off between the number of patterns employed and spatial resolution.

This topic has been explored previously, (e.g., \cite{Ceddia2018, gureyev2018, lane2020advantages}), where various types of experimental noise were considered. In Ref.~\cite{kingston2021inherent}, the degradation of an $n \times n$ pixel adjoint recovered ghost image, $\widehat{T}$, compared with the ideal image, $T$, was quantified by the root mean square error (RMSE) as follows:
\begin{equation}\label{eq:rmse}
    \mbox{RMSE}(\widehat{T},T) = \sqrt{ \frac{1}{n^2}\sum_x \sum_y \left(\widehat{T}(x,y) - T(x,y)\right)^2 }.
\end{equation}

There are several main contributions to this metric: artifacts from insufficient patterns or measurements, denoted $\mbox{RMSE}_0(\widehat{T},T)$ and $\mbox{RMSE}^\perp_0(\widehat{T},T)$; photon shot-noise represented by a Poisson distribution, denoted $\mbox{RMSE}_p(\widehat{T},T)$; and per-measurement noise modeled as a Gaussian distribution, denoted as $\mbox{RMSE}_m(\widehat{T},T)$. These contributions are added in quadrature, i.e.,
\begin{equation}
    \mbox{RMSE}(\widehat{T},T) = \sqrt{ \mbox{RMSE}^2_0(\widehat{T},T) + \mbox{RMSE}^2_p(\widehat{T},T) + \mbox{RMSE}^2_m(\widehat{T},T)}.
\end{equation}
A model for each type of error introduced was presented in Ref.~\cite{kingston2021inherent} as a function of pattern, image, and experiment properties. It was observed that $\mbox{MSE}_0$ decreases with $J$, while $\mbox{MSE}_p$ remains constant for a fixed experiment time, thus given photon shot-noise only, the more patterns the lower the RMSE. This aligns with the dose fractionation theorem for CT \cite{hegerl_zn_1976}. However, a per-measurement noise introduces a consequence for each additional measurement and $\mbox{MSE}_m$ increases with $J$. When this noise is present, a trade-off must occur and an optimum number of measurements can be calculated.

Assuming GI illumination patterns are produced by scanning subsets of a larger mask, either a designed or naturally random mask, one question that requires future research is that of subset-mask position dependence. Is it sufficient to scan subsets over a local region of a mask? Conversely, is it always better to distribute the subset positions across the entire mask as much as possible? Is a grid of positions desirable? Conversely, are pseudo- or quasi-random positions preferable in order to minimize artifacts?

\subsubsection{Mask fabrication effects}
\label{sec:fabErrs}

Given a design for a binary mask, the actual fabrication process can produce something very close to that designed, but there will be some discrepancies due to the physical processes involved. The available fabrication methods introduce constraints on the lateral mask resolution, the choice of material, the aspect ratio and the uniformity of the mask patterns. Additionally, each fabrication step can produce an error in the final result and reduce the efficiency of the mask. A process with the minimum fabrication steps is desirable.

An efficient process (in terms of cost, time, and fabrication error) for making a binary mask with micrometer resolution is to use a combination of photolithography and electroplating techniques. A summary of the technique is as follows:
\begin{enumerate}
    \item A thin metal film, e.g., Au, is first deposited on a substrate (which can be Si or SiO$_2$). This seed, or conductive, layer is used as a conductive layer for a subsequent  electroplating process. 
    \item A photolithography process is then applied in three steps: 
        \begin{itemize}
        \item A photoresist (positive or negative) is coated on the metal film. The thickness of the photoresist depends on the actual thickness (height) of the design mask.
        \item The pattern of the mask is transferred from a hard mask into the photoresist layer under UV exposure. 
        \item A developing solution is used to remove the exposed area (positive photoresist case) or unexposed area (negative photoresist case). 
        \end{itemize}
    \item After this step the trenches have been created and are then filled by a metal, e.g., Au, through the electroplating process. 
    \item In the final step, the remainder of the photoresist is removed through a lift-off process.
\end{enumerate}

Each fabrication step can have effects on the fabricated mask patterns. For instance, (i) development time is different for different structure sizes and densities, and (ii) if a given mask design is fractal in nature or contains different masks with different resolution requirements (i.e., different feature sizes), some parts of the pattern might be over developed. This causes the lateral pattern resolution to be slightly larger than the actual design. This effect can be minimized by designing a mask with consistent feature sizes. For example, the effect would be insignificant if the resolution of the masks varies from 10 $\mu$m to 15 $\mu$m. This effect can also be ignored for random masks, where the requirements on mask structure are not critical. Given these two considerations, these effects have not been simulated below.

A more important fabrication defect may accrue in the electroplating process. First of all, there is a limitation on the choice of material. Electroplating is only readily available for a limited number of materials, such as Ni and Au. Ni is cheaper and more available compared to Au, however it is not always the best option, especially when using hard x rays. The thickness (or height) of the electroplated material should be adjusted for a desired photon energy. For instance, to achieve 40\% transmission of 25keV x rays, a mask requires a thickness of 75 $\mu$m of Ni or 14 $\mu$m of Au. While gold is more expensive, it is less challenging in terms of aspect ratio and mechanical stability.

Beside these challenges and limitations, there are a range of electroplating parameters that affect the uniformity of the electroplated patterns, including the bath size and temperature, the solution pH and concentration, the movement of electrolyte, and the distance between cathode and anode. All of these parameters should be adjusted to obtain uniform electroplated structures. In addition, higher current is recorded toward the edge of the substrate/wafer compared to the center \cite{tan2003understanding}. This means that, if we have a full wafer of structures, their height is slightly higher at the edges than the center of the wafer. This is modeled below as a broad Gaussian profile, with the edges being slightly thicker than the center. As a result, the transmission through the material across the wafer is not uniform.

Another factor, that affects total transmission through the mask, is the choice and thickness of the seed layer. Its thickness reduces the transmission throughout the wafer. This layer should be thin enough to maximize the transmission and thick enough to be a strong conductive layer for the electroplating process. Generally, an Au film with a thickness of 50 nm to 100 nm is sufficient for this purpose. In addition, since the adhesion between gold and most substrates is low, a thin adhesion layer (usually Cr or Ti) is deposited between the substrate and the gold layer. This adhesion layer can be as thin as 10 nm, which is almost transparent for high-energy x rays. Thus, it can be neglected in hard x-ray transmission calculations.

\begin{figure}
    \centering
    \begin{minipage}{0.2\textwidth}
        \centering
        \scriptsize{(a-i)}\\
        \includegraphics[width=0.8\textwidth]{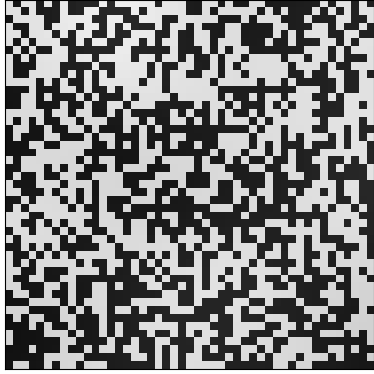}\\[1ex]
        \scriptsize{(a-ii)}\\
        \includegraphics[width=0.8\textwidth]{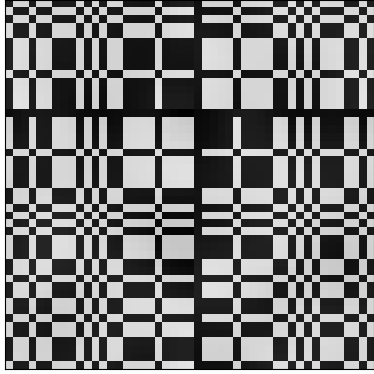}
    \end{minipage}%
    \begin{minipage}{0.2\textwidth}
        \centering
        \scriptsize{(b-i)}\\
        \includegraphics[width=0.8\textwidth]{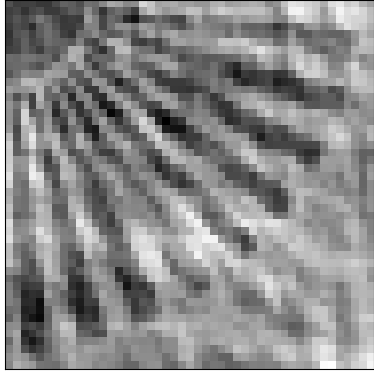}\\[1ex]
        \scriptsize{(b-ii)}\\
        \includegraphics[width=0.8\textwidth]{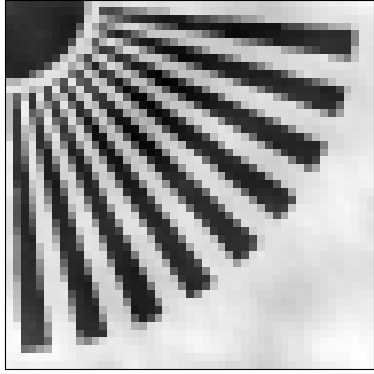}
    \end{minipage}%
    \begin{minipage}{0.2\textwidth}
        \centering
        \scriptsize{(c-i)}\\
        \includegraphics[width=0.8\textwidth]{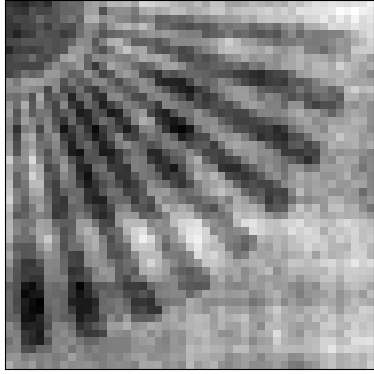}\\[1ex]
        \scriptsize{(c-ii)}\\
        \includegraphics[width=0.8\textwidth]{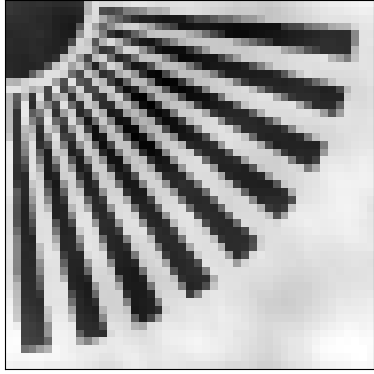}
    \end{minipage}%
    \begin{minipage}{0.2\textwidth}
        \centering
        \scriptsize{(d-i)}\\
        \includegraphics[width=0.8\textwidth]{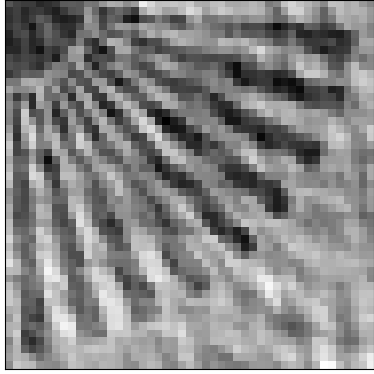}\\[1ex]
        \scriptsize{(d-ii)}\\
        \includegraphics[width=0.8\textwidth]{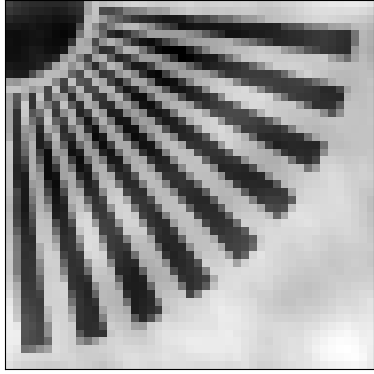}
    \end{minipage}%
    \begin{minipage}{0.2\textwidth}
        \centering
        \scriptsize{(e-i)}\\
        \includegraphics[width=0.8\textwidth]{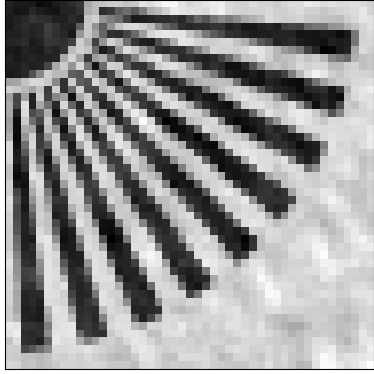}\\[1ex]
        \scriptsize{(e-ii)}\\
        \includegraphics[width=0.8\textwidth]{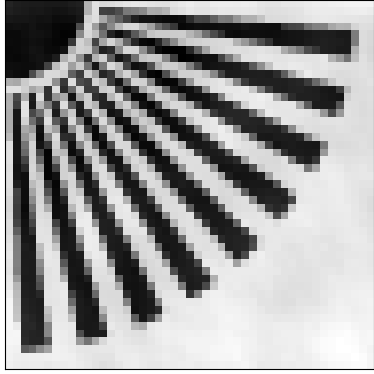}
    \end{minipage}
    \caption{The effects of including mask fabrication details, as described in the text, for (i) a scanning binary mask using $4418$ measurements, and (ii) a scanning orthogonal URA mask using all $2209$ measurements. (a) Example $47 \times 47$ pixel illumination patterns, (b) adjoint GI assuming the patterns are perfect, (c) 4 Kaczmarz iterations assuming the patterns are perfect, (d) adjoint GI having measured the patterns, and (e) 4 Kaczmarz iterations having measured the patterns. }
    \label{fig:fabError}
\end{figure}

As an example case study, we calculated total transmission through all layers for a fabricated mask and included fabrication defects in simulation. Given a mask fabricated on a 500 $\mu$m SiO$_2$ substrate with a 20 nm Cr adhesion layer and a 100 nm Au seed layer,  the total transmission of 25 keV x rays is 0.834. If the patterns are fabricated with 30 $\mu$m thick electroplated Au, the transmission through the patterns would be 0.0778. Hence, the maximum and the minimum transmission throughout the fabricated mask would be approximately 83.4\% and 6.5\%, respectively. These transmission effects along with the effect of nonuniformity from the electroplating process are shown in Fig.~\ref{fig:fabError}. Here the maximum and the minimum transmission through a random mask and a URA mask are set to 85\% and 5\%, respectively. A Gaussian profile is applied to the transmission of the mask, increasing transmission by 8\% in the center of the overall mask. Moreover, a slow variation to the mask transmission with 2\% standard deviation is added to the simulation, to compensate for the nonuniformity of the electroplating process.


The results in Fig.~\ref{fig:fabError} demonstrate the effects in GI recovery when a binary mask is assumed (Fig.~\ref{fig:fabError}b-c) and when the defects are characterized (Fig.~\ref{fig:fabError}d-e). For the binary masks GI is still possible when incorrectly assuming perfect binary patterns and Kaczmarz iteration is able to improve fidelity over adjoint GI, however, artifacts have become prominent. Since random masks structure is not critical, inverse GI after mask characterization works well and gives the best NMSE overall. For the URA masks that are orthogonal under translation (given ideal fabrication), assuming perfect binary patterns still exhibits robust GI recovery. Some low-frequency artifacts are present, but these results have greater fidelity than for the random masks. This is also predicted by Eq.~\eqref{eq_perturb}, where the errors in the mask $\epsilon$ are magnified by the condition number in  reconstruction (and orthogonal masks such as URAs have minimal condition number). The majority of the orthogonality properties remain despite fabrication errors. It is interesting to note that mask characterization does not improve the result significantly, since the orthogonality properties that are lost are not recoverable by characterization. We do note that Kaczmarz iteration was able to reduce some of the artifacts and produced the best result overall.

\section{Summary of findings}
\label{sec:summary}

We explored several factors that are important when considering potential masks, either natural or designed, to be used for producing structured illumination in classical computational ghost imaging. Our main motivation, for such a mask-based approach, is classical ghost imaging using radiation and matter wavefields---such as x rays, neutrons, electrons, muons, etc.---for which dynamically configurable high-resolution beamshaping elements (i.e.~the analog of a visible-light spatial light modulator) do not exist. Below, we summarize our key findings. 

\begin{itemize}
    \item Given a translating mask, the translation (or stride)---used between each pattern subset illuminated---should be selected to be greater than the resolution to which the mask is characterized.
    \item In an ideal scenario, feature size (to a certain extent) is not important. The important attribute is the sharpness of the features in the patterns. 
    \item The robustness of adjoint reconstruction can be improved by making the set of patterns closer to orthogonal. This is achieved by setting the pattern translation to be greater than or equal to the minimum feature size.
    \item The stable rank of a set of patterns was defined, giving an indication of the relative performance of the sets of patterns. This quantity can be used to compare masks or patterns, with preference to the masks with higher rank.
    \item Several methods to estimate the optimal mask stride were presented. These methods explored the power spectrum, as well as simulated GI MSE and pattern stable rank as a function of stride.
    \item Translating masks are optimal in terms of mask footprint. Fabricated masks that are orthogonal under translation only require 4 times the area of the imaging FOV.  This may be compared with footprints that are thousands of times the area of the FOV, for sets of unique orthogonal patterns such as the Walsh-Hadamard set. Differential masks remain difficult to fabricate and perform experiments with, even for translating masks.
    \item When considering robustness to photon shot-noise, the mask variance is a key indicator, with greater variance giving better results. This is linked to the magnitude of unnormalized SVD curves of the set of patterns. These curves provide a good indication of the noise level at which full resolution is no longer feasible, as well as the degradation severity as a function of increased noise.
    \item The L2 norm of the mask (or pattern) image gradient is a good indication of sensitivity to mask misalignment errors. A mask with a larger L2 norm is more susceptible. This was verified with simulations of decreasing alignment precision.
    \item Differential GI (as defined in Ref.~\cite{ferri2010differential}), where the illumination patterns and bucket values are recorded simultaneously, is a good method to deal with slowly varying illumination intensity or background illumination.
    \item Creating positive/negative mask pairs for differential ghost imaging \cite{Sun--DifferentialComputationalGhostImaging2013,welsh2013fast} significantly improves GI robustness in scenarios with slowly varying illumination intensity, background illumination, or beam position. Each positive/negative pattern must be imaged as a pair in sequence.
    \item The concept of mask {\it scalability} was introduced and exemplified through random fractal masks. Fractal masks perform well at several magnifications and we believe this type of mask should be useful in GI development.
    \item When considering the number of masks to use for a given experiment duration in GI---i.e., the question of dose fractionation---more masks are better, given photon shot-noise only. However, if each measurement includes a significant cost in terms of Gaussian noise, then an optimal number of masks does exist.
    \item The degradation in GI quality due to mask fabrication defects does not seem to be too significant, provided the patterns are characterized experimentally. Orthogonal masks perform reasonably well, even if assuming ideal patterns in GI recovery.
    
\end{itemize}

We emphasize that in this work, we have assumed no prior knowledge of the sample or object to be imaged. It is likely that many of the preceding conclusions will be different in situations where one has prior knowledge of the sample.  This latter case enables tools such as compressed sensing, maximum {\em a posteriori} methods, or deep-learning methods to be employed.

\section{Recommendations}
\label{sec:recommendations}

We now interpret the above summary in practical terms. In particular, below we make recommendations as to how to employ this information, using some example scenarios.    We organize these recommendations under the three key questions that are given as headings below. 

{\bf Which mask, or set of patterns, should I choose and why?}

{\em (A) Selecting a natural mask or designing and fabricating a mask:} Natural masks are a practical choice and are sufficient for initial GI experiments. However, to push the limits of GI in areas such as resolution or dose reduction, fabricated masks are recommended.

{\em (B) Imaging resolution requirement:} If the resolution required is unknown beforehand, we recommend the use of fractal masks as they function well over a range of length scales. The possibilities for natural masks with fractal properties is outlined in Sec.~\ref{sec:patterns}. If the resolution requirement is known, use a a mask with features on the order of this size, although they can be slightly larger. Avoid a mask with features that are too regular (or crystalline), e.g., using a periodic grid or monodisperse beads, as this will not enable sufficient unique patterns. We can assess this property by comparing the stable rank of the pattern set or a {\it ring} appearing in the PSF with a radius related to the characteristic length of the mask.

{\bf What other experimental conditions affect, or further refine, my choice of mask?}

{\em (C) Unstable illumination flux, beam profile, or background illumination:} Synchrotrons provide an example of this instability, wherein the accelerator electron beam current (and thus x-ray flux) slowly reduces over time and is periodically ``topped-up''. The change in flux over time can be significant, e.g.~on the order of 10\%. If one is recording illumination patterns simultaneously with the bucket values, employing differential GI image reconstruction \cite{ferri2010differential} is able to tolerate this instability in illumination. If one is undertaking a form of computational GI, then either (i) use differential masks (i.e., positive/negative pairs \cite{Sun--DifferentialComputationalGhostImaging2013,welsh2013fast}), or (ii) monitor these changes somehow. Some possibilities for implementing differential masks are discussed in detail in Sect. \ref{sec:fluxVar}.

{\em (D) Significant noise:} An example of this is when limited experiment time is allocated or dose reduction is being explored. In this case, choose masks with the highest variance, i.e., maximal contrast. Binary masks are optimal from this point of view. For a moderate amount of noise, choose the mask with the greatest area under the unnormalized SVD curve. For the case of extremely high noise levels, choose masks with the largest unnormalized singular values, such as a fractal mask.

{\em (E) Inaccurate mask positioning:} If the accuracy of the x-y translation or rotation stages used to move the mask about the imaging FOV is poor, choose masks that have a lower L2 norm of the pattern gradient, i.e., smoother masks or masks with larger features.

{\bf What are the experimental requirements for my selected mask?}

{\em (F) Determining mask translation or stride:} Transverse mask translation between illumination patterns should be greater than the characteristic length scale of the patterns. This stride can be found by identifying the maximum in the Fourier power spectrum of an example pattern image (or the other methods presented in Sec.~\ref{sec:psf}).

{\em (G) Determining the number of patterns to measure in an allocated time:} If per-measurement noise, e.g., electronic readout noise, is not significant then more patterns produce a better GI result (subject to practical considerations like dead time between measurements). Otherwise, there is an optimal number of measurements that should be determined, e.g., using the methods described in Ref.~\cite{kingston2021inherent}.


\section{Conclusion}
\label{sec:conclusion}

The quality of a ghost image recovered from an experiment is highly subject to the properties of the illumination structures used, i.e., the set of patterns employed. For optical GI, it is straightforward to use a spatial light modulator to produce any set of patterns, whereby desirable properties such as orthogonality can be ensured. Given weakly interacting radiation such as x rays and neutrons, the patterns are typically produced through the use of an attenuation or refraction mask. In this paper we have outlined methods to determine which patterns are preferable over others, to aid mask selection or which mask design to favor, subject to imaging requirements and experimental conditions. We explored questions such as resolution requirements, noise levels, mask positioning accuracy, illumination intensity stability, optimal mask translation, optimal number of mask positions, and scalability. The findings have been summarized and the practical process of selecting a mask has been proposed in the discussion section. Note, when applying these recommendations, that this study was undertaken assuming no knowledge of the object to be imaged. Different rules-of-thumb will likely apply when knowledge of the object is leveraged to minimize the number of measurements recorded. A similar study for this scenario is the subject of future work.

\section{Acknowledgments}
AMK, DP and DMP acknowledge funding via Australian Research Council Discovery Project ARC DP210101312.

\bibliographystyle{unsrt}
\bibliography{GiMasks}

\end{document}